\documentclass[fleqn,usenatbib]{mnras}

\usepackage{mathptmx}

\usepackage[T1]{fontenc}
\usepackage{ae,aecompl}

\usepackage{graphicx}	
\usepackage{amsmath}	
\usepackage{amssymb}	
\usepackage{afterpage}
\usepackage{multirow}
\usepackage{subfigure}
\usepackage{csvsimple}	
\usepackage{pdflscape}	
\usepackage{xcolor}

\newcommand{\hersc}{{\it Herschel}}
\newcommand{\spitz}{{\it Spitzer}}
\newcommand{\chand}{{\it Chandra}}

\newcommand\CpercOne{41} 		
\newcommand\CnumOne{29}			

\newcommand\CnumPWNd{4}			
\newcommand\CnumShell{23}		
\newcommand\CnumCentral{8}		

\newcommand\CnumTypeOnea{1}		
\newcommand\CnumCoreCollapse{12} 
\newcommand\CnumTypeUnknown{16}	

\newcommand\CnumLessOne{2}		
\newcommand\CnumOnetTen{12}		
\newcommand\CnumTentTwenty{1}	
\newcommand\CnumTwentyPlus{4}	
\newcommand\CnumAgeUnknown{9}	

\newcommand\numStudied{119}		
\newcommand\percOne{8} 		
\newcommand\numOne{10}			

\newcommand\numPWNd{0}			
\newcommand\numShell{8}		
\newcommand\numCentral{4}		
\newcommand\numInteracting{1}

\newcommand\numTypeOnea{1}		
\newcommand\numCoreCollapse{1} 
\newcommand\numTypeUnknown{8}	

\newcommand\numLessOne{1}		
\newcommand\numOnetTen{2}		
\newcommand\numTentTwenty{0}	
\newcommand\numTwentyPlus{0}	
\newcommand\numAgeUnknown{7}	

\graphicspath{{SubPlotCol/}{New_Detections/}{AllSNRs/}{Temperature_Maps/}{SEDPlots/}}

\title[Dusty Supernova remnants II]{A Complete Galactic Plane Catalogue of Far-infrared Supernova Remnants with {\it Herschel}: an inventory of dusty supernovae in the Milky Way}

\title[Dusty Supernova remnants II]{A Complete Catalogue of Dusty Supernova Remnants in the Galactic Plane}

\author[H. Chawner et al.]{
H. Chawner$^{1}$\thanks{E-mail: ChawnerHS@cardiff.ac.uk},
H.L. Gomez$^{1}$,
M. Matsuura$^{1}$,
M.W.L. Smith$^{1}$,
A. Papageorgiou$^{1}$,
\newauthor
J. Rho$^{2, 3}$,
A. Noriega-Crespo$^{4}$,
I. De Looze$^{5, 6}$,
M. J. Barlow$^{5}$,
P. Cigan$^{1}$,
L. Dunne$^{1}$ and
\newauthor{K. Marsh$^{7}$}
\\
$^{1}$School of Physics and Astronomy, Cardiff University, Queens Buildings, The Parade, Cardiff, CF24 3AA, UK\\
$^{2}$SETI Institute, 189 N. Bernardo Ave, Suite 100, Mountain View, CA 94043, USA\\
$^{3}$SOFIA Science Center, NASA Ames Research Center, MS 232, Moffett Field, CA 94035, USA\\
$^{4}$Space Telescope Science Institute, 3700 San Martin Drive, Baltimore, MD 21218, USA\\
$^{5}$Department of Physics and Astronomy, University College London, Gower Street, London WC1E 6BT, UK\\
$^{6}$Sterrenkundig Observatorium, Ghent University, Krijgslaan 281\,--\,S9, 9000 Ghent, Belgium\\
$^{7}$IPAC, Caltech, 1200 E California Blvd, Pasadena, CA 91125, USA
}

\date{Accepted XXX. Received YYY; in original form ZZZ}

\pubyear{2020}

\begin{document}
\label{firstpage}
\pagerange{\pageref{firstpage}--\pageref{lastpage}}
\maketitle

\begin{abstract}
We search for far-infrared (FIR) counterparts of known supernova remnants (SNRs) in the Galactic plane (360$^{\circ}$ in longitude and b = $\pm\,1^{\circ}$) at 70\,--\,500\,$\mu$m with \hersc.  We detect dust signatures in 39 SNRs out of 190, made up of 13 core-collapse supernovae (CCSNe), including 4 Pulsar Wind Nebulae (PWNe), and 2 Type Ia SNe.  A further 24 FIR detected SNRs have unknown types.  We confirm the FIR detection of ejecta dust within G350.1$-$0.3, adding to the known sample of $\sim$\,10 SNRs containing ejecta dust. We discover dust features at the location of a radio core at the centre of G351.2$+$0.1, indicating FIR emission coincident with a possible Crab-like compact object, with dust temperature and mass of T$_d$ = 45.8\,K and M$_d$ = 0.18\,M$_{\odot}$, similar to the PWN G54.1$+$0.3. We show that the detection rate is higher among young SNRs.
We produce dust temperature maps of 11 SNRs and mass maps of those with distance estimates, finding dust at temperatures $15\,\lesssim\,T_d\,\lesssim\,40$\,K. If the dust is heated by shock interactions the shocked gas must be relatively cool and/or have a low density to explain the observed low grain temperatures.
\end{abstract}


\section{Introduction}
\label{sec:intro}
Whether or not supernovae are an important contributor to the dust budget of galaxies at all cosmic epochs remains an unresolved question \citep{Morgan2003,Dwek2007,Matsuura2009,Gall2011,Rowlands2014,Mancini2015,DeVis2017}. One way to resolve this is to search for infrared (IR) and submillimetre (submm) emission from cold ejecta dust in nearby supernovae (SNe) and supernova remnants (SNR), where the sources are close enough to resolve dust structures in the FIR.   There now exists a handful of individual SNRs with reported detections of SN-associated dust structures, containing significant quantities of dust. These include Cassiopeia A \citep{Dunne2003,Krause2004,Dunne2009,Barlow2010,DeLooze2017}, the Crab Nebula \citep{Temim2012,Gomez2012b,DeLooze2019}, SNR1987A \citep{Indebetouw2014, Matsuura2015}, G54.1$+$0.3 \citep{Temim2017,Rho2018} and also G0.0$+$0.0 \citep{Lau2015}, G292.0$+$1.8 \citep{Ghavamian2016}, G350.1$-$0.3 \citep{Lovchinsky2011} and G357.7$+$0.3 \citep{Phillips2009}. The majority of SNe with reported high ejecta dust masses to date have been the remnants of core-collapse SNe.

We can now move on from small numbers of individual studies thanks to the availability of large surveys of the Galactic Plane with the {\it Spitzer} Space Telescope \citep[hereafter \spitz,][]{Werner2004} and the {\it Herschel Space Observatory} \citep[hereafter \hersc,][]{Pilbratt2010}; these provide images of hundreds of SNRs from the near- to far-infrared (NIR\,--\,FIR, 3.6\,--\,500\,$\mu$m).
Previous works have used areas within these surveys to detect dust signatures associated with SNRs including the {\it Spitzer} catalogues of \citet{Reach2006} and \citet{Goncalves2011}, and the recent \hersc\ sample of \citet[hereafter C19, drawn from the \hersc\ Infrared Galactic Plane Survey I, Hi-GAL I over $10^{\circ}< \mid l \mid <60^{\circ}$ at 70--500\,\micron]{Chawner2019}. Those works reported the detection of dust emission associated with SNR structures in 18, 39 and \CnumOne\ SNRs respectively over a region of the Galactic Plane covering $\sim$100 square degrees, suggesting SN-related dust emission is rather common.
Prior to the \spitz\ and \hersc\ era, FIR surveys of Galactic SNRs with IRAS by \citet{Arendt1989} and \citet{Saken1992} detected emission from dust in 11 and 17 SNRs respectively within the Galactic Plane (b\,$\leq\,\mid\,1^{\circ}\,\mid$), based on the \citet{Green2014} catalogue. These FIR surveys used observations centred at 12, 25, 60, and 100\,\micron\ at respective resolutions of 0.5\,$^\prime$ at 12\,\micron\ and 2\,$^\prime$ at 100\,\micron. The more recent whole sky survey with {\it AKARI}, at wavelengths of 2.4\,--\,160\,\micron\ and angular resolution of 0.5$^\prime$ in the 65\,\micron\ band, studied 20 Galactic SNRs \citep{Koo2016}.
With the higher resolution of {\it Herschel} (6$^{\prime\prime}$ at 70\,\micron) it should be possible to more easily disentangle SNR and unrelated interstellar dust near to the remnant, and to confirm or update the detection levels of these previous studies.

C19 further discovered clear evidence for dust formation in the ejecta of the pulsar wind nebulae (PWNe) G11.2$-$0.3, G21.5$-$0.9 and G29.7$-$0.3 alongside the previously discovered G54.1$+$0.3, with a detection rate of $\sim$45\,per\,cent for this class of remnant within the area studied. Such a high detection rate implies that many PWNe may contain dust, some of which can be freshly formed ejecta dust \citep[see also][]{Omand2019}. Though they report that the high detection rates in this SN class may simply be due to observational bias: the dust in the PWNe appears hotter than the surrounding regions and it is easier to disentangle warmer/heated dust structures from unrelated dust in the surrounding cooler ISM.

One remaining critical question is whether or not Type Ia supernovae could also produce dust in their ejecta.  \hersc\ observations of the historical Tycho and Kepler SNRs by \citet{Gomez2012a} showed that although dust structures associated with the SNR material (as traced by X-ray and radio emission) were clearly detected, these dust structures originated from the swept up circumstellar and interstellar medium respectively (see also \citealp{Ishihara2010}) and not from freshly-formed ejecta dust.

Using \hersc\ in the FIR to search for dust signatures is not always superior to using MIR observations from, for example, \spitz\ (particularly at 24\,$\mu$m where searching for dust signatures can be easier since the hotter dust emission can be more clearly disentangled from interstellar material). However, the increased sensitivity and resolution with \hersc\ at wavelengths $\ge$\,70\,\micron\ compared to lower resolution FIR observatories can be important for disentangling interstellar and SN-related structures in confused regions compared to e.g. \spitz\ at the same wavelengths, IRAS, and AKARI \citep{Murakami2007}.  The longer \hersc\ wavebands have the added advantage of being able to reveal the presence of any existing colder dust potentially missed by shorter wavelength observations \citep[e.g.][]{Barlow2010,Matsuura2011,Gomez2012b,Temim2017,Rho2018}.

Unfortunately, confusion with the ISM is the major limitation when studying dust in SNRs (e.g. in Cas A). Indeed in the \hersc\ Galactic SNRs studied by \citet{Chawner2019}, confusion meant that SN dust structures could not be separated from unrelated ISM in more than 60\,per\,cent of the sample, and even for those sources with clear FIR dust signatures coincident with known X-ray and radio emission originating from the SNRs, confusion still persists in a large fraction (also noted by \citealt{Goncalves2011}).

Therefore, in order to verify the importance (or not) of SNe as dominant dust producers in galaxies and to determine which SNe form dust in their ejecta material, a larger statistical sample is needed.  In this paper we supplement the first-look catalogue from C19 and present a {\emph{full \hersc\ Galactic Plane survey} of SNRs detected at FIR wavelengths ($\geq$70\,$\mu$m). This study covers an area roughly twice as large as that of \cite{Chawner2019}, and reveals dusty SNRs that are older than the previous targeted \hersc\ SNRs such as Cas A and Tycho. In Section~\ref{sec:FIRSurvey} we introduce the survey and classify Galactic SNRs with detection levels based on whether there are associated dust features in the \hersc\ images.  Section~\ref{sec:investigate} investigates the properties of the dust for those sources where clear dust features were found. Our results are discussed in Section~\ref{sec:furtherwork}, and our summary is listed in Section~\ref{sec:Summary}.

\section{Survey for Far Infrared Supernova Remnant Emission} \label{sec:FIRSurvey}

We study the remaining SNRs in the Galactic Plane \citep{Green2004} using Hi-Gal \citep{Molinari2011,Molinari2016} covering $360^{\circ}$ in longitude and $\mid b \mid \leq 1$. There are a total of 200 SNRs in this area, although \hersc\ images for 10 are unavailable, this work therefore includes an additional 119 sources compared to C19, and an additional 92 and 66 sources compared to the previous dusty catalogues of Galactic SNRs studied using \spitz\ in \citet[MIR, $\lambda \le 24\,\mu$m]{Reach2006} and \citet[$\lambda \le 70\,\mu$m]{Goncalves2011} respectively (an additional 163 and 137 SNRs when this work is combined with the \hersc\ catalog in C19).

In brief, each remnant was first inspected as a false colour image combining the 70, 160, and 250\,$\mu$m \hersc\ wavebands, regridded and convolved to the resolution of the 250\,$\mu$m band (see the Appendix at {\url {https://github.com/hanchawn/Arxiv}} for the three colour \hersc\ images, Figure~A1). FIR emission from SNRs can be thermal or non-thermal. Thermal FIR will arise from warm dust such as warm ejecta material, shock-heated dust in the forward or reverse blast waves, photo-ionised material near a PWN, or swept-up ISM. Non-thermal emission originates from synchrotron emission from shocks or PWNe, although this is not expected to be significant except in the case of plerion objects such as the Crab where synchrotron emission contributes between 19 and 88 per cent of the flux in the various \hersc\ wavebands \citep{DeLooze2019}. We compare with multi-waveband images to identify dust emission associated with each SNR which, in principle, should allow us to identify the emission mechanism. For the most part we compare with radio and, where available, X-ray images, both of which trace shock-heated material, see individual notes in Section~\ref{subsec:IndividualResults} for details. Where radio and X-ray structures originate from a PWN, incident FIR emission may indicate dust in a photo-ionised region. We can also use X-ray and optical images to identify ejecta material or swept-up/shocked ISM and in some cases, any correlation observed between FIR and MIR or NIR emission can help to identify a variety of FIR sources such as shock-heated dust or swept up ISM.

The method for determining if dust signatures related to SNe material are present in the \hersc\ images is described in full in C19 and is the same classification criteria as used in the {\it Spitzer} SNR catalogues of \citet{Reach2006} and \citet{Goncalves2011}.  Following the definition of detection levels from the MIR \spitz\ SNR catalogue in \citet[see also C19]{Reach2006}, we define detections as: 1 = detection (FIR emission which is clearly correlated with radio, MIR, or X-ray structure and can be distinguished from the ISM), 2 = possible detection (FIR emission in the region of the SNR, potentially related to radio, MIR, or X-ray structure but confused with ISM), 3 = unlikely detection (detection of FIR emission which is probably unrelated to the SNR), and 4 = no detection of FIR emission. We also introduce a new category in this work, `i', to account for the fact that in some cases, the only dust signatures associated with the SNR (via radio and/or X-ray emission) are found in regions where the shell is known to be interacting with a molecular cloud.
Table~A1 ({\url {https://github.com/hanchawn/Arxiv}}) lists the full 117 Galactic SNRs from HiGAL studied in this survey along with the 71 sources studied by C19, with the 39 detected level 1 SNRs summarised in \autoref{tab:DetectionComparisonLevel1}. Individual details provided for each SNR assigned a detection rate of 1 or 2 in this work (10 level 1 and 11 level 2) are listed in the following Section. We note that the availability and quality of ancillary data and the depth of those observations is not the same for all SNRs, which makes it impossible to automate the classification of the detection levels, and thus requires manual intervention and interpretation. This could introduce a bias towards bright, well studied sources that we classify as detections here, similarly this could also be a factor in the previously quoted rates for SNRs detected in near-MIR catalogues with Spitzer \citep{Reach2006,Goncalves2011}.

\subsection{Results for Individual Remnants} \label{subsec:IndividualResults}
\begin{table*} 
	\caption{A summary of the supernova remnants in the Hi-GAL Survey with detection level = 1: FIR emission which is clearly correlated with radio, MIR, or X-ray structure and can be distinguished from the ISM.  \textsuperscript{a} `Y' indicates that a source contains an associated PWN, `?' indicates an unconfirmed PWN candidate. The FIR detection of the PWN is indicated in the brackets. \textsuperscript{b} Location of FIR detected dust features. \textsuperscript{c} Waveband of previous detection to which FIR structure is compared: O = optical, R = radio, X = X-ray. References for age and SN type for this table are:
\textsuperscript{1}\citealp{Pihlstrom2014},
\textsuperscript{2}\citealp{Rho2002},
\textsuperscript{3}\citealp{Borkowski2016},
\textsuperscript{4}\citealp{Case1998},
\textsuperscript{5}\citealp{Reynolds2006},
\textsuperscript{6}\citealp{Klochkov2016},
\textsuperscript{7}\citealp{Voisin2016},
\textsuperscript{8}\citealp{Bietenholz2008},
\textsuperscript{9}\citealp{Bocchino2005},
\textsuperscript{10}\citealp{Leahy2008},
\textsuperscript{11}\citealp{Morton2007},
\textsuperscript{12}\citealp{Chen2004},
\textsuperscript{13}\citealp{Wolszczan1991},
\textsuperscript{14}\citealp{Zhu2013},
\textsuperscript{15}\citealp{Harrus1999},
\textsuperscript{16}\citealp{Leahy2016},
\textsuperscript{17}\citealp{Pye1984},
\textsuperscript{18}\citealp{Smith1985},
\textsuperscript{19}\citealp{Hwang2000},
\textsuperscript{20}\citealp{Lopez2013},
\textsuperscript{21}\citealp{Bocchino2010},
\textsuperscript{22}\citealp{Park2013},
\textsuperscript{23}\citealp{Shan2018},
\textsuperscript{24}\citealp{Bamba2016},
\textsuperscript{25}\citealp{Combi2010b},
\textsuperscript{26}\citealp{Reynolds2013},
\textsuperscript{27}\citealp{Combi2016},
\textsuperscript{28}\citealp{Andersen2011},
\textsuperscript{29}\citealp{Pannuti2014},
\textsuperscript{30}\citealp{Nugent1984},
\textsuperscript{31}\citealp{Frank2015},
\textsuperscript{32}\citealp{Caswell1983},
\textsuperscript{33}\citealp{Giacani2011},
\textsuperscript{34}\citealp{Yamaguchi2012},
\textsuperscript{35}\citealp{Pannuti2014},
\textsuperscript{36}\citealp{Tian2012},
\textsuperscript{37}\citealp{HESSCollaboration2008b},
\textsuperscript{38}\citealp{Tian2014},
\textsuperscript{39}\citealp{Yasumi2014},
\textsuperscript{40}\citealp{Gaensler2008},
\textsuperscript{41}\citealp{Lovchinsky2011},
\textsuperscript{42}\citealp{Dubner1993}.} 
	\csvreader[tabular= l c c c c c c c c,
				late after last line=\\\hline,
				table head=\hline SNR & Name & \parbox{1.5cm}{\centering PWN \textsuperscript{a} \\ (FIR)} & \parbox{1cm}{\centering Age \\ (kyr)} & SN Type & Dust features \textsuperscript{b} & Comparison \textsuperscript{c} & References\\\hline\hline] 
	{Level1_Detections.csv}{SNR=\snr, Name=\name, Size=\size, Morphology=\morph, PWN=\pwn, Age=\age, Type=\type, Distance=\dist, References=\refs, GLIMPSE=\irac, MIPSGAL=\mips, HiGal=\herschel, C18=\c18, Region=\region, Comparison=\comp} 
	{\snr & \name & \pwn & \age & \type & \region & \comp & \refs} 
	\\[1.5pt]
	\label{tab:DetectionComparisonLevel1}  
\end{table*}

Notes on individual sources classed as {\bf detection level 1} (FIR emission which is clearly correlated with radio, MIR, or X-ray SNR structure and can be distinguished from the ISM) and {\it detection level 2} (FIR emission in the region of the SNR, potentially related to radio, MIR, or X-ray structure but confused with ISM) are provided in this section.

\bigskip

\begin{figure}
	\subfigure{
	\includegraphics[width=1.0\linewidth, trim = 0cm 0.5cm 0cm 0cm, clip]{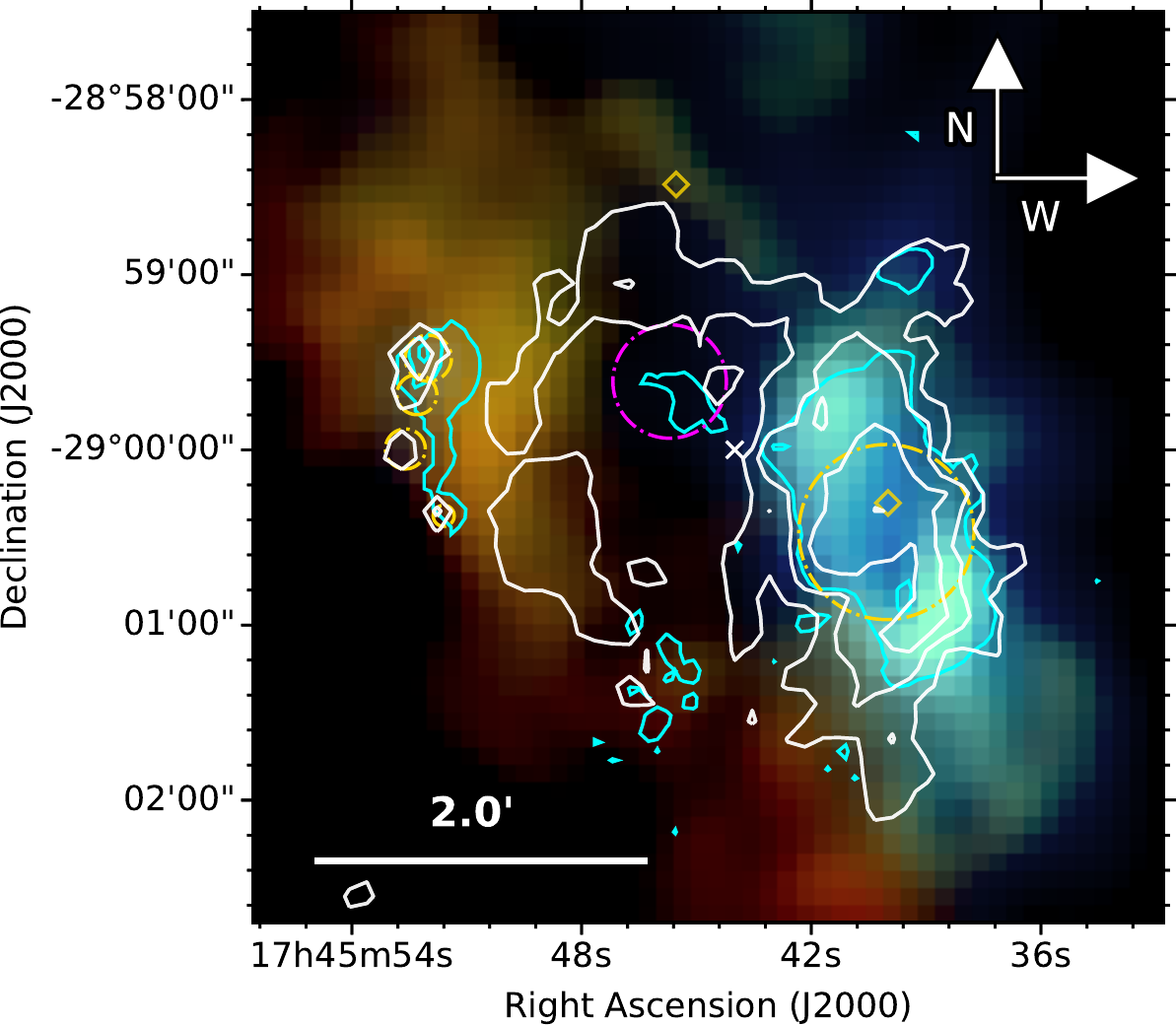}}
	\subfigure{
	\includegraphics[width=1.0\linewidth]{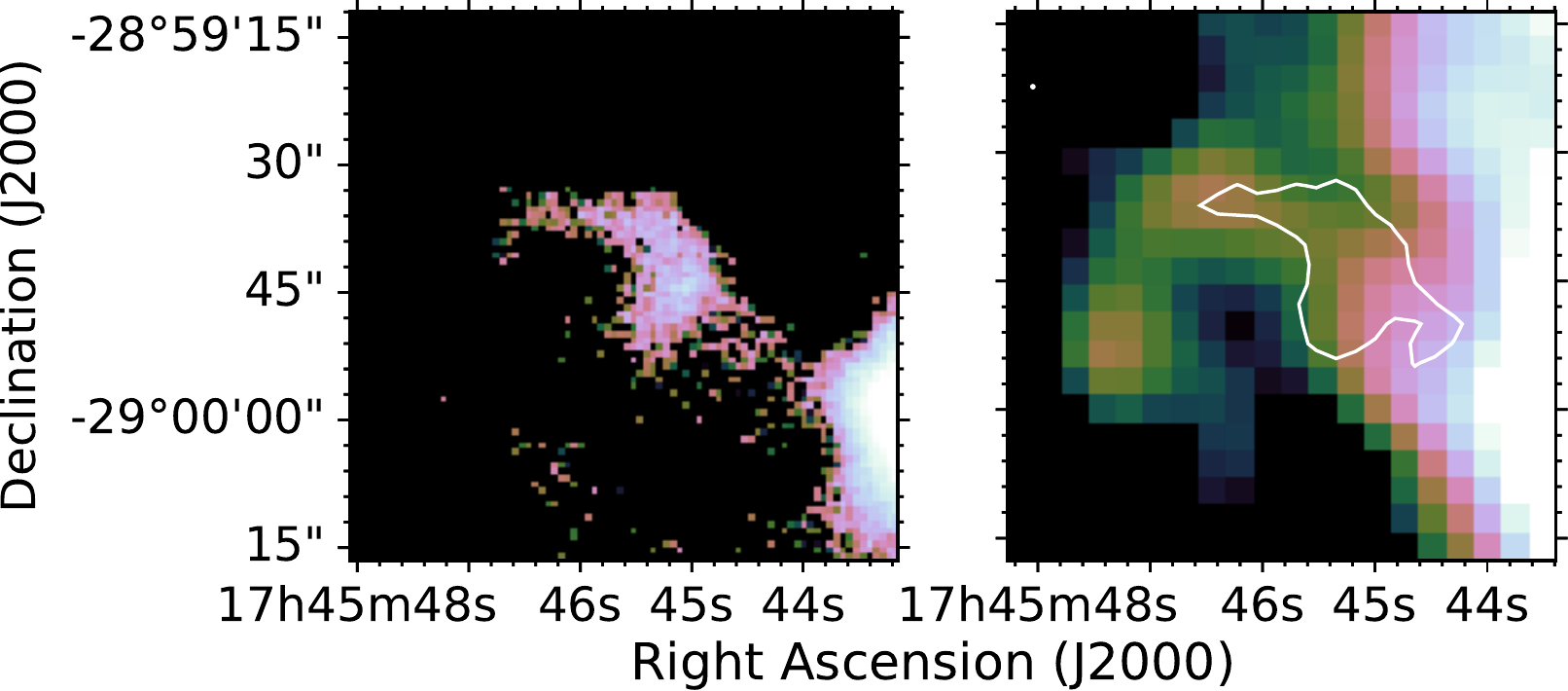}}
	\caption{{\it G0.0$+$0.0, Sgr A East} -
	{\it top}: \hersc\ three colour image with JVLA 5.5\,GHz contours overlaid in white and SOFIA 37.1\,\micron\ contours overlaid in cyan, colours are red = 250\,$\mu$m, green = 160\,$\mu$m, and blue = 70\,$\mu$m.
	{\it Bottom:} A zoom in on the dust factory detected within the magenta circle (top) by \citet{Lau2015}. {\it left:} the SOFIA 37.1\,\micron\ detection from their work, and {\it right:} the \hersc\ 70\,\micron\ image with SOFIA 37.1\,\micron\ contours overlaid. Here the \hersc\ and SOFIA data have been scaled considerably to show this faint feature. (For the bottom panels we use the cube helix colour scheme \citep{Green2011}.)
	The gold circles to the east indicate the locations of H{\sc ii} regions \citep{Goss1985} and the gold circle to the west indicates the location of Sgr A West. The diamond to the north indicates the location of the associated neutron star, `The Cannonball', and the diamond to the west indicates the location of Sgr A*.
	The white cross shows the radio coordinates of the SNR centre from \citet{Green2014}. The same \hersc\ colour combinations and orientation are used in Figures~\ref{fig:G6.4-0.1Image}\,--\,\ref{fig:G357.7+0.3Image}.
	}
	\label{fig:G0.0+0.0Image}
\end{figure}

\begin{figure}
	\includegraphics[width=1.0\linewidth, trim = 0cm 0cm 2.6cm 0cm, clip]{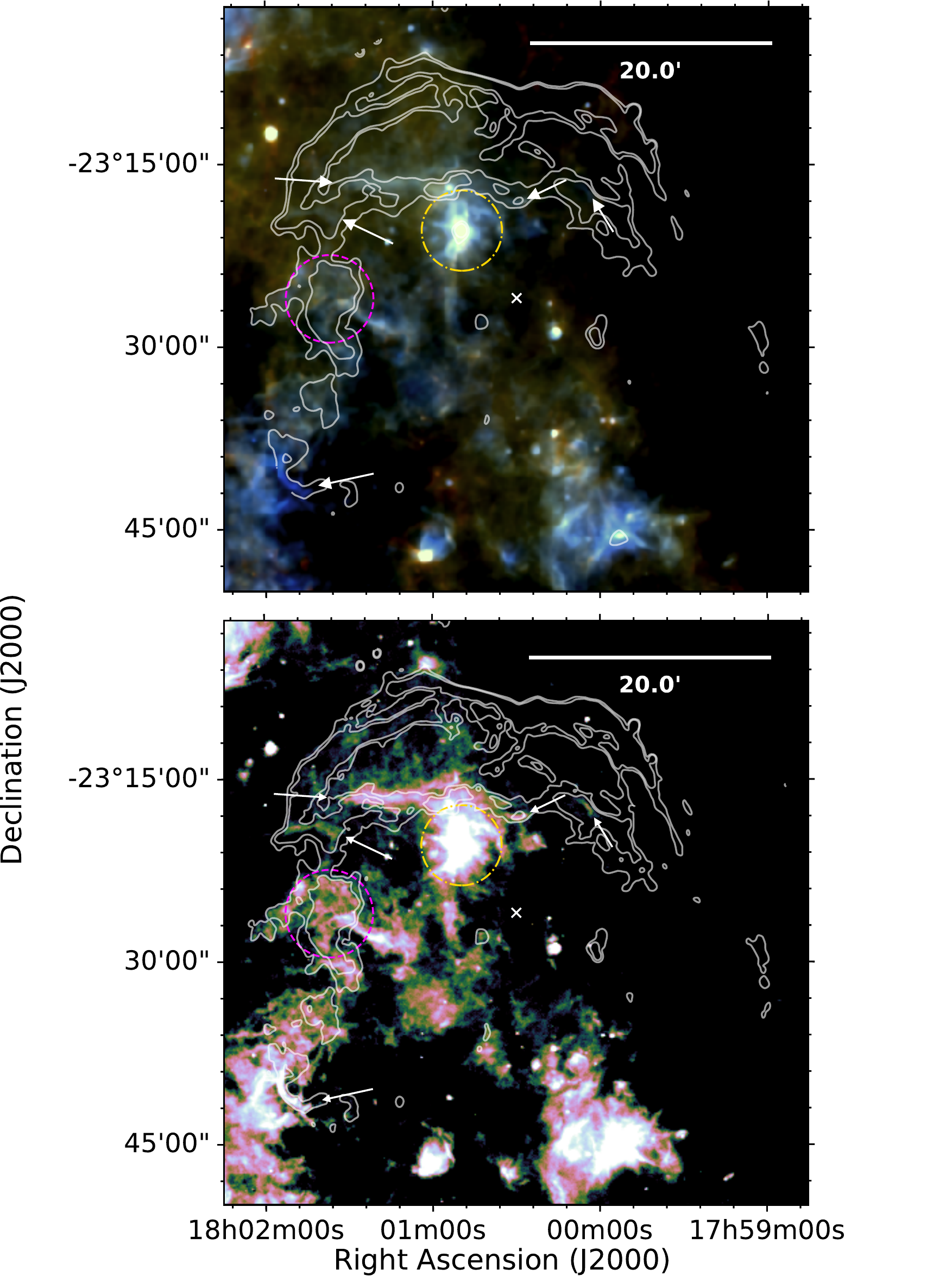}{}
	\caption{{\bf G6.4$-$0.1, W28} - {\it Top:} \hersc\ three colour image and {\it bottom:} \hersc\ 70\,\micron\ image, both with radio (90\,cm) contours overlaid. The arrows and magenta circle indicate filaments of 70\,\micron\, emission which are coincident with radio structure. There is FIR emission at the location of other radio emission, although the structure is different and association is unclear. The gold circle indicates the location of an unrelated H{\sc ii} region.
	The white cross shows the radio coordinates of the SNR centre from \citet{Green2014}.
	}
	\label{fig:G6.4-0.1Image}
\end{figure}

\begin{figure}
	\centering
	\includegraphics[width=1.0\linewidth]{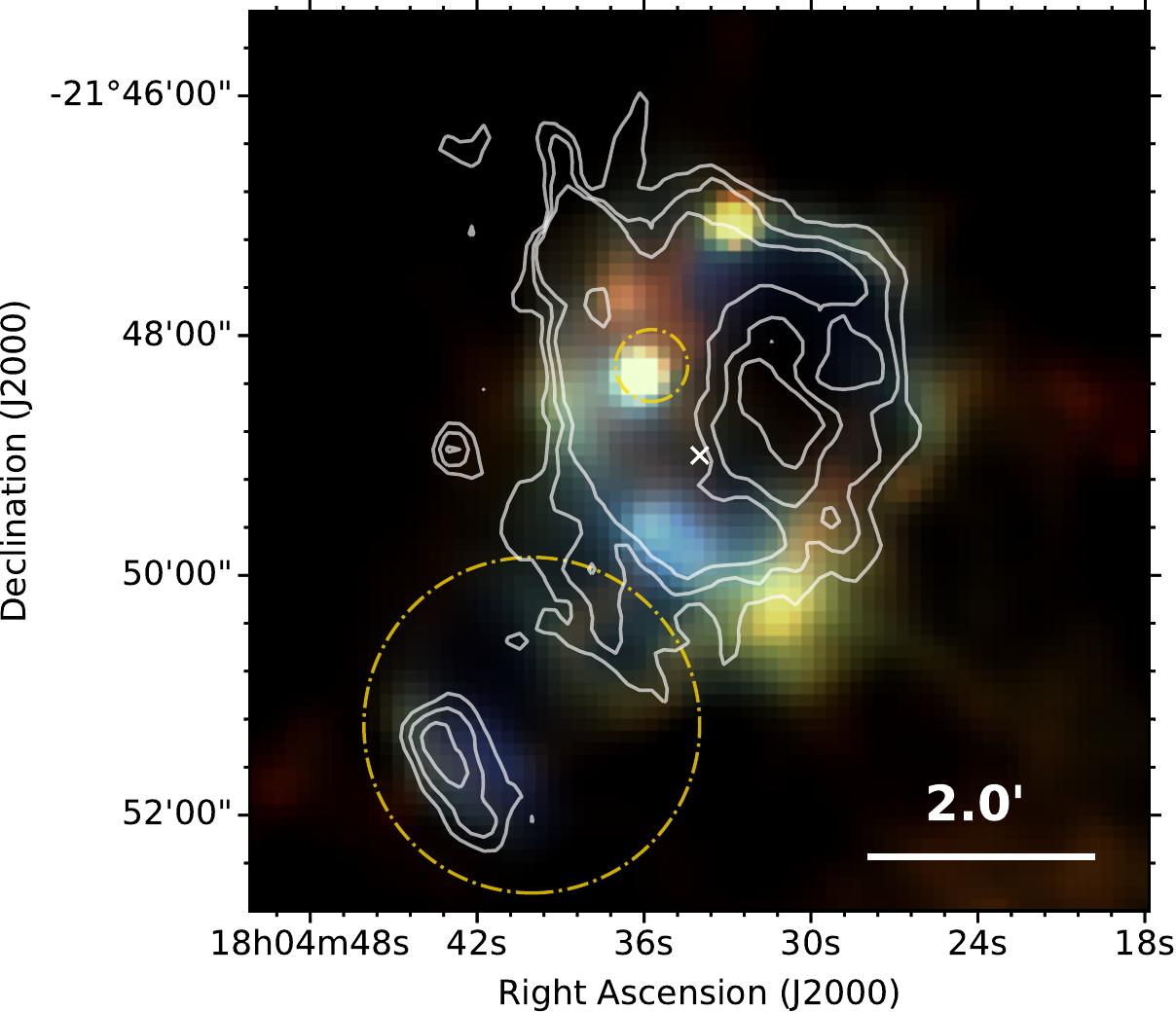}{}
	\caption{{\bf G8.3$-$0.0} - \hersc\ three colour image with VLA 20\,cm contours overlaid. A shell structure is detected across all \hersc\ wavebands. There is some variation in the FIR structure compared with the radio, however there is coincidence between the two.
	The large gold circle indicates the location of an IR bubble.
	The white crosses shows the X-ray coordinates of the SNR centre and the small gold circle indicates the location of a maser.}
	\label{fig:G8.3-0.0Image}
\end{figure}

{\it G0.0$+$0.0 (Sgr A East):}
This CCSN remnant (\autoref{fig:G0.0+0.0Image}) has an elliptical radio shell with a plume of radio and X-ray emission at the northern edge. There is an associated runaway neutron star, the Cannonball, and a PWN in front of this plume \citep[e.g.][]{Nynka2013, Zhao2013}. There are numerous unrelated H{\sc ii} regions in this complex; the largest, Sgr A West, is observed within the SNR shell on the western side and has a spiral structure. Four smaller regions are found to the east, just outside of the SNR radio shell. Observations of 327\,MHz absorption place Sgr A West in front of the SNR \citep{YusefZadeh1987}, although their relative distances are uncertain. There is some evidence that the two sources are at a similar distance and there has been some interaction between the expanding shell and Sgr A West \citep{Yusef-Zadeh1999, Yusef-Zadeh2000a, Maeda2002b}, although this is controversial.

The SNR age and X-ray observations of hot ejecta at its centre both suggest that the reverse shock has already reached the centre of this source \citep{Maeda2002b}. \citet{Lau2015} completed a detailed analysis of the central region using data in the range 5.8--70\,\micron\ from \spitz\ IRAC, the Faint Object Infrared Camera for the SOFIA Telescope (FORCAST), and \hersc\ PACS (70\,\micron). These observations indicate the presence of dust within the SNR central region, north-east of Sgr A West (indicated by the magenta circle in the top panel of \autoref{fig:G0.0+0.0Image}).
They studied the 5.8--37.1\,\micron\ emission from several regions in and around the location of the SNR dust, using the 70\,\micron\ flux as an upper limit. They assumed it is composed of amorphous carbon grains with two grain sizes (large grains of $\sim$\,0.04\,\micron\ and very small grains of $\sim$\,0.001\,$\mu$m) which is heated by a nearby cluster of massive young stars. This gave a total ejecta mass of $\sim\,0.02\,M_\odot$ of warm ($\sim$\,100\,K) dust which has survived the passage of the reverse shock.

We compare \hersc\ maps with the JVLA 5.5\,GHz image \citep{Zhao2013}. We reduced the VLA data, using the Common Astronomy Software Applications (CASA) of the National Radio Astronomy Observatory (NRAO). The C-band (4.5--6.4\,GHz) continuum image of Sgr A* was produced by combining multiple VLA observations taken between March and July 2012. The data were processed in the AIPS software package \citep{Greisen2003}, following standard calibration and wide-band imaging procedures.

Using the five bands of \hersc\ data from HiGAL we search for dust features associated with the SNR and with the \citet{Lau2015} dust discovery (see the bottom panel of \autoref{fig:G0.0+0.0Image}).  We find that the entire region inside the SNR shell is extremely confused in the \hersc\ bands (as originally noted by \citealt{Lau2015}). The flux at 70\,\micron\ due to background emission (determined by placing multiple apertures across the map outside of the bright dust features) is of order 1000\,Jy. Integrating the emission within the \citet{Lau2015} dust region (cyan contours in \autoref{fig:G0.0+0.0Image}) and subtracting the background we estimate that $\sim$\,95\,per\,cent of the flux in that area originates from unrelated background flux i.e. there is no significant detection above the background noise in the \hersc\ 70\,--\,500\,\micron\ bands. The 3\,$\sigma$ upper limit we derive for the dust region from the 70\,\micron\ map is $\sim$12\,Jy which is consistent with the \citet{Lau2015} estimate. This region is too heavily confused to constrain if there is cooler dust at this location using the \hersc\ images and therefore we assign this a detection level 2.

There is bright emission to the west in the 70\,--\,250\,\micron\ bands at the location of the H{\sc ii} region Sgr A West which appears bluer compared to the surrounding dust emission (\autoref{fig:G0.0+0.0Image}), and which is coincident with the brightest region in the radio.
There is also an IR structure to the east of the source in all \hersc\ wavebands (orange structure in the top panel \autoref{fig:G0.0+0.0Image}). It is thought that the SNR is interacting with a molecular cloud on this side \citep[e.g.][]{Sjouwerman2010, Tsuboi2015}. This continuum emission may originate from shock-heated dust, however, it does not correspond to the radio morphology and it is more likely that it is unrelated.
\bigskip

{\bf G6.4$-$0.1 (W28):}
	W28 is a well-studied mixed morphology remnant with evidence of a molecular cloud-SN blast wave interaction \citep[e.g.][]{Wooten1981, Reach2005}. We see many filaments of 70\,\micron\, emission which follow the radio contours in \autoref{fig:G6.4-0.1Image}. Although the proximity of H{\sc ii} regions makes this region confused, we propose that the structure in the \hersc\ data (arrows and magenta circle in \autoref{fig:G6.4-0.1Image}) on the eastern side of the remnant, and to the north is associated with the SNR due to its close correlation with radio emission.  The 70\,$\mu$m emission on the eastern side is also coincident with shocked $\rm H_2$ emission \citep{Reach2005} from the interaction border between the SNR and molecular clouds. Furthermore, the dusty region in the magenta circle and the northernmost part of the dust feature indicated by the arrows overlap with shocked molecular gas ($^{12}$CO(3-2), Figure~1b in \citep{Zhou2014}, see also \citealp{Arikawa1999}).
Although the region is confused in the \hersc\ images, given the strong association of the 70\,\micron\ filaments with the radio structures, we classify this as a level 1 detection.
\bigskip

\begin{figure}
	\centering
	\includegraphics[width=1.0\linewidth]{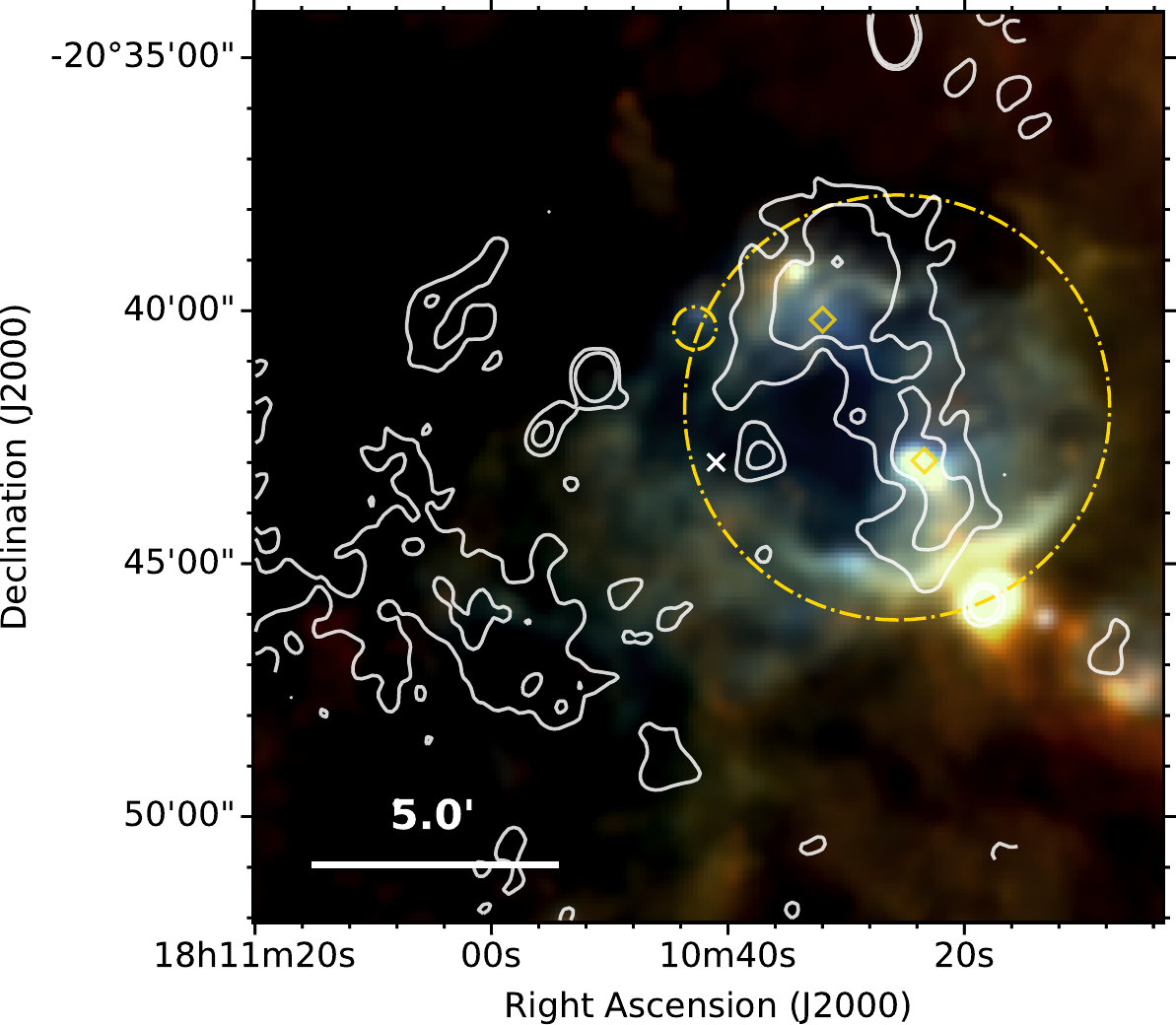}{}
	\caption{{\it G9.9$-$0.8} - \hersc\ three colour image with VLA 90\,cm contours overlaid. There is FIR emission in a shell structure at the western/north-western edge of the radio shell. However, the morphology of the structures are different and there are multiple H{\sc ii} regions (gold diamonds) and FIR bubbles (gold circles) \citep{Simpson2012}, as such, we can not confirm if this is associated with the SN.
	The white crosses shows the radio coordinates of the SNR centre.}
	\label{fig:G9.9-0.8Image}
\end{figure}

\begin{figure}
	\centering
	\includegraphics[width=1.0\linewidth]{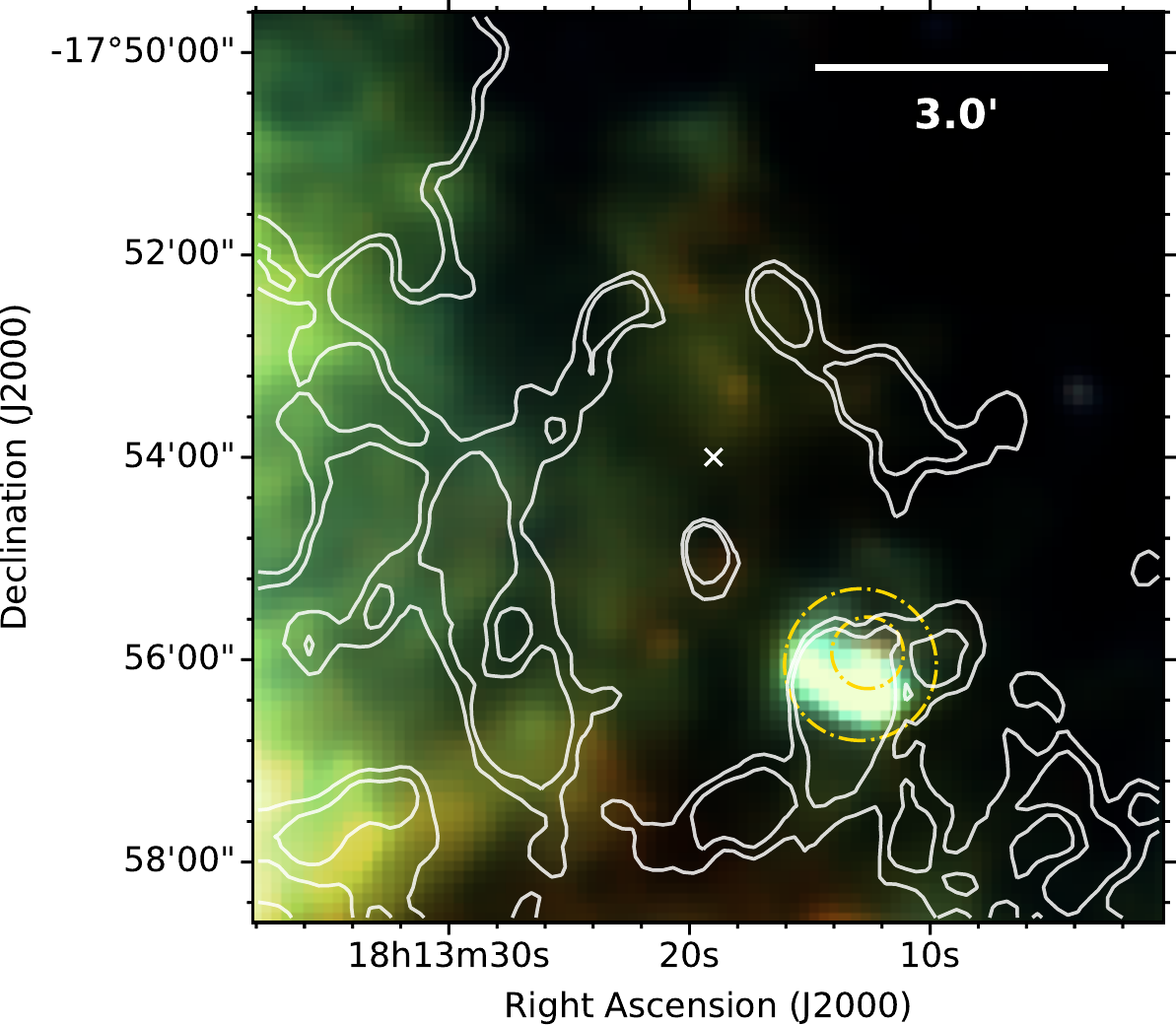}{}
	\caption{{\it G12.7$-$0.0} - \hersc\ three colour image with VLA 20\,cm contours overlaid. The region is very confused to the east, there is FIR emission in the south-western region of the SNR shell. There are also FIR bubbles at this location \citep{Simpson2012}, as indicated by the gold circles.
	The white crosses shows the radio coordinates of the SNR centre.}
	\label{fig:G12.7-0.0Image}
\end{figure}

\begin{figure}
	\centering
	\includegraphics[width=1.0\linewidth]{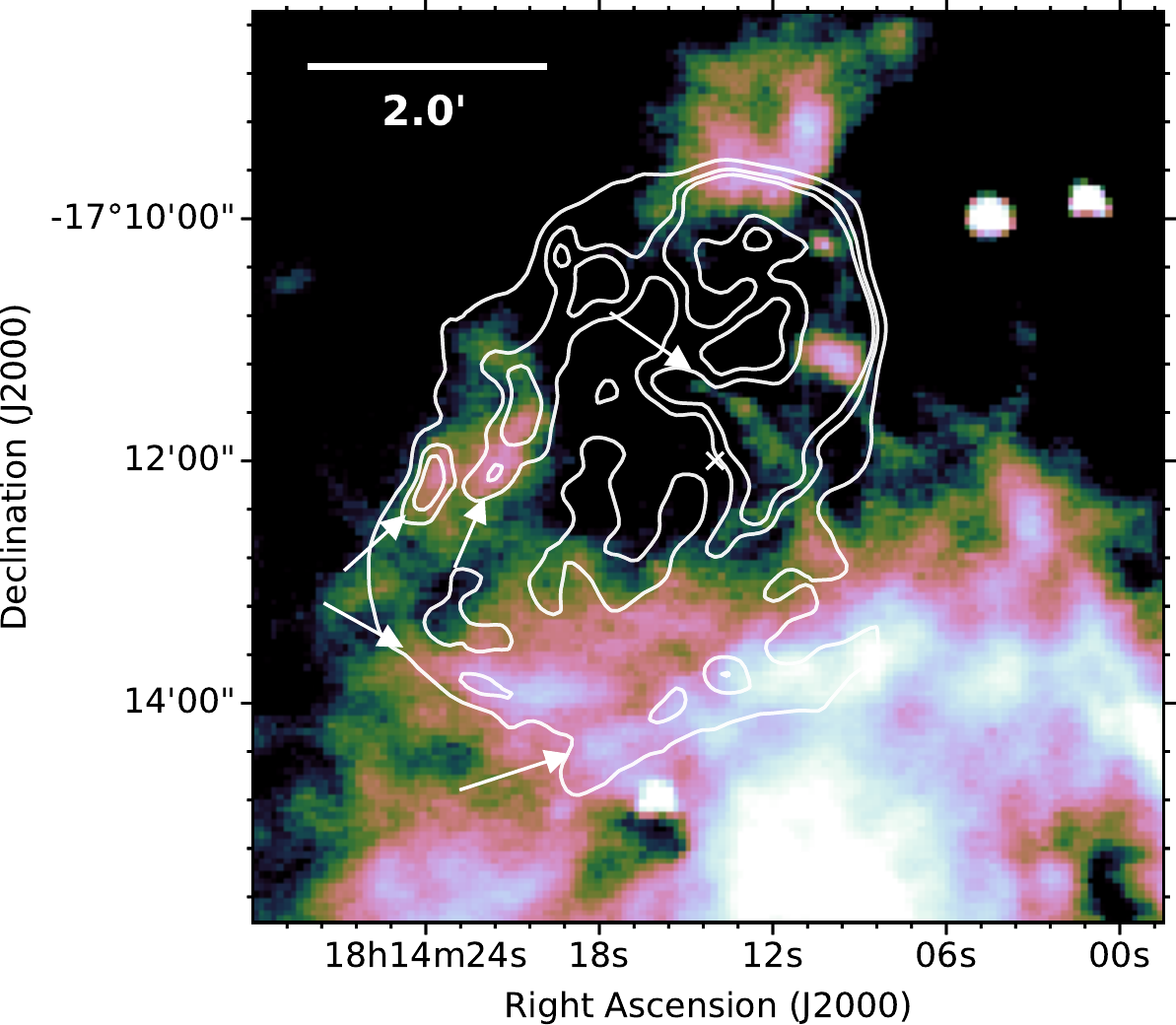}{}
	\caption{{\bf G13.5$+$0.2} - \hersc\ 70\,\micron\ image with VLA 20\,cm contours overlaid. Filaments are detected at 70\,$\mu$m which are coincident with radio emission, and are indicated by the arrows. These are more confused at 160\,\micron, and cannot be distinguished from surrounding ISM in the longer \hersc\ wavebands.
	The white cross shows the radio coordinates of the SNR centre from \citet{Green2014}.
	}
	\label{fig:G13.5+0.2Image}
\end{figure}

\begin{figure}
	\centering
	\includegraphics[width=1.0\linewidth]{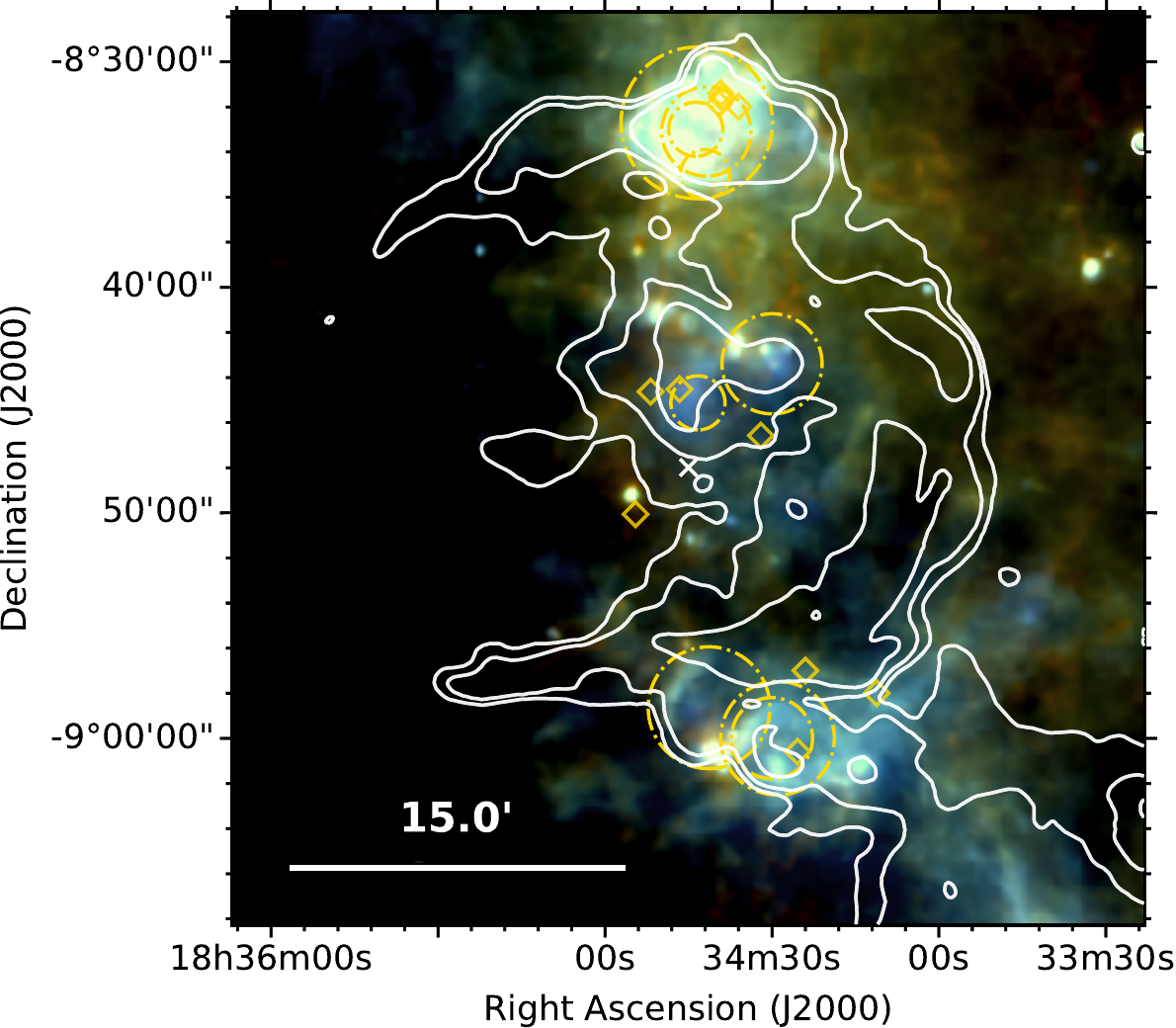}{}
	\caption{{\it G23.3$-$0.3 (W41)} - \hersc\ three colour image with VGPS 21\,cm contours overlaid.
	The gold circles indicate the locations of some of the unrelated sources in this region, such as molecular clouds and bubbles.
	The gold diamonds at $\alpha = 18^\text{h}34^\text{m}51.8^\text{s}, \delta = -08^\circ44^\prime38.2''$ and $\alpha = 18^\text{h}34^\text{m}46.6^\text{s}, \delta = -08^\circ44^\prime31.4''$ indicate the locations of a gamma-ray source (HESS J1834--087) and OH (1720\,MHz) maser lines.
	The other gold diamonds indicate the locations of masers.
	}
	\label{fig:G23.3-0.3Image}
\end{figure}

{\bf G8.3$-$0.0:}
\citet{Brogan2006} detected a radio shell at 90\,cm with a very bright region at $\alpha = 18^\text{h}04^\text{m}38^\text{s},\,\delta = -21^\circ47^\prime17''$.
There is a shell of emission detected across all \hersc\ wavebands which correlates with the radio structure as shown in \autoref{fig:G8.3-0.0Image}. As such, this satisfies our criteria for a level 1 detection. There is a marked colour variation within this shell.
At wavelengths longer than 70\,$\mu$m there is a bright patch to the south-west, and emission is detected at all wavelengths from the location of a maser at $\alpha = 18^\text{h}04^\text{m}36.02^\text{s}, \delta = -21^\circ48^\prime19.6''$ \citep{Green2010, Hewitt2009a}.
There is also FIR emission across all \hersc\ wavebands to the south-east of the SNR, near $\alpha = 18^\text{h}04^\text{m}43.5^\text{s}, \delta = -21^\circ51^\prime29.5''$, although this region is at the edge of a FIR bubble (\citealp{Simpson2012}, \autoref{fig:G8.3-0.0Image}) making its association with the SNR unclear.
\bigskip

{\it G9.9$-$0.8:} This SNR has a shell structure detected in radio \citep[90\,cm,][]{Brogan2006} and optical \citep[H$\alpha$,][]{Stupar2011} wavebands. As seen in \autoref{fig:G9.9-0.8Image}, there is FIR emission detected at the location of the western edge of the radio shell. However, there are multiple H{\sc ii} regions and FIR bubbles \citep{Simpson2012} at this location, and the FIR structure is distributed differently to that of the radio.  Although molecular hydrogen from an interaction between the SNR and molecular clouds was discovered in \citet{Hewitt2009a}, there is no accompanying image to compare with the dust features. We therefore argue that the FIR features are potentially related to the SNR, as such this satisfies our criteria for a level 2 detection.
\bigskip

{\it G12.7$-$0.0:} \citet{Brogan2006} detected a radio shell from this SNR. There is FIR emission in \autoref{fig:G12.7-0.0Image} coincident with the south-western region of the radio shell, although there is a FIR bubble \citep{Simpson2012} at this location and therefore we cannot be sure of the origin of this emission. We therefore classify this as a level 2 detection.
\bigskip

{\bf G13.5$+$0.2:} We detect filaments of emission at 70 and 160\,$\mu$m which correlate with VLA 20\,cm radio structure \citep{Helfand1989} as indicated by the arrows in \autoref{fig:G13.5+0.2Image}. There is 70\,\micron\, emission along the south-western edge, within the VLA contours. It is thought that this may be a ridge of shock-heated dust resulting from the interaction between the SNR and surrounding molecular clouds, which is in agreement with the detection of $\rm H_2$ from this source \citep{Froebrich2015}.
There is more confusion in the longer wavebands and there does not seem to be any associated structure in those bands. Based on the coincidence with the 70\,\micron\ filaments and radio features, we therefore classify this as detection level 1.
\bigskip

\begin{figure}
	\centering
	\includegraphics[width=1.0\linewidth, trim = 0cm 0cm 2.6cm 0cm, clip]{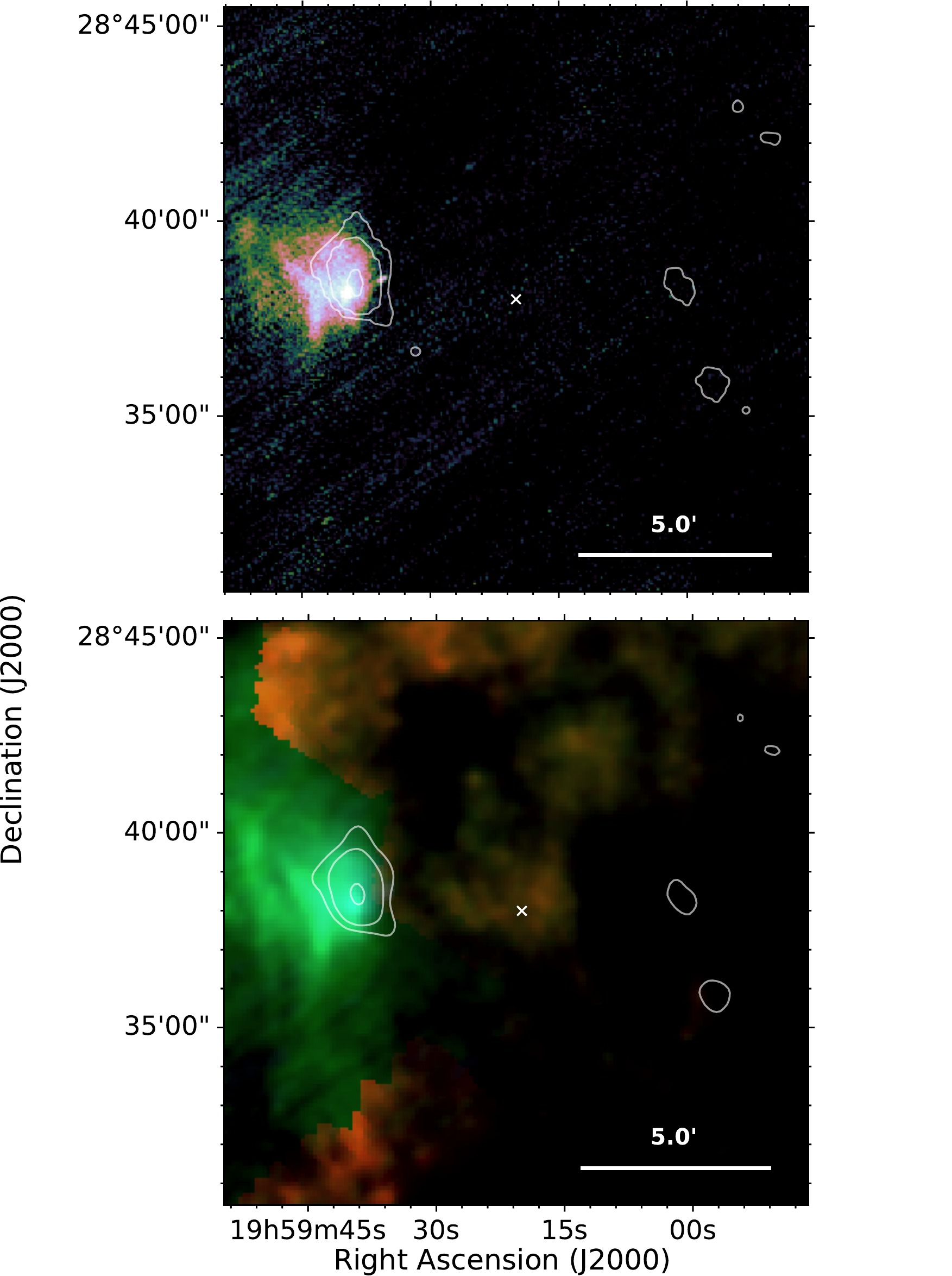}{}
	\caption{{\bf G65.8$-$0.5} -
	{\it Top:} \hersc\ 70\,\micron\, image and {\it bottom:} \hersc\ three colour image, both with radio NVSS 1.4\,GHz contours overlaid.
	FIR structure is detected at the eastern edge of the SNR, at the location of radio and H$\alpha$ structure \citep{Sabin2013}.
	An unidentified region of FIR emission is detected at the centre of the source at 160\,--\,500\,\micron.
	The white cross shows the radio coordinates of the SNR centre from \citet{Green2014}.
	}
	\label{fig:G65.8-0.5Image}
\end{figure}

\begin{figure}
	\centering
	\includegraphics[width=1.0\linewidth]{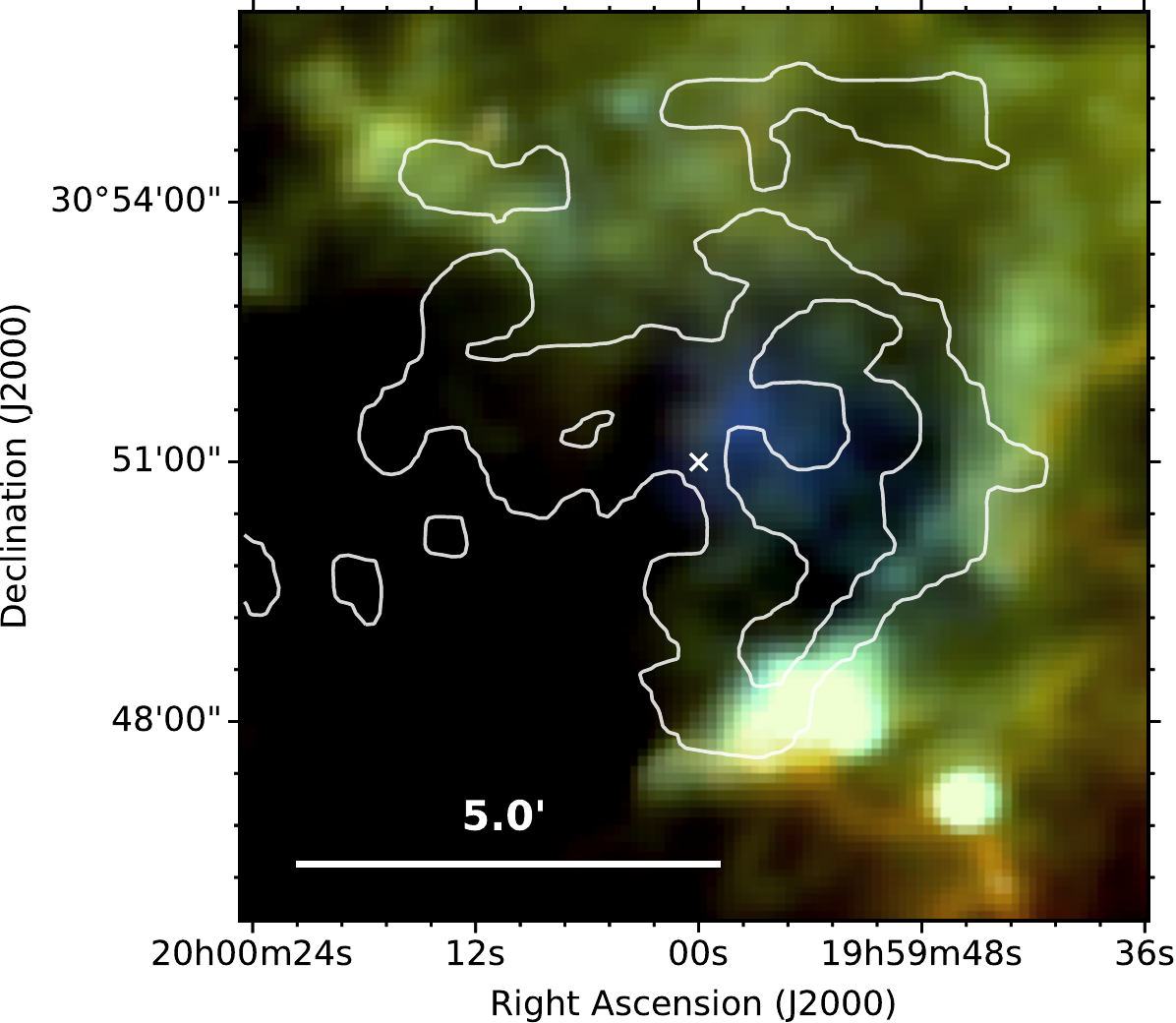}{}
	\caption{{\bf G67.8$+$0.5} - \hersc\ three colour image with NVSS 1.4\,GHz contours overlaid. Dust emission is detected at 70\,\micron\ from a faint radio source, and at all \hersc\ wavelengths from a partial shell along the western and south-western edges of the SNR.
	The white cross shows the radio coordinates of the SNR centre from \citet{Green2014}.
	}
	\label{fig:G67.8+0.5Image}
\end{figure}

{\it G23.3$-$0.3 (W41):} This SNR is in a very crowded region of the Galactic Plane, where there are multiple dense cores, YSOs, H{\sc ii} regions, and bubbles. The presence of a gamma-ray source (HESS J1834--087) and OH (1720\,MHz) maser lines near the centre of the source (indicated in \autoref{fig:G23.3-0.3Image}) suggest that the SNR may be interacting with a molecular cloud \citep{HESSCollaboration2015}.
Several molecular clouds are labelled in \autoref{fig:G23.3-0.3Image} with gold circles. There is bright FIR emission at the location of the radio structure, however the clearest FIR is also coincident with molecular clouds. Although there may be SNR-related dust in this region, we cannot determine its association because of confusion, this therefore satisfies our criteria for a level 2 detection.
\bigskip

{\bf G65.8$-$0.5:} This object was only recently classified as a SNR by \citet{Sabin2013} who used H$\alpha$ to identify its nature.
We compare H$\alpha$ and NVSS images with \hersc, finding bright 70 and 160\,\micron\ emission at the east of the source at the location of the \citet{Sabin2013} H$\alpha$ structure.
We propose that this 70 and 160\,\micron\ feature is associated with the SNR, although the structure is offset to, and more extended than the radio emission seen in the NVSS 1.4\,GHz image (\autoref{fig:G65.8-0.5Image}). Unfortunately the \hersc\ SPIRE maps at 250\,--\,500\,\micron\ do not cover the area where the H$\alpha$ and radio structure is seen.
Additionally, at 160\,--\,500\,\micron\, there is a region of emission detected at the radio centre of the SNR, although this does not correlate with radio emission and we cannot determine if this is associated. Given the close spatial coincidence between the \hersc\ emission, H$\alpha$ and radio structures to the east of the SNR, we classify this as a level 1 detection, though we note that here we do not have the full colour information in the \hersc\ bands.
\bigskip

\begin{figure}
	\includegraphics[width=1.0\linewidth]{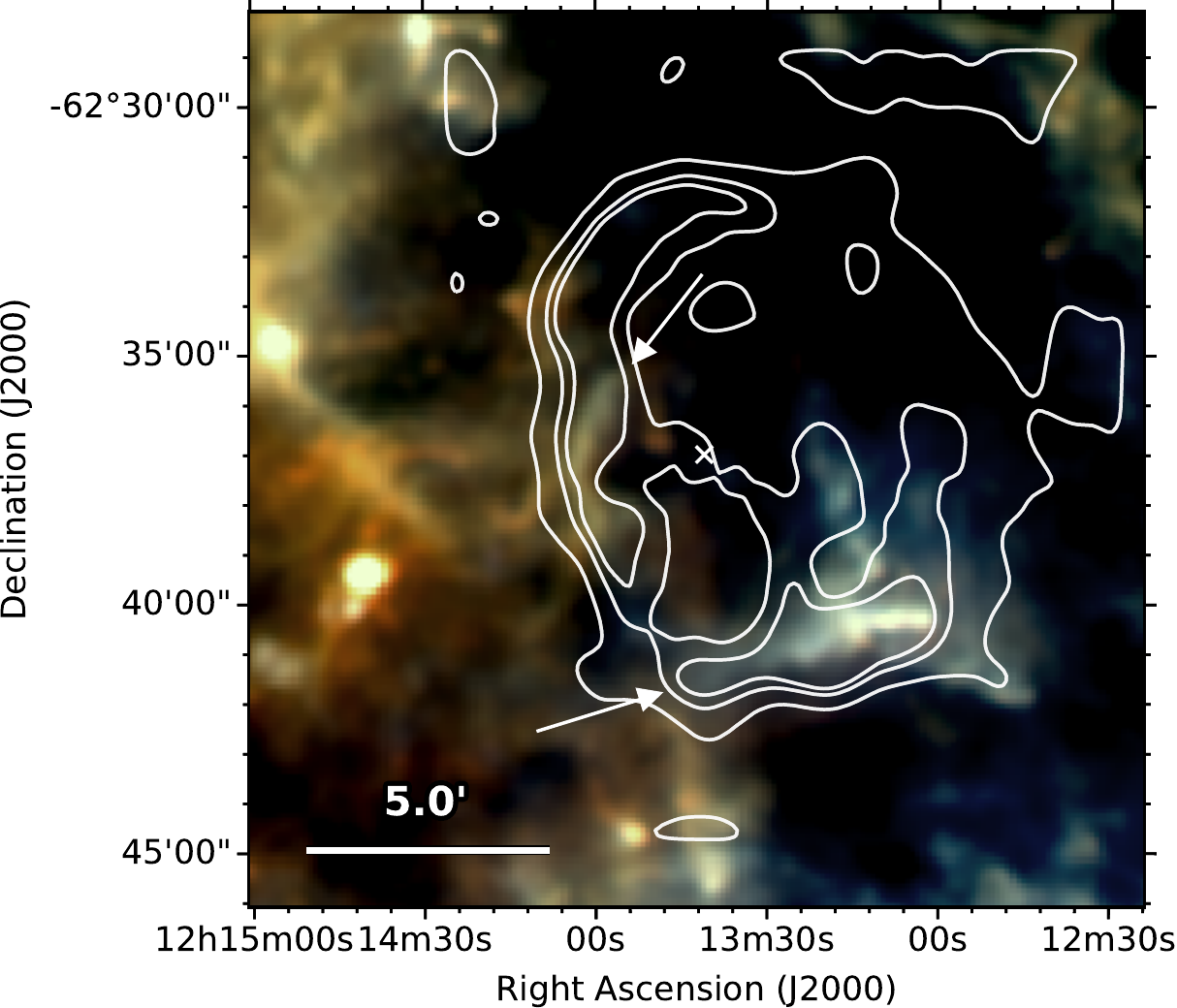}{}
	\caption{{\bf G298.6$-$0.0} - \hersc\ three colour image with 843\,MHz (Molonglo Observatory Synthesis Telescope, MOST) contours overlaid. Dust is detected in filaments at the east of the source, and in a ridge along the southern edge as indicated by the arrows.
	The white cross shows the radio coordinates of the SNR centre from \citet{Green2014}.
	}
	\label{fig:G298.6-0.0Image}
\end{figure}

\begin{figure}
	\centering
	\includegraphics[width=1.0\linewidth]{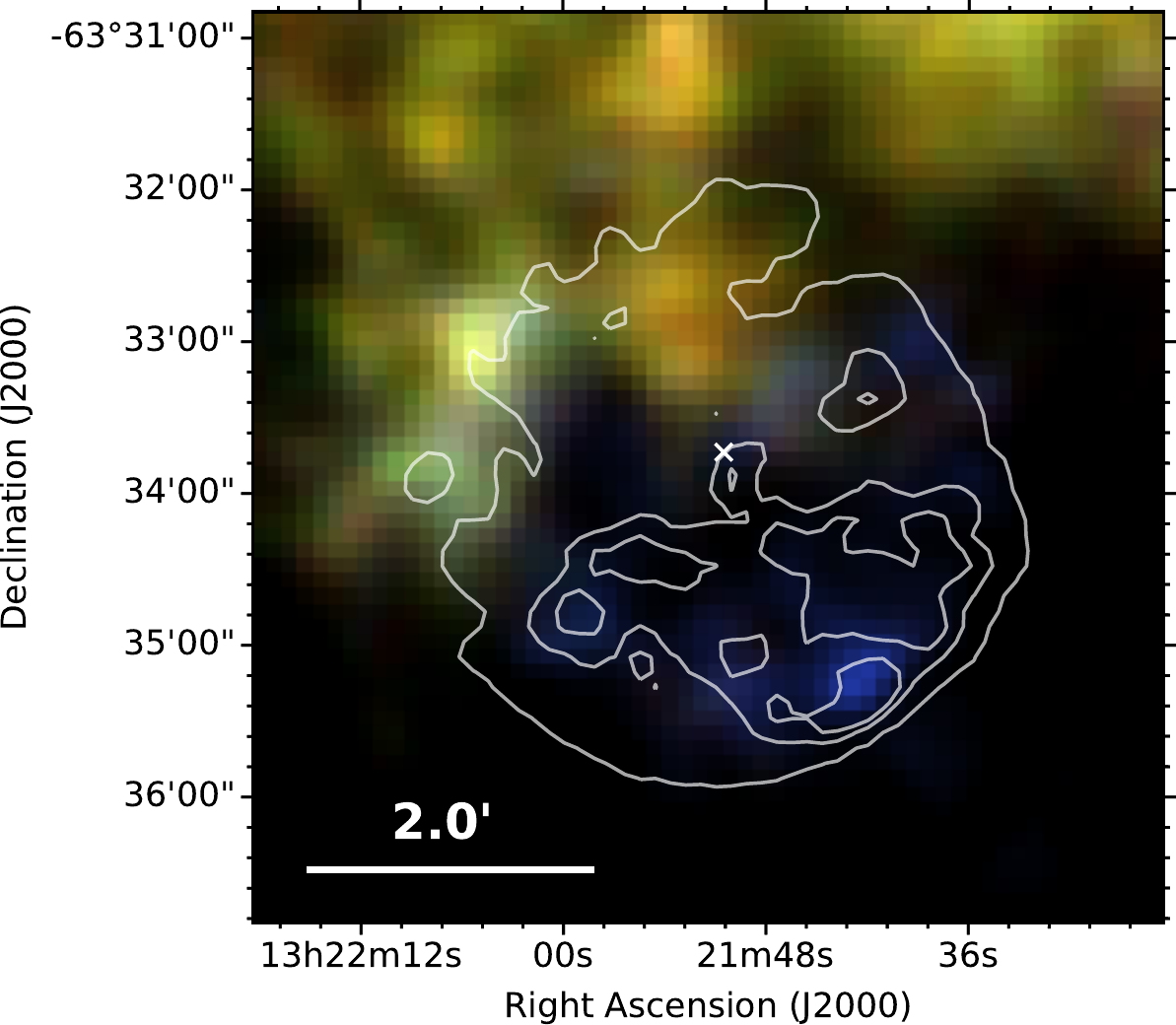}{}
	\includegraphics[width=0.9\linewidth, trim = 0cm 0cm 15cm 0cm, clip]{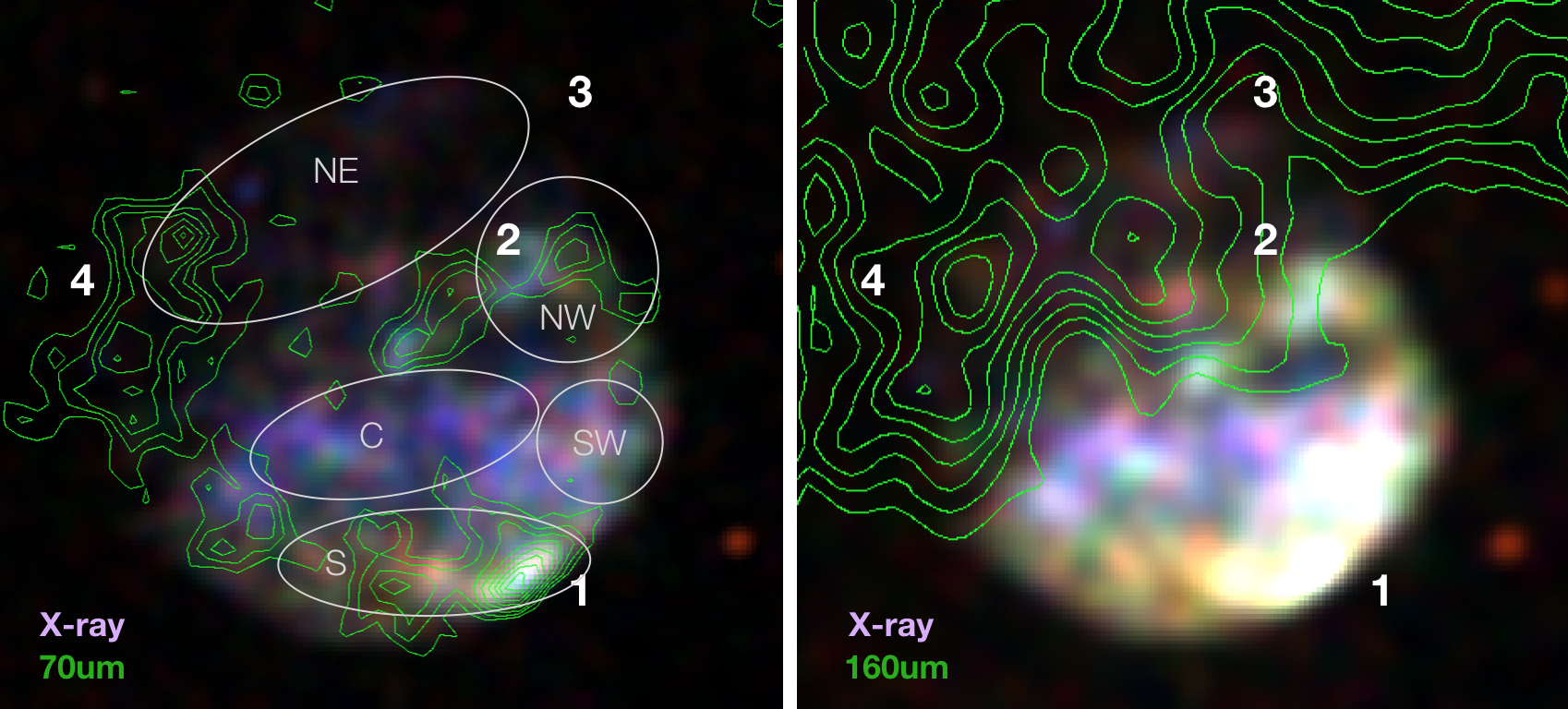}{}
	\caption{
	{\bf G306.3$-$0.9} - {\it Top:} \hersc\ three colour image with \chand\ contours overlaid. Dust emission is detected at 70\,$\mu$m inside the X-ray contours, especially towards the south-west where there is the brightest X-ray emission. The white cross shows the X-ray coordinates of the SNR centre.
	{\it Bottom:} three colour X-ray image using \chand\ archive images at 0.73-1.47\,keV (red), 1.47-2.2\,keV (green) and 2.2-3.3\,keV (blue) with {\it Herschel} 70\,$\mu$m contours overlaid. See \citet{Reynolds2013,Combi2016} for further details of the X-ray features. X-ray `distinct' regions from \citet{Combi2016} are labeled, where regions NE, C and SW are attributed to ejecta material, and regions NW and S are consistent with foreground shock material.  Dust features seen in the \hersc\ images are labeled 1--4.}

	\label{fig:G306.3-0.9Image}
\end{figure}

{\bf G67.8$+$0.5:}
Like G65.8$-$0.5, this source was only recently identified as a SNR by \citet{Sabin2013}.
There is bright 70\,\micron\ emission at $\alpha\,=\,19^\text{h}59^\text{m}57.5^\text{s}, \delta\,=\,30^\circ51^\prime26''$, coinciding with the location of an unidentified IRAS source (IRAS 19579$+$3043) and with emission in the WISE 22\,$\mu$m archive image (though at lower resolution than \hersc\ 70\,$\mu$m). Here, the FIR is anti-correlated with the radio as the \hersc\ FIR peak is located in a gap in the radio structure. There is also brighter 70\,\micron\ emission than the surroundings (bluer FIR than the surrounding emission) indicating the presence of warmer dust emission. Although this doesn't align with the detection criteria defined in Section~\ref{sec:FIRSurvey}, we expect that this emission is associated with the SNR due to the FIR colours and the coincidence with the radio centre.
There is also a bright shell across all \hersc\ wavebands to the west and south-western edge of the SNR.
We propose that the central source and outer shell are associated with the SNR and therefore classify this SNR as a level 1 detection. It is possible that the central emission, which lacks radio emission, could be cold SN ejecta dust \citep[as seen in Cas A,][]{Barlow2010}, while the shell emission is from the ISM. Further follow up observations in other wavelengths would be needed to investigate this.
The temperature map of this SNR is discussed in \autoref{fig:dusttempmaps}.
\bigskip

{\bf G298.6$-$0.0:} This is a mixed-morphology SNR, with an X-ray filled centre and radio shell \citep{Rho1998,Bamba2016}. A filament of dust is detected in all \hersc\ wavebands at the eastern edge of this source at the location of a bright radio filament detected by MOST (Molonglo Observatory Synthesis Telescope) at 843\,MHz  \citep{Whiteoak1996}. There is also dust detected along the south of the radio structure. At 70\,\micron\ this extends along the length of the radio contours and is brightest at the western edge, where emission at the other \hersc\ wavebands is found. We therefore classify this as a level 1 detection. Comparing the Suzaku X-ray image from \citet{Bamba2016}, these features are not correlated with the centrally filled X-ray emission, and therefore are likely to originate from swept up interstellar material given the coincidence with the edge of the radio shell.
\bigskip

\begin{figure}
	\centering
	\includegraphics[width=1.0\linewidth]{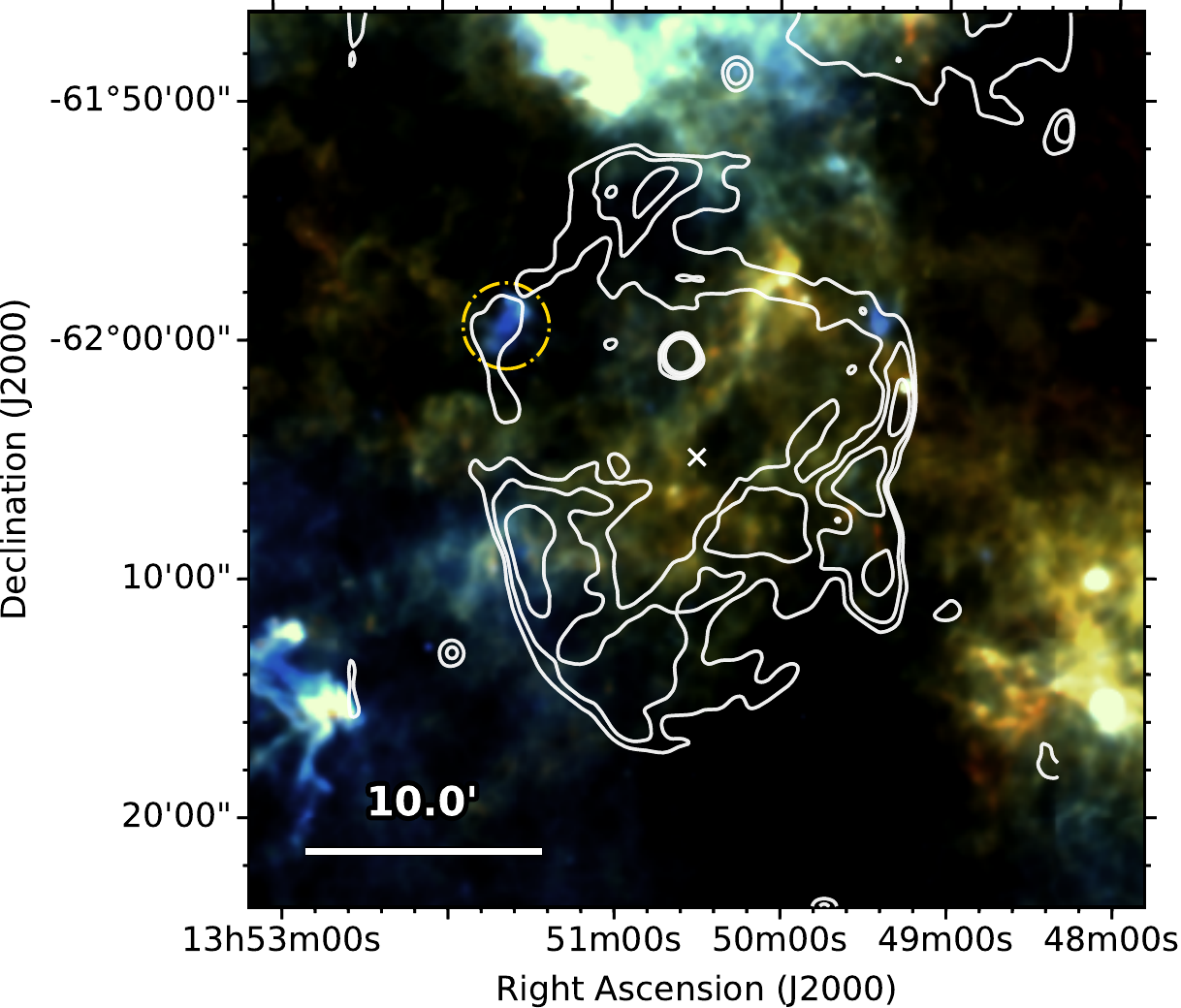}{}
	\caption{{\it G309.8$+$0.0} - \hersc\ three colour image with \chand\ contours overlaid. The gold circle indicates a bright knot of 70\,\micron\ emission which may be associated with the SNR, although its location within the radio contours may be coincidental. 	The white cross shows the X-ray coordinates of the SNR centre.}
	\label{fig:G309.8+0.0Image}
\end{figure}

{\bf G306.3$-$0.9:} This source is a recently discovered remnant of a Type Ia SN \citep{Reynolds2013} in which the reverse shock has heated the central ejecta region \citep{Combi2016}. The three colour \hersc\ image is displayed in the top panel of \autoref{fig:G306.3-0.9Image}.   The overall morphology in the X-ray \citep{Reynolds2013,Combi2016} is a semi-circular shape with strong radio and X-ray emission in the southwest, suggestive of an interaction with an interstellar cloud or an asymmetric SN explosion.   Both X-rays and radio emission are fainter in the north. \citet{Combi2016} showed that the SNR is brighter in softer X-rays towards the centre and found distinct shock-heated ejecta regions in the northeast, central and southwestern areas (see their Figure~7) with enhanced abundances and temperatures.  Similarly, they saw distinct X-ray regions in the northwest and south of the SNR that they attributed to shockfronts interacting with the interstellar medium.

Comparisons with the X-ray, radio, and hot dust emitting at 24\,$\mu$m using {\it Spitzer} observations were presented in \citet{Reynolds2013} and \citet{Combi2016}.  In the latter study, they noted that the 24\,$\mu$m morphology from shocked dust grains is shell-like, matching the distribution observed in hard X-rays. The former study noted that an additional ridge feature to the south seen in radio and X-ray images is also correlated with the 24\,$\mu$m emission, commenting that this could originate from swept up dust or ejecta dust.  \citet{Reynolds2013} also mention that \hersc\ 160\,--\
,500\,$\mu$m images show an over density of dust in the north where the X-ray emission is the faintest.

In the \hersc\ image, the dust emission is concentrated in the south, with very little dust at 70\,$\mu$m in the north-east.  The \hersc\ dust features that are bright at 70\,$\mu$m (blue clumps in \autoref{fig:G306.3-0.9Image}) overlap with the known 24\,$\mu$m features, therefore this is a clear level 1 detection.  The bottom panel of \autoref{fig:G306.3-0.9Image} compares the {\it Herschel} emission at 70\,$\mu$m with the distinct X-ray regions observed by \citet{Combi2016}. NE, C, and SW show the location of the reverse shock heated ejecta regions and NW and S show regions where the X-ray originates from forward shock material. 
We see a new dust feature in the \hersc\ maps (labeled `4' in \autoref{fig:G306.3-0.9Image}), to the east of the centre of the SNR, which lies slightly ahead of the radio shell.  This is a bright elongated structure to the east of the SNR which has two peaks, and lies slightly ahead of the radio shell. It appears to overlap with the eastern-most part of the NE X-ray region from \citet{Combi2016} but has very little associated X-ray and 24\,$\mu$m emission.  This dust feature has a very different colour to features 1 and 2, and is similar to feature 3, suggesting the dust temperature is cooler.  Given the lack of X-ray and radio emission associated with features 3 and 4 and their different colour, we suggest they are unrelated to the SNR even though feature 4, at first glance, appears to be a continuation of the southern dust shell along the eastern side. Instead we propose these are part of a dust cloud that lies in front of the remnant in the north (as seen clearly in the 160\,$\mu$m emission).   The remaining SN-related dust features (1 and 2) originally seen in the 24\,$\mu$m {\it Spitzer} data are spatially coincident with the forward shock material regions NW and S in \citet{Combi2016}.   All of the dust features that are bright in the mid-FIR are anti-correlated with the X-ray ejecta regions indicating the dust in the south and across the centre is simply swept up dust in a shell, with emission appearing to lie interior to the shock wave simply being due to projected emission from the shell (as we also saw in the Tycho SNR, \citealp{Gomez2012a}).

We note that C19 also found dust associated with another candidate Type Ia SNe G344.7$-$0.1 (see also \citealp{Combi2010a,Giacani2011,Yamaguchi2012}), where the dust features are also consistent with a non-supernova origin.

\bigskip

\begin{figure}
	\centering
	\includegraphics[width=1.0\linewidth, trim = 0cm 0cm 2.6cm 0cm, clip]{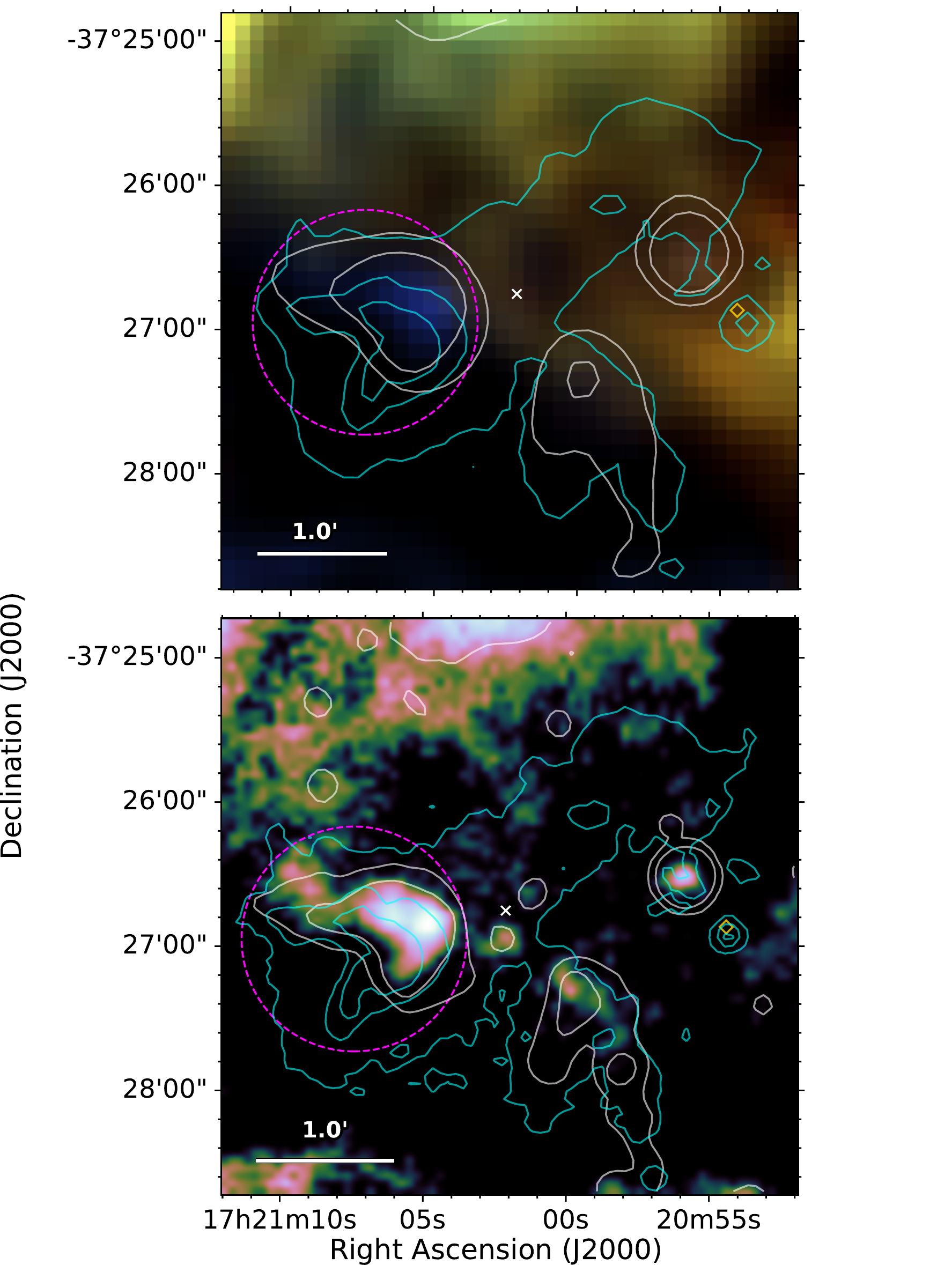}{}
	\caption{{\bf G350.1$-$0.3} -
	{\it Top:} \hersc\ three colour image (smoothed to the 250\,\micron\ resolution).
	{\it Bottom:} \hersc\ 70\,\micron\ image.
	The white contours are 24\,$\mu$m \spitz\ MIPS \citep{Lovchinsky2011} and the cyan contours are X-Ray \chand. Dust emission is detected to the eastern side of this remnant, within the magenta circle. This region correlates with SNR emission in the X-ray, radio, and MIR \citep{Gaensler2008, Lovchinsky2011}. The gold diamond indicates the location of the associated neutron star, XMMU J172054.5$-$372652 \citep{Gaensler2008}.
	The white cross shows the X-ray coordinates of the SNR centre.
	}
	\label{fig:G350.1-0.3Image}
\end{figure}

{\it G309.8$+$0.0:}
There is lots of FIR emission across the region, although the majority seems to be unrelated extended emission from interstellar clouds. We detect a bright knot of 70\,$\mu$m emission to the east which has a similar morphology to the radio structure (indicated by the circle in \autoref{fig:G309.8+0.0Image}). However, the FIR extends slightly beyond the radio contour, and the location of this emission may be coincidental. \citet{Reach2006} listed this as a level 3 detection based on NIR {\it Spitzer} observations (3.6\,--\,8.0\,$\mu$m). The SNR is listed in the table of OH detections from \citet{Green1997} but whether this originates from a maser has not been confirmed. A bright radio source, thought to be extragalactic in origin \citep{Whiteoak1996}, is observed to the north of the SN centre at $\alpha = 13^\text{h}50^\text{m}35.2^\text{s}, \delta = -62^\circ00^\prime42''$; no related FIR emission is observed at this location. Given the uncertainty as to whether the dust feature within the gold circle is associated with the ejecta due to the multiple sources seen in the field with similar \hersc\ colours, we classify this SNR as a level 2 detection.
\bigskip

\begin{figure*}
	\centering
	\includegraphics[width=0.9\linewidth, trim = 0cm 3.6cm 0cm 3.9cm, clip]{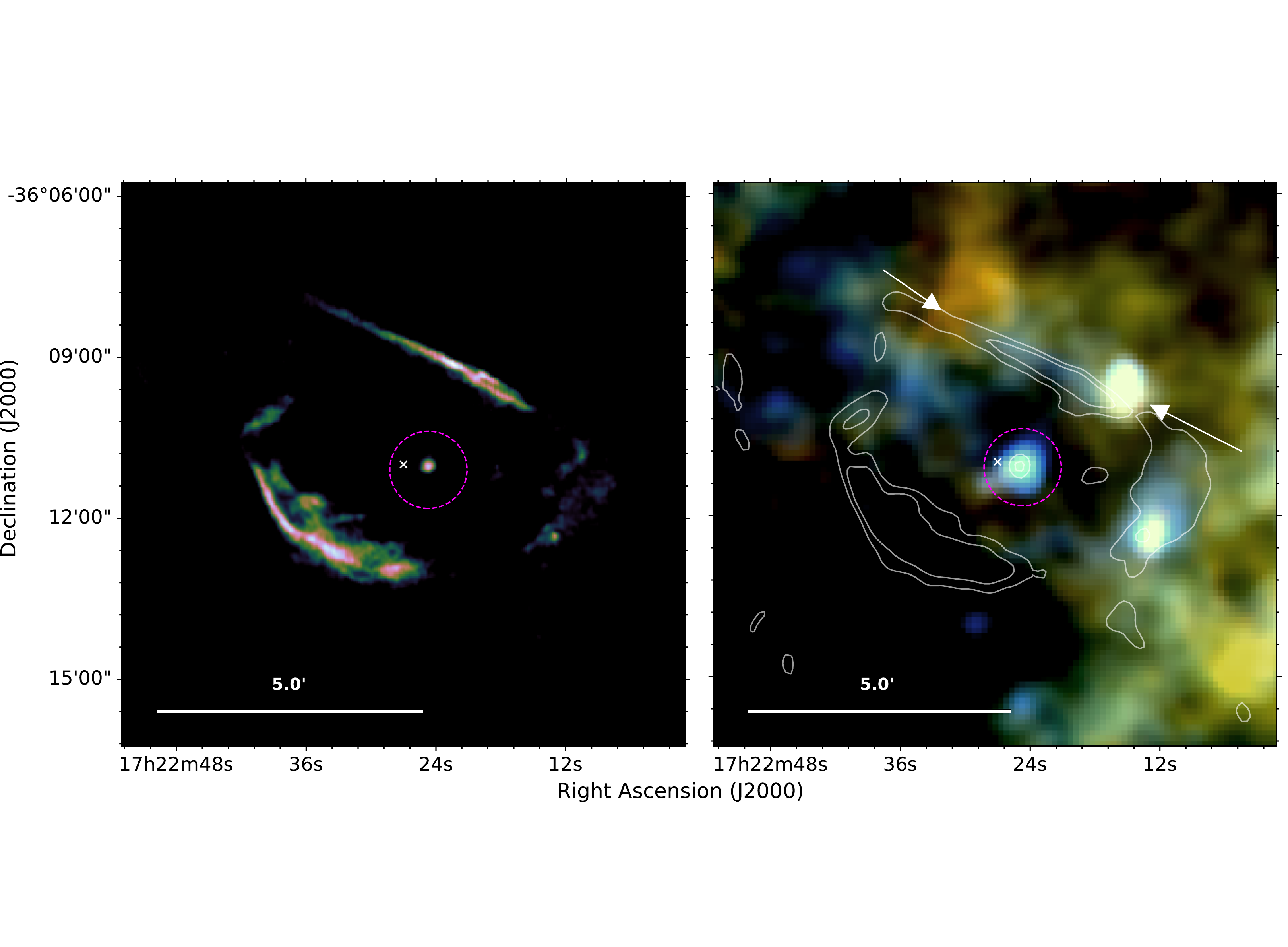}{}
	\includegraphics[width=0.9\linewidth]{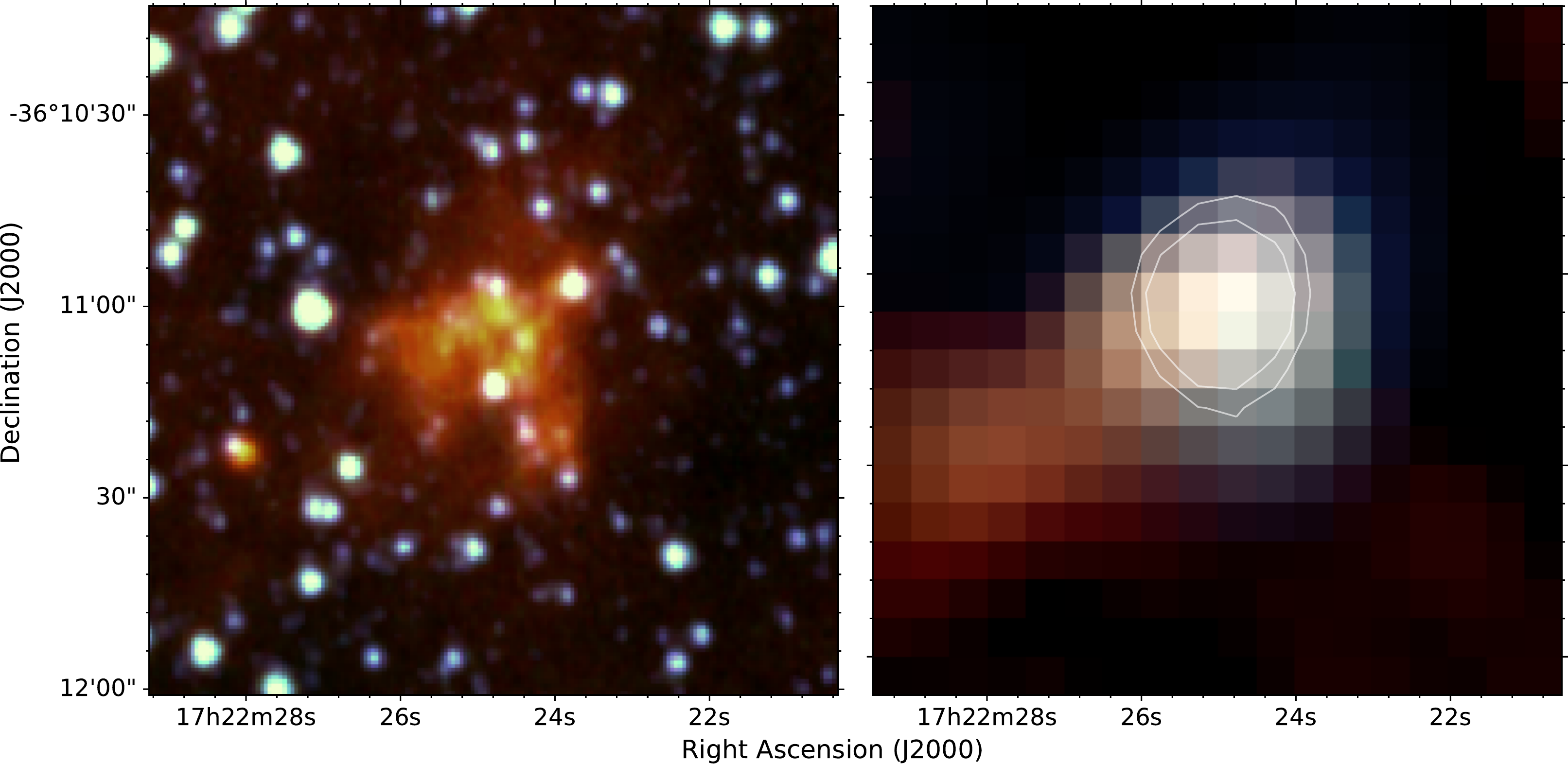}{}
	\caption{{\bf G351.2$+$0.1} -
	{\it Top left:} VLA C-band (4.5\,--\,6.4\,GHz) image and
	{\it Top right:} \hersc\ three colour image with VLA C-band (4.5\,--\,6.4\,GHz) contours overlaid.
	FIR emission is detected in a northern bar, between the two arrows, and more faintly from the south-western shell at 70\,$\mu$m. Emission from the central region within the magenta circle is detected across all \hersc\ wavebands.   The white cross shows the radio coordinates of the SNR centre from \citet{Green2014}.  {\it Bottom left:} Three colour IRAC image using bands 1 (3.6\,$\mu$m - blue), 3 (5.8\,$\mu$m - green) and 4 (8\,$\mu$m - red). This image was made using archive data from the {\it Spitzer} Enhanced Imaging Products Super Mosaic Pipeline. {\it Bottom right:} \hersc\ three colour image with VLA radio contours (see top panel), note we have changed the colour scaling compared to the top right panel in order to highlight the central features more clearly.
	}
	\label{fig:G351.2+0.1Image}
\end{figure*}

{\bf G350.1$-$0.3:} This source is thought to be a very young core-collapse SNR ($\leq$\,1\,kyr) with an associated neutron star \citep{Gaensler2008, Lovchinsky2011}. It has a distorted clumpy structure with X-ray emission from four regions.
\spitz\ MIPS observations revealed 24\,\micron\ emission arising from two of these X-ray regions, which was suggested as likely originating from ejecta dust heated by the reverse shock \citep{Lovchinsky2011}. Although \spitz\ MIPS did not reveal 70\,\micron\ emission associated with the SNR structures \citep{Lovchinsky2011}, with the higher resolution and more sensitive \hersc\ PACS, we have detected dust emission at 70\,\micron\ in one of the 24\,\micron-emitting regions to the east, as indicated by the magenta circle in \autoref{fig:G350.1-0.3Image}. We do not see any evidence of associated emission in the longer \hersc\ wavebands. Since the {\it Herschel} 70\,\micron\ feature overlaps with the 24\,\micron\ dust attributed to the SNR and with X-rays, we classify this source as a level 1 detection. We do not detect FIR emission at the location of the proposed compact object (CCO) \citep{Gaensler2008} (indicated by the gold diamond).  The strong 70\,\micron\ emission from the dust region is similar to the properties of the ejecta dust discovered in SNRs G11.2$-$0.3, G21.5$-$0.9, G29.7$-$0.3 (C19), and G54.1$+$0.3 \citep{Temim2017,Rho2018}. Therefore G350.1$-$0.3 can be added to the growing list of SNRs that contain dust within their ejecta.
\bigskip

\begin{figure}
\includegraphics[width=1.0\linewidth]{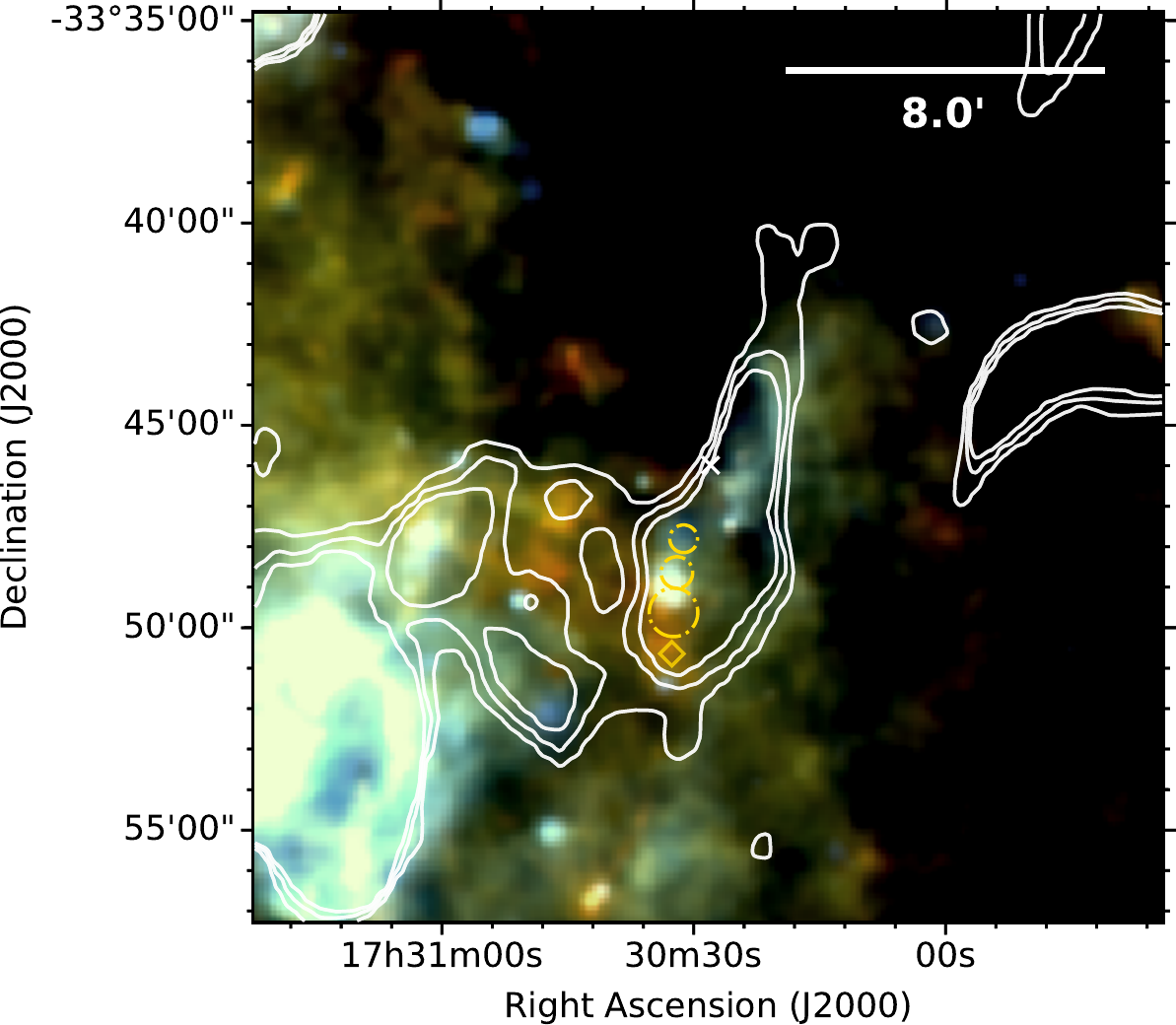}{}
	\caption{{\it G354.1$+$0.1} - \hersc\ three colour image with MOST 843\,MHz radio contours overlaid. FIR emission is detected from within the radio contours although association with the SNR is unclear. The gold circles indicate the locations of bubbles \citep{Simpson2012} and a dense core \citep{Purcell2012}.
	The white cross shows the radio coordinates of the SNR centre from \citet{Green2014}.
	}
	\label{fig:G354.1+0.1Image}
\end{figure}

{\bf G351.2$+$0.1:} This SNR has a well defined radio shell \citep{Caswell1983_g351}, which appears flattened in the north \citep{Dubner1993} indicating that the shockfront is interacting with dense ISM and that the SNR is relatively young.  The SNR has a compact ($15^{\prime \prime}$ diameter) central core in the radio (within the magenta circle in \autoref{fig:G351.2+0.1Image}) and this is attributed to synchrotron emission powered by a pulsar \citep{Becker1988}.  \autoref{fig:G351.2+0.1Image} compares the radio emission at 4.8\,GHz (top left panel) with the \hersc\ three colour image (top right panel).  We processed the radio data presented in \citet{Becker1988} in AIPS, following standard calibration and imaging procedures.

In the \hersc\ images, we detect emission from the northern shell of the SNR coincident with the thin radio shell seen in the VLA image, as indicated by the arrows in \autoref{fig:G351.2+0.1Image}. There is also a bright 70\,$\mu$m clump on the south-western side of the SNR that coincides with a radio peak. We see very little 70\,$\mu$m emission at the location of the south-east shell where the radio is bright. The brightest 70\,$\mu$m emission observed in the SNR is from the central radio core (within the magenta circle, this is also visible in archival {\it Spitzer} MIPS 24\,$\mu$m data and the \spitz\ IRAC image in \autoref{fig:G351.2+0.1Image}).   A further bright 70\,$\mu$m source can be seen just north of the flattened northern shell, and a fainter point source that lies to the immediate east of the central core are also detected; these are detected in the longer \hersc\ wavebands and are likely not associated with the SNR.  We classify this source as a level 1 detection due to the spatial coincidence of the central FIR core and northern FIR shell with corresponding radio features.

The potential FIR detection of a compact object, or PWN is interesting to explore, although the central core is atypical of its class. It is very faint (10\,mJy at 6\,cm) \citep{Becker1988} with spectral slope $\alpha=+0.27$, where flux varies with frequency as $S_{\nu} \propto \nu^{-\alpha}$ suggestive of a very weak source (or possibly old source), compared to Crab-like remnants where typically $\alpha$ ranges from $\sim -0.25$ to $-0.3$.  The oddness of this source led \citet{Becker1988} to discuss whether it could be a H{\sc ii} region or stellar wind, where the spectral slope would be predicted to be $-0.1$ ($+2.0$) for optically thin (optically thick) emission or $+0.6$ respectively.  The source was not detected by the MOST radio survey \citep{Whiteoak1996}, though it lies below their sensitivity level. 
Notably, the central core is not detected in \chand\ archive observations\footnote{\chand\ maps (obsid 3844; 4591) covering the SNR footprint are presented in the massive star forming region omnibus of \citet{Townsley2018}.}.  We note that a hard spectrum X-ray source attributed to a pulsar was observed within a $2 \times 4$\,deg region with GINGA \citep{Tawara1988} overlapping the location of G351.2$+$0.1 and with a previously detected X-ray source (GPS1722-363). The pulsation rate of this source is 414\,s, oscillating between 1\,--\,4\,milliCrab (where 1 Crab is equivalent to $15\,\rm keV\,cm^{-2}\,s^{-1}$ in the X-ray range 2\,--\,10\,keV).

The compact core region is clearly detected in NIR archive images of the region. \autoref{fig:G351.2+0.1Image} (bottom panel) shows a close up view of the NIR and FIR emission in the central core, with NIR {\it Spitzer} IRAC three colour image on left (3.6, 5.8 and 8\,$\mu$m), and the FIR \hersc\ three colour image on the right.  The central radio core is an extended feature in the NIR bands, and is very similar to the IRAC images of the PWN G54.1$+$0.3 and compact PWN in G21.5$-$0.9 \citep{Zajczyk2012}. In G54.1$+$0.3, {\it Spitzer} IRS spectra revealed that the IRAC feature was from argon line emission from the ejecta and a broad silicate dust feature peaking at 9\,$\mu$m \citep{Temim2017,Rho2018}.  The similarity of the NIR emission features could therefore suggest that G351.2$+$0.1 adds to the growing list of sources of SN ejecta dust detected from a PWN.  We return to this source in Section~\ref{sec:g351}.
\bigskip

\begin{figure}
\includegraphics[width=1.0\linewidth]{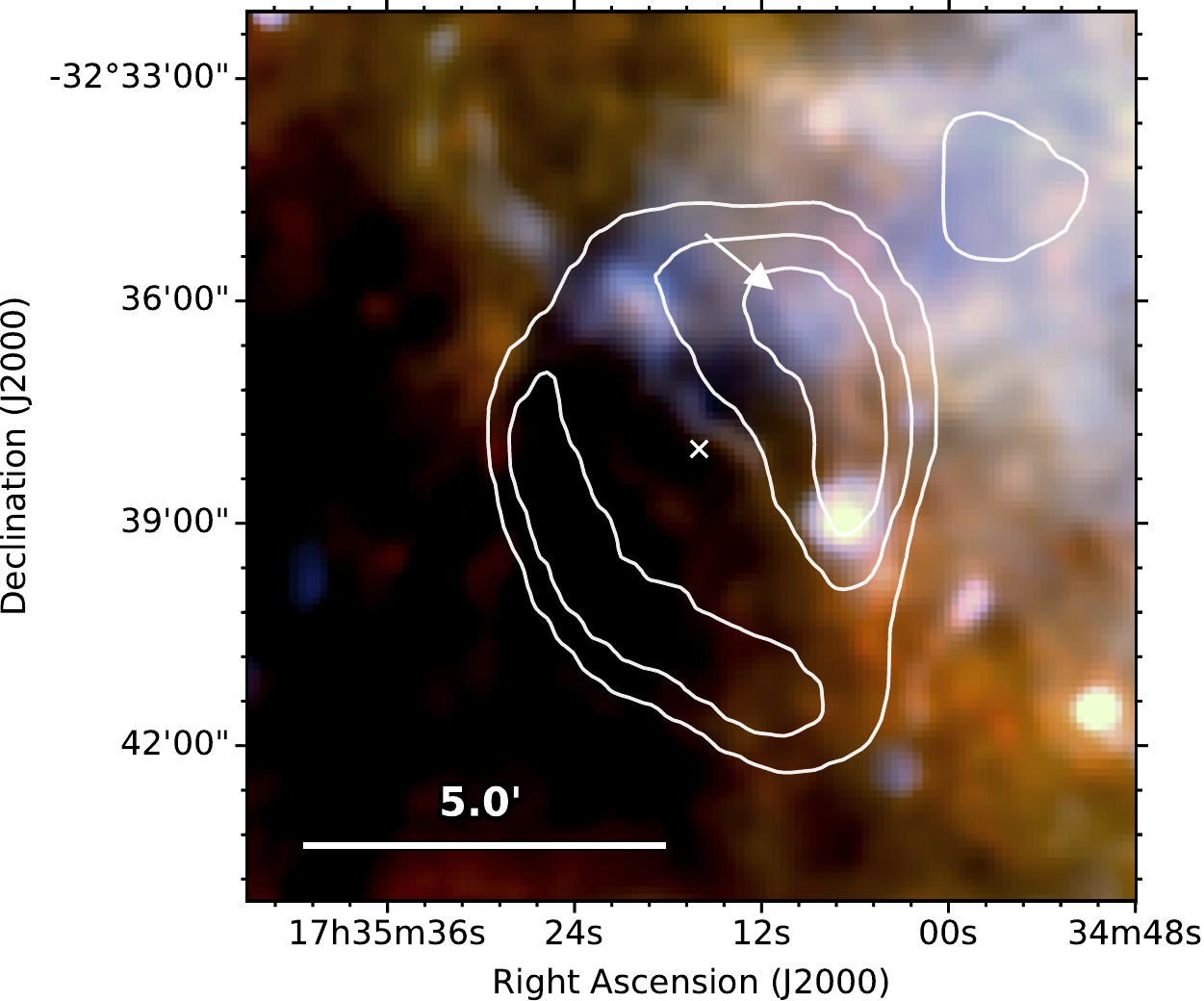}{}
	\caption{{\it G355.6$-$0.0} - \hersc\ three colour image with MOST 843\,MHz radio contours overlaid. There is FIR emission, with strong 70\,\micron\ flux indicated by the arrow, which may be related with SNR radio structure, although the region is extremely confused and any SNR emission is difficult to disentangle.
	The white cross shows the radio coordinates of the SNR centre from \citet{Green2014}.
	}
	\label{fig:G355.6-0.0Image}
\end{figure}

{\it G354.1$+$0.1:} There are multiple radio features in this region and there are FIR filaments of emission observed in the same regions as radio structures. 
Assuming that the radio emission from the SNR is the region encompassing the pulsar  (see \autoref{fig:G354.1+0.1Image}), the FIR may be associated with the SNR, although we cannot definitively determine an association due to the numerous unrelated FIR sources in this region (gold circles in \autoref{fig:G354.1+0.1Image}). Thus we classify this as a level 2 detection. We do not detect emission at the location of the associated pulsar, PSR 1727$-$33, indicated by the gold diamond in \autoref{fig:G354.1+0.1Image} \citep{Frail1994}.
\bigskip

{\it G355.6$-$0.0:} This SNR is another mixed morphology source \citep{Rho1998}. As seen in \autoref{fig:G355.6-0.0Image}, radio observations show a clear shell in the MOST 843\,MHz radio image (\citealt{Gray1994III}, their Figure 2) with enhanced emission to the west. There are multiple FIR filaments in the region of this SNR, although the majority do not seem to correlate with the radio. There is a filament of emission near $\alpha = 17^\text{h}35^\text{m}08^\text{s}, \delta = -32^\circ36^\prime14''$ which may be associated with the radio peak at the western edge of the SNR, although the region is confused and we cannot confidently determine whether this is related. This satisfies our criteria for a level 2 detection in that there is potential FIR emission related to SN features.
\bigskip

\begin{figure}
\includegraphics[width=1.0\linewidth]{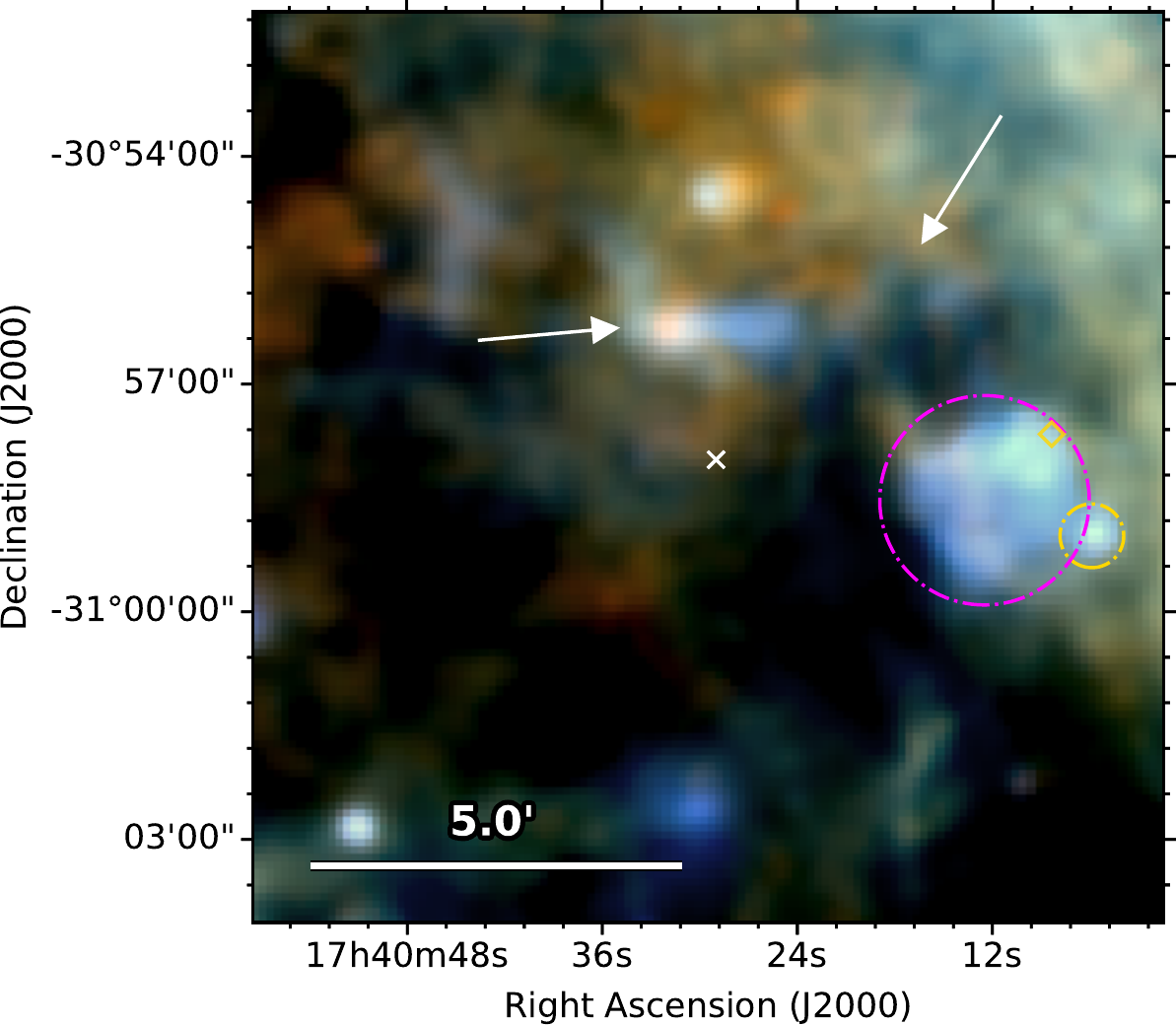}{}
	\caption{{\bf G357.7$-$0.1, The Tornado} -
\hersc\ three colour image.
	We detect dust emission across all \hersc\ wavebands for the `head' of The Tornado, within the magenta circle. We also detect FIR emission from the `tail' of The Tornado, and from a fainter filament extending around the head, as indicated by the arrows.
	The H{\sc ii} region known as the `eye' of The Tornado is detected to the west of the `head', within the gold circle. The gold diamond indicates the location of an OH (1720\,MHz) maser.	The white cross shows the radio coordinates of the SNR centre from \citet{Green2014}.
	}
	\label{fig:G357.7-0.1Image}
\end{figure}

{\bf G357.7$-$0.1 (MSH 17-39):} Known as `the Tornado', this unusual SNR candidate comprises a `head', which appears as a shell- or ring-like feature in the radio \citep{Shaver1985} and an unresolved extended clump in X-rays \citep{Gaensler2003}, a larger extended radio shell/filamentary structure, and an elongated `tail'. These features are marked with a magenta circle, contours and arrows respectively in \autoref{fig:G357.7-0.1Image}. The compact, bright radio source seen to the west of the head at $\alpha = 17^\text{h}40^\text{m}05.9^\text{s}, \delta = -30^\circ59^\prime00''$ is the so-called `eye' of the Tornado. However, this has been shown to be unrelated to the SNR structure and is proposed to be an isolated core embedded in a foreground H{\sc ii} region \citep{Brogan2003, Burton2004}.

Although this region is confused in the FIR (\autoref{fig:G357.7-0.1Image}), the \hersc\ images show clear emission from dust, with bluer colours than the surrounding interstellar material, in the \hersc\ images at the location of the head and the tail. The coincidence of FIR emission with the radio-bright head makes this a level 1 detected source. Further analysis of this source can be found in Chawner et al. {\it in prep}.
\bigskip

\begin{figure}
	\includegraphics[width=1.0\linewidth, trim = 0cm 0cm 2.6cm 0cm, clip]{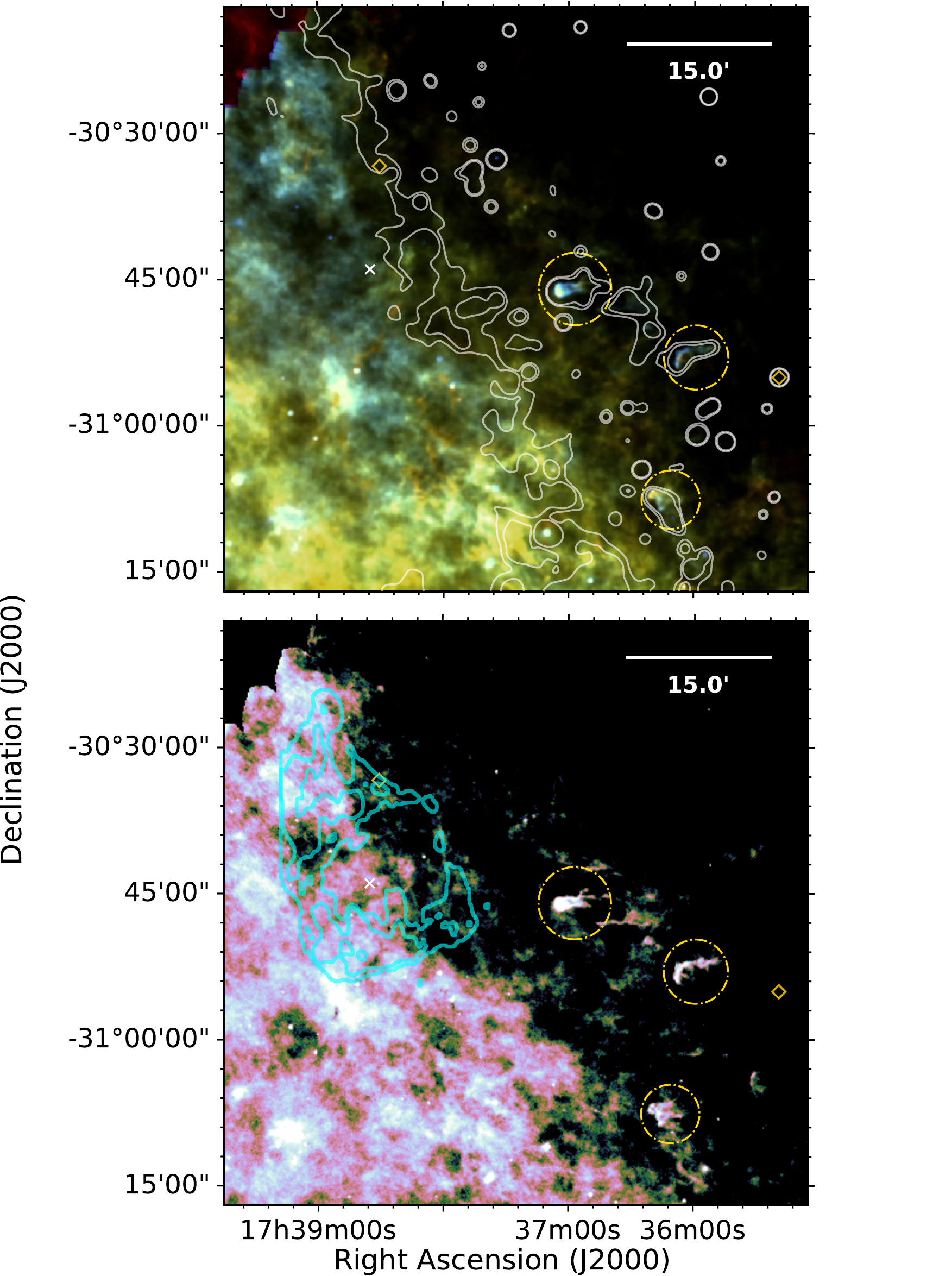}{}
	\caption{{\it G357.7$+$0.3} -
	{\it Top}: \hersc\ three colour image with MIPS 24\,\micron\ contours overlaid.
	{\it Bottom}: \hersc\ 70\,\micron\ image with MOST 843\,MHz contours overlaid (radio image is shown in Figure\,2 from \citet{Gray1994II} and Figure\,1 from \citet{Rho2017}).
	FIR emission is observed within the gold circles from `comet' like structures to the west detected by \citet{Phillips2009}, we do not expect that these structures are associated with the SNR.
	The white cross shows the radio coordinates of the SNR centre from \citet{Green2014}.
	The gold diamond to the west indicates the location of the Mira star V1139 Sco and the diamond in the north indicates the location of an OH (1720\,MHz) maser.
	}
	\label{fig:G357.7+0.3Image}
\end{figure}

\begin{figure}
\includegraphics[width=1.0\linewidth]{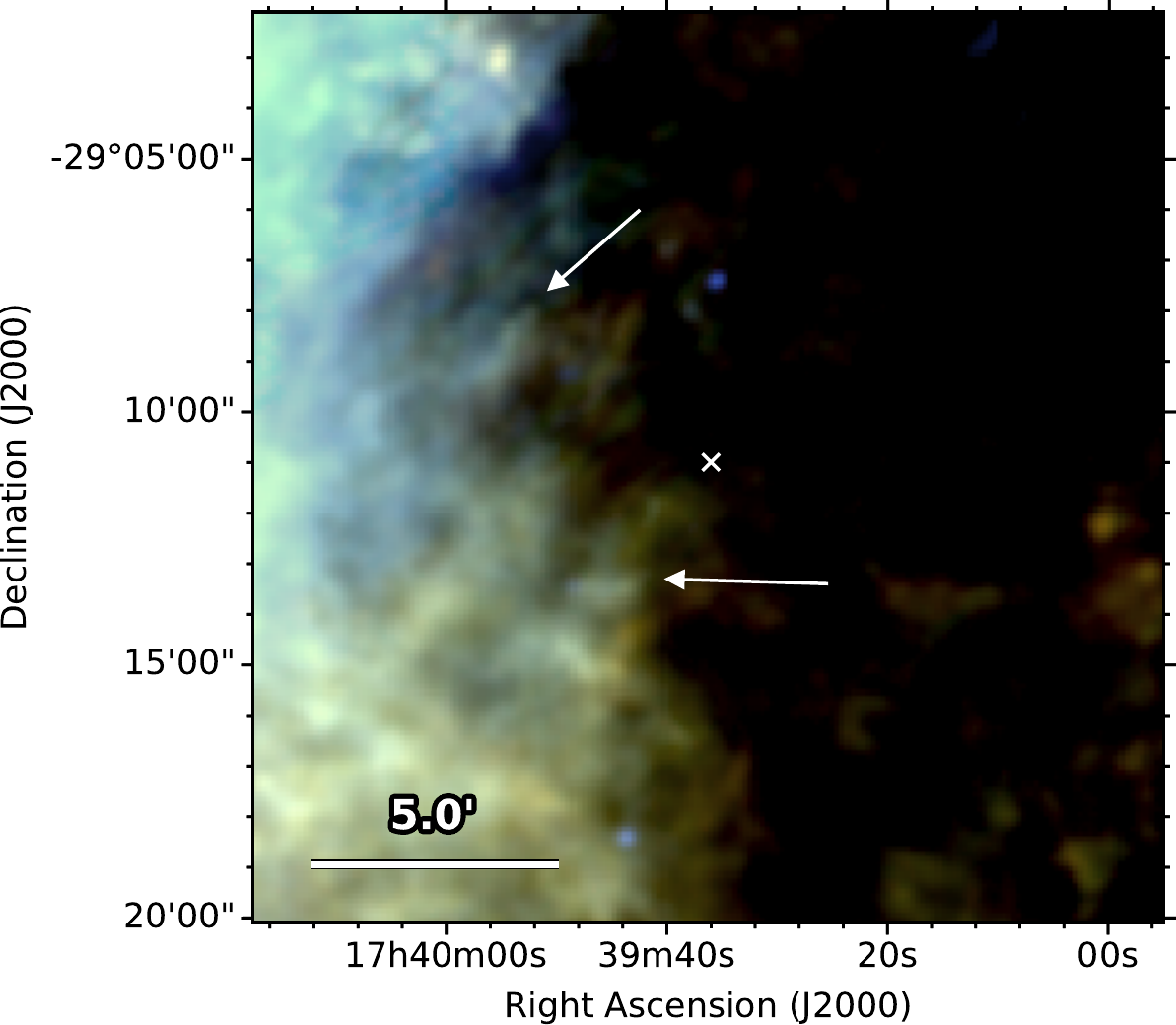}{}
	\caption{{\it G359.1$+$0.9} - \hersc\ three colour image. There are FIR filaments towards the western edge of the SNR which may be associated although the region is very confused.
	The white cross shows the radio coordinates of the SNR centre from \citet{Green2014}.
	}
	\label{fig:G359.1+0.9Image}
\end{figure}

{\it G357.7$+$0.3:} This source has a distinctive square morphology giving rising to the name `the Square Nebula'. There is extended OH (1720\,MHz) emission along the western edge of the SNR \citep{Yusef-Zadeh1999} and the recent discovery of shocked molecules, broad Co in millimetre, shocked H$_2$ lines with \spitz, and 158\,\micron\ [CII] with SOFIA in \citet{Rho2017} provides unambiguous evidence that this SNR is interacting with a molecular cloud. Although there is lots of extended FIR structure making associated emission difficult to distinguish, there are filaments at the western edge near to the OH (1720\,MHz) emission which may be related. We therefore classify this as a level 2 detection.

\citet{Phillips2009} observed comet-like structures to the west of this source which they suggest result from interactions between the SNR and nearby molecular clouds. We detect FIR emission across all \hersc\ wavebands from these structures near to $\alpha = 17^\text{h}36^\text{m}57^\text{s}, \delta = -30^\circ46^\prime02''$, $\alpha = 17^\text{h}35^\text{m}59^\text{s}, \delta = -30^\circ53^\prime03''$, and $\alpha = 17^\text{h}36^\text{m}11^\text{s}, \delta = -31^\circ07^\prime40''$, indicated by the gold circles in \autoref{fig:G357.7+0.3Image}.
We also see a plume of dust emission near the Mira star V1139 Sco, originally seen at 24\,\micron\ in \citet{Phillips2009}. The origin of this emission is unclear; \citet{Phillips2009} propose that the wind is being swept behind the star by an interaction with the SNR, although the Mira star may be unrelated and its location coincidental.
We do not expect that any of these structures are associated with the SNR. If the structures were the result of interactions with the expanding SNR we would expect them to point away from it and the brightest emission would be at the furthest point from the remnant centre. The morphology of these objects is more similar to evaporating gaseous globules (EGGs) or protoplanetary discs, it is therefore more likely that the knots are associated with star formation, which may have been triggered by the SN.
In order to check whether the comet-like structures could be a result of the SNR interaction, we roughly estimate the distance to these regions and compare with the distance travelled by the SNR shock. We note that little is known about the progenitor for this source.
We focus on the structures labelled A to D in Figure 2 of \citet{Phillips2009} and the closest source to the SNR, structure G357.46+0.60. To be consistent with \citet{Phillips2009} we assume that the SNR is at 4 kpc, and then assume that the comet-like structures are at the same distance. Thus the closest edge of the structures are at distances between 22 and 51 pc from the radio centre of the SNR (from \citealp{Green2014}). The distance reached by the shock front is estimated initially by assuming the most optimistic case, whereby the SNR is freely expanding since the explosion (i.e. it is not yet in the Sedov-Taylor phase). Assuming an ejecta mass between 5\,--\,15\,M$_\odot$ and an age of <\,6400 years \citep{Phillips2009}, the shock wave would have reached a maximum distance of $\sim$29.3\,pc in this time. We find that only the closest globule (G357.46+0.60) is within range of the shockwave, and only if the ejected mass is smaller than 8.5\,M$_\odot$. If we assume the SNR is in the Sedov-Taylor expansion phase, the SF globules are well outside of the estimated shock wave radius (5.0\,pc) for a typical ISM n(H) of ~1\,cm$^{-3}$. Higher interstellar densities would further reduce the reach of the SNR. Thus we do not expect that any of these structures are associated with the SNR.

\citet{Phillips2009} used MIPS 24\,\micron\ combined with low (arcmin) resolution IRAS data to propose that the northern-most of these structures contains dust at $\sim$\,30\,K. Fitting a modified blackbody to background subtracted \hersc\ fluxes (see C19 for details) confirms that it contains dust at $\sim$\,33\,K.
\bigskip

{\it G359.1$+$0.9:} This SNR has a shell-like structure in the radio (\citealp{Gray1994III}, their Figure 10).  Comparing the \hersc\ emission by eye with their figure\footnote{unfortunately, the radio fits data for this source or region was not available for download online.}, we see bright extended emission to the east of this source and multiple FIR filaments at the location of the SNR. Some of these filaments may be associated with the SNR, although \autoref{fig:G359.1+0.9Image} shows that this region is very confused and we cannot definitively distinguish SNR and unrelated emission making this a level 2 detection.

\bigskip
In the following section, we look at some of the properties of the dust emission detected in the level 1 SNRs. 

\section{Investigating dust properties in Galactic Supernova Remnants} \label{sec:investigate}
Here we investigate the properties of the dust within a sample of our level\,1 detected SNRs from both this work and C19. We search for detection level 1 SNRs that are not in overly complex/crowded regions of the Galactic Plane, and sources that have not previously been studied in detail (this excludes G11.2$-$0.3, G21.5$-$0.9, G29.7$-$0.3 \citep{Chawner2019}, G54.1$+$0.3 \citep{Temim2017,Rho2018}, and G357.7$-$0.1 (Chawner et al. {\it in prep}). This leaves us with a total of 11 sources out of the 39 level 1 detected SNRs.

First we investigate the dust temperatures in the SNRs. Dust heated by the PWN, a reverse shock, or a forward shock in a supernova remnant should lead to differences in dust temperatures compared to that in the surrounding interstellar medium since shocks lead to high temperature electrons which can collisionally heat dust grains \citep{Dwek1992}. We create temperature maps of the dust in these SNRs (simulations of the temperature maps are discussed in full in Appendix~B1 after estimating the background flux using the Nebuliser routine\footnote{{\url {http://casu.ast.cam.ac.uk/surveys-projects/software-release/background-filtering}}}, which subtracts medium to large scale variations in the background of astronomical images.

\subsection{Background Subtraction} \label{sec:background_subtraction}
A large number of our level 1 detected sample are in complicated regions of the ISM and have large variations in the source flux. We minimise this issue by subtracting contamination from the ISM using the Nebuliser routine.
Nebuliser estimates medium to large scale variations in the background which can be subtracted from our SNR regions to give a more accurate background-subtracted map\footnote{In brief, the Nebuliser algorithm takes a square of $N$ by $N$ pixels around a pixel and estimates the median of the intensities centred on that pixel to estimate the background at that position. The background map is smoothed using a box-car mean filter with box size chosen to be $N/2\,\times\,N/2$ \citep{Irwin2010}. The value of N used varies  between 75 and 350, depending on fluctuations in the background and size of the SNR, and in some cases we mask bright sources to avoid overestimating the background.}. As demonstrated in Appendix~B1, this can be used to accurately determine the properties of source-related structures.

Many of our level 1 detected SNRs are in crowded fields and are close to or interacting with unrelated structures such as H{\sc ii} regions. We cannot subtract these using Nebuliser without also losing information about the SNR and our resultant temperature maps are sensitive to some contamination from such sources.
This difficulty is aggravated as the ISM is bright at \hersc\ wavelengths, especially 160\,--\,500\,\micron, making some SNRs difficult to distinguish in any other than the 70\,\micron\ waveband.
Consequently, we find that the Nebuliser background subtraction at 160\,\micron\ is too harsh in some regions, although this effect is much reduced compared with for example, if we were to simply subtract a constant value across the entire region.

\subsection{Flux Estimates} \label{sec:SNRfluxes}
\begin{table*} 
	\csvreader[tabular= l c c c c c c,
				late after last line=\\\hline,
				table head=\hline
				SNR & \multicolumn{5}{c}{Flux (Jy)}\\
				& \parbox{1cm}{\centering 70\,\micron} & \parbox{1.5cm}{\centering 160\,\micron} & \parbox{1cm}{\centering 250\,\micron} & \parbox{1cm}{\centering 350\,\micron} & \parbox{1cm}{\centering 500\,\micron} \\\hline\hline] 
	{SNR_Fluxes_Nebuliser.csv}{SNR=\snr, Flux_70=\fluxone, Flux_160=\fluxtwo, Flux_250=\fluxthree, Flux_350=\fluxfour, Flux_500=\fluxfive, Distance=\dist, Sigma_Clipped_Mass=\mass, Temperature=\temp, Ref_Num=\refnum} 
	{\snr & \fluxone & \fluxtwo & \fluxthree & \fluxfour & \fluxfive} 
	\\[1.5pt]
	\caption{
	The SNR flux is measured within the dashed white circle in \autoref{fig:dusttempmaps} from background subtracted maps convolved to the resolution of the 500\,\micron\ \hersc\ map.
	In some cases point sources within the aperture are masked before measuring the flux.
	Uncertainties are estimated from the standard deviation of the background map within the aperture and the \hersc\ calibration uncertainty.
	$^*$Sources which may be confused with a H{\sc ii} region \citep{Anderson2017, Gao2019}.
	}
	\label{tab:SNRFluxes}
\end{table*}

After subtracting the medium-large scale background with Nebuliser we estimate the FIR flux from each SNR in five {\it Herschel} wavebands. For several of our sample there is some contamination remaining in the background-subtracted maps as there is structure of a similar scale to the SNR which is difficult to remove. In these cases we apply a secondary subtraction estimated from the average level using an annulus. The fluxes for our 11 SNRs are shown in \autoref{tab:SNRFluxes}.

The uncertainty associated with the flux is estimated using a combination of the \hersc\ calibration uncertainty, variation of the Nebuliser background map within the aperture, and, where applicable, variation within the annulus used for a secondary background subtraction. In these sources the contaminating ISM flux may act to decrease our estimates of the dust temperature, nevertheless, this effect will be much reduced compared with temperatures estimated without any background subtraction, as seen in \autoref{fig:example_t_sub}.

\subsection{Temperature  and Mass Maps of Milky Way SNRs} \label{sec:tempmaps}
\begin{figure}
	\centering
  	\includegraphics[width=\linewidth]{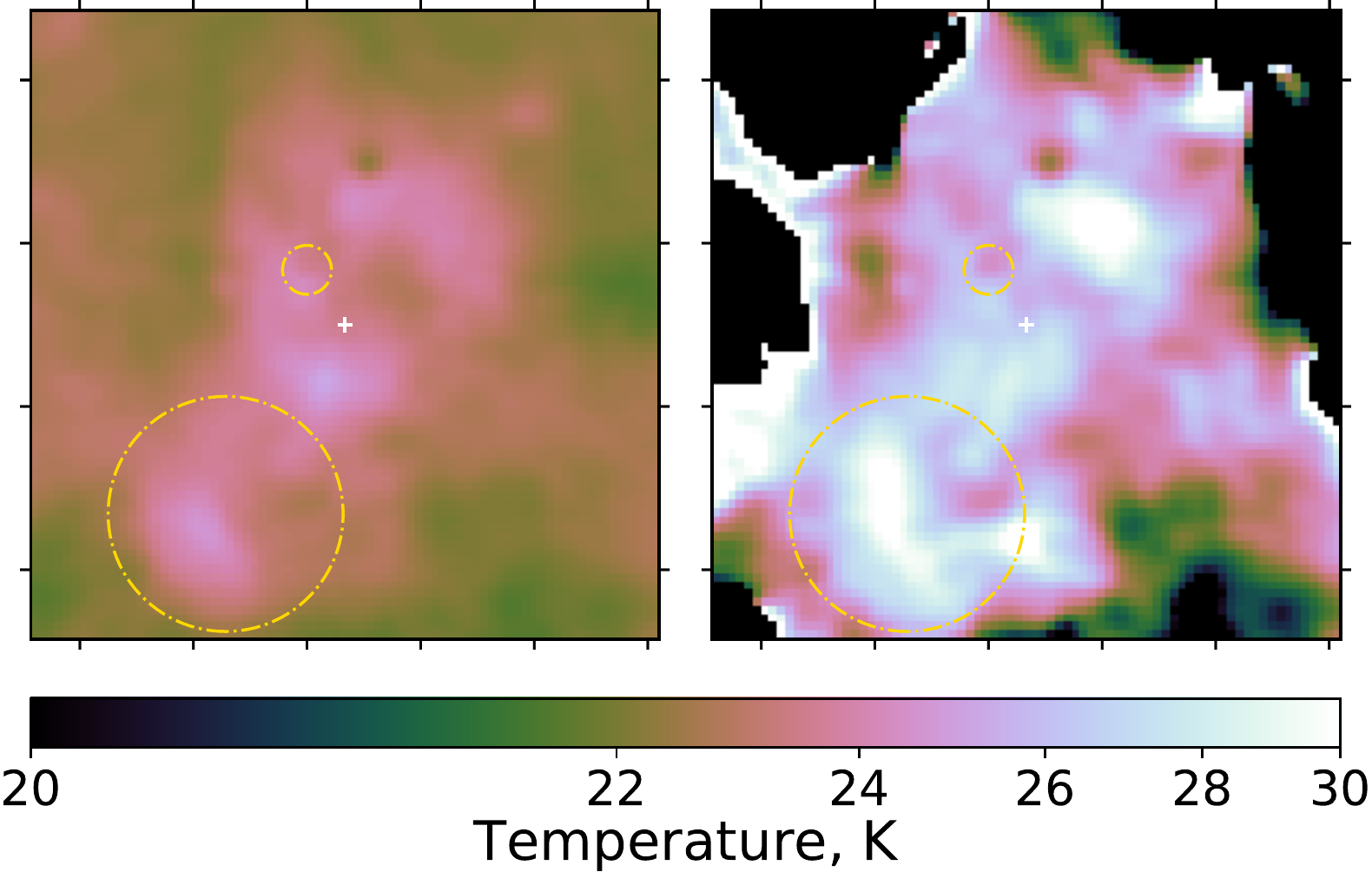}{}
	\caption{
	{\it Left:} Dust temperature map of the level 1 detected SNR G8.3$-$0.0 (\autoref{fig:G8.3-0.0Image}), derived using \autoref{eqn:dusttemp} without background subtraction.
	{\it Right:} Dust temperature map derived after using Nebuliser to subtract the background.
	The {\it Herschel} images used to create the right-hand figure have been convolved to the 250\,\micron\ resolution to improve the signal.
	We are clearly able to identify a wider range of temperatures after applying Nebuliser background subtraction.
	The large and small gold circles indicate the locations of an IR bubble and a maser respectively.}
	\label{fig:example_t_sub}
\end{figure}

The three colour images created out of the \hersc\ images shown in Section~\ref{subsec:IndividualResults} and Figure~A1 reveal dusty structures and features that appear to be associated with other known SNR tracers e.g. radio, X-ray or optical emission (i.e. det\,=\,1) or potentially related to the SN (i.e. det\,=\,2). The colour of the dust emission in these images is, in effect, tracing the dust temperature in the environs of the SNR, however the colour scales chosen earlier are arbitrary, selected only to highlight dusty features. Here we attempt to investigate the temperature of SNR-related dust and potential heating sources using the FIR flux ratio (70\,\micron~/~160\,\micron) seen by \hersc\ for the detected level 1 sources (following similar studies that used 24\,\micron~/~70\,\micron\ flux ratios,  \citealp{Sankrit2010,Williams2011,Goncalves2011,Lakicevic2015,Temim2015,Koo2016}). We select this colour ratio as this should be sensitive to dust emission from grains at temperatures of $\sim$\,15\,--\,70K. It should also be less affected by contamination from ionic line emission, which can account for a non-negligible proportion of the 24\,$\mu$m flux \citep[e.g.][]{Koo2016,DeLooze2019} and less affected by unrelated point sources seen in 24\,$\mu$m images.  However, this ratio, and particularly the 160\,$\mu$m band, may be affected by confusion with unrelated interstellar dust and line emission ([CII] 158\,\micron).  The temperature maps are derived using the 70 and 160~\micron\ Nebuliser subtracted maps and assuming a single temperature dust component with power law dust opacity index $\beta=1.9$ (see C19 for more details) using the equation:

\begin{equation} \label{eqn:dusttemp}
	\frac{F_{70} } {F_{160}} = \bigg(\frac{\lambda_{160}}{\lambda_{70}}\bigg)^{\beta+3} ~ \frac{{\rm exp}\left({\frac{hc}{\lambda_{160} kT}}\right)-1}{{\rm exp}\left({\frac{hc}{\lambda_{70} kT}}\right)-1}.
\end{equation}

where $F_{70}$ and $F_{160}$ are the fluxes at 70 and 160\,\micron, $\lambda$ is the wavelength, $h$ is the Planck constant, $c$ if the speed of light, $k$ is the Boltzmann constant, and $T$ is the dust temperature.

\begin{figure*}
	\centering
	\includegraphics[width=0.24\linewidth]{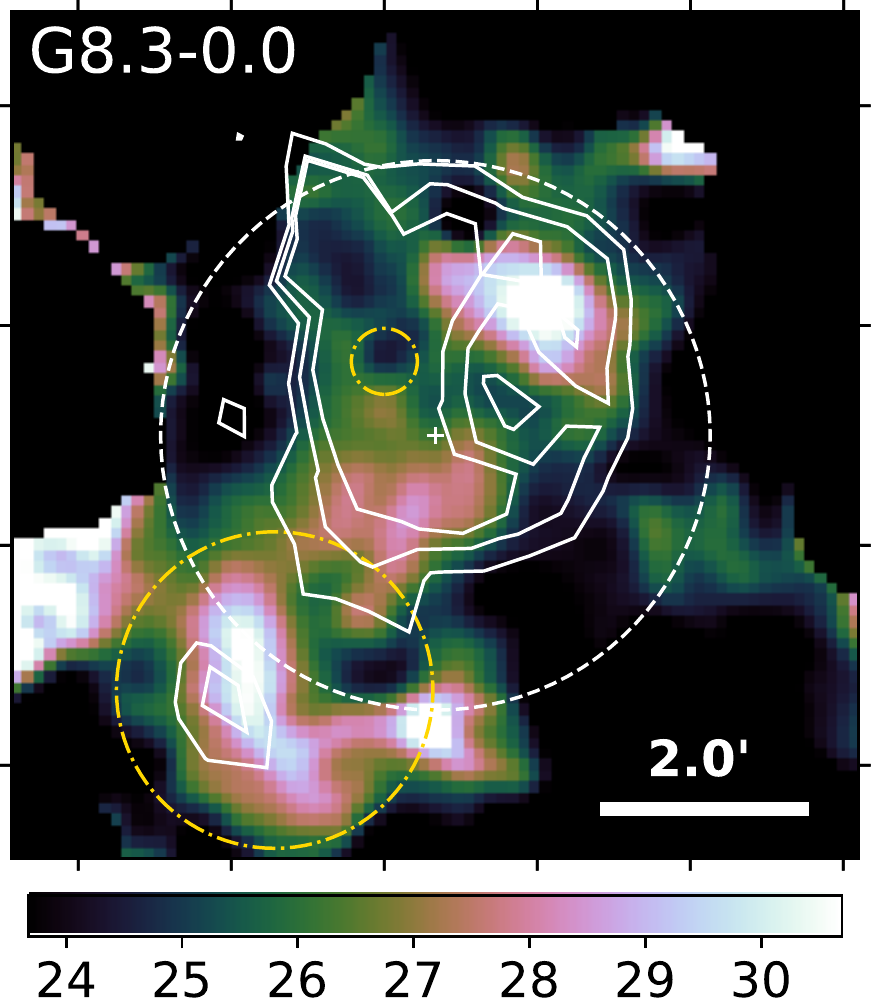}
	\includegraphics[width=0.24\linewidth]{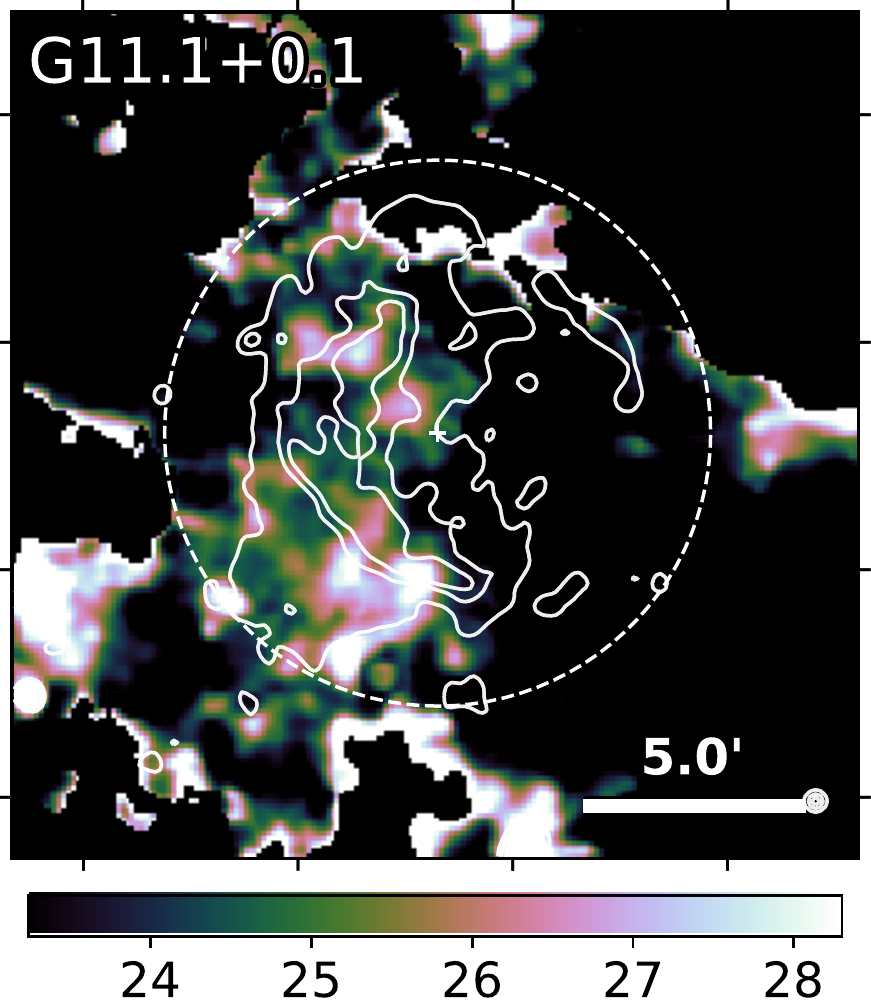}
	\includegraphics[width=0.24\linewidth]{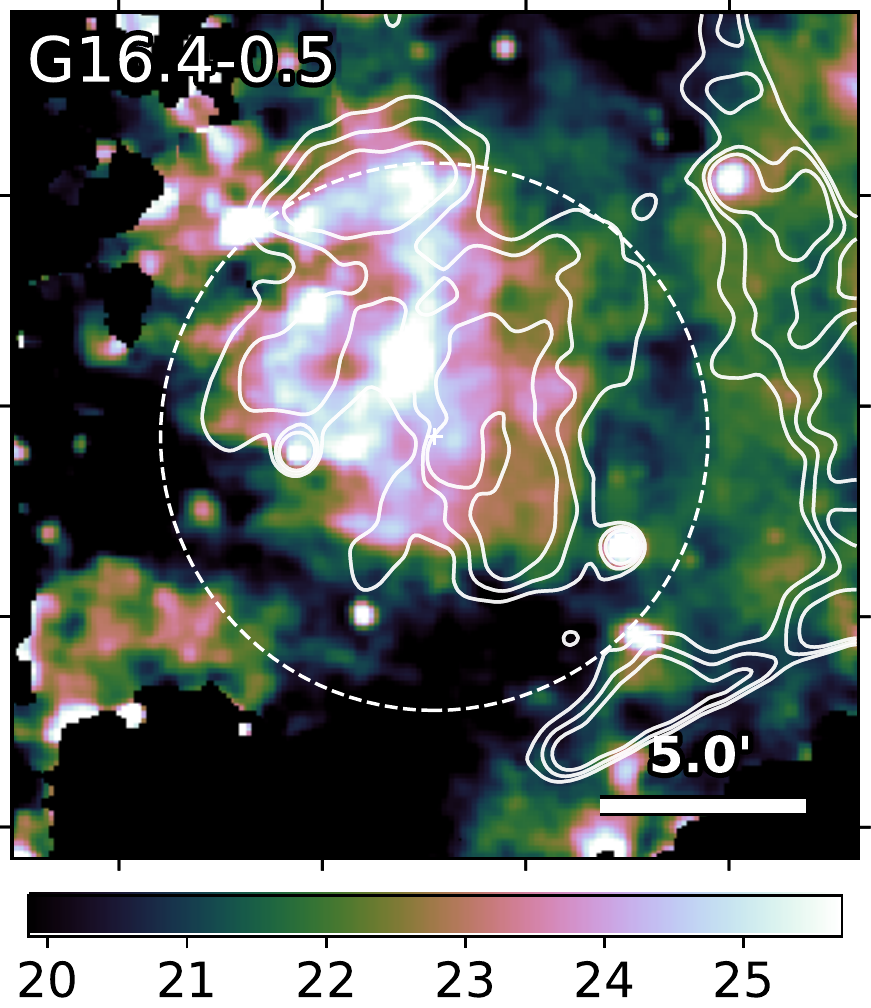}
	\includegraphics[width=0.24\linewidth]{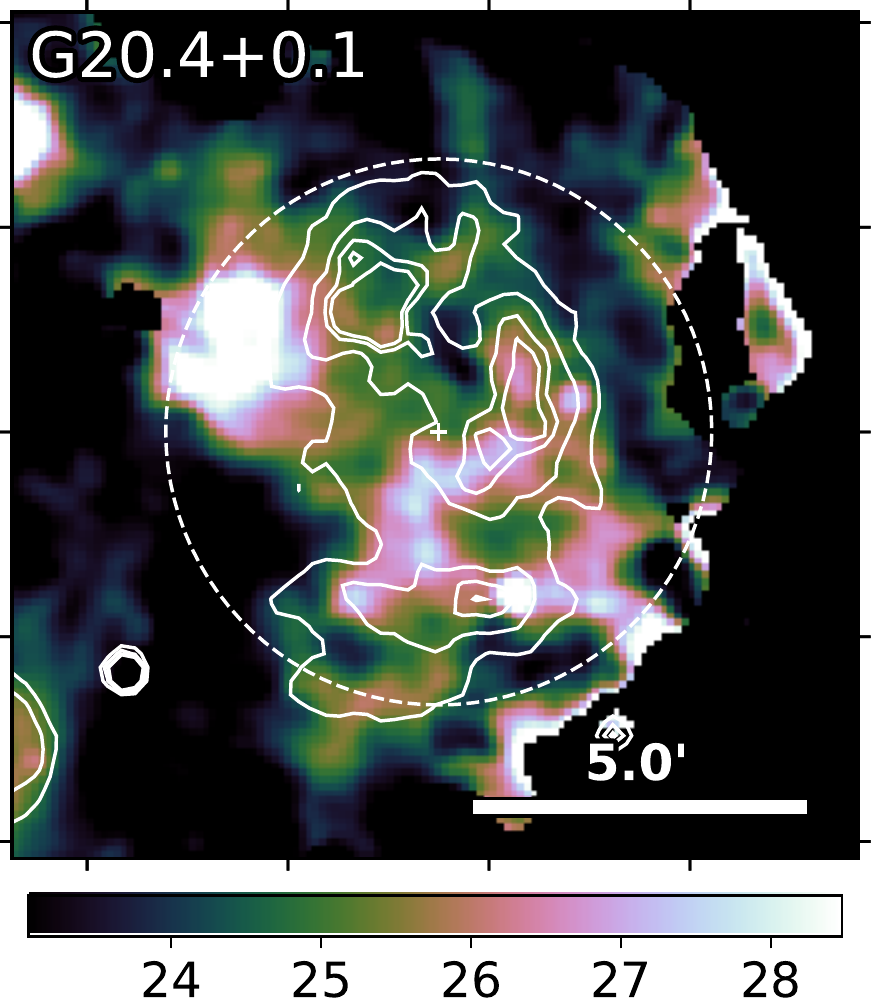}
	\includegraphics[width=0.24\linewidth]{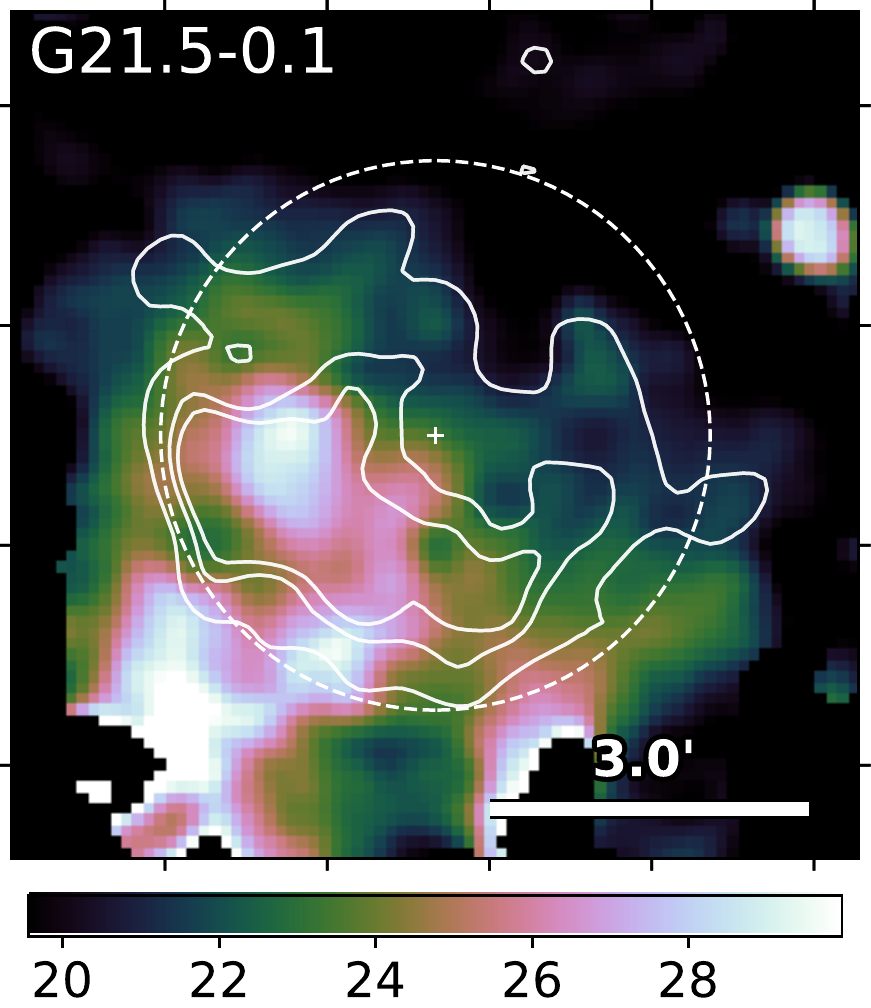}
	\includegraphics[width=0.24\linewidth]{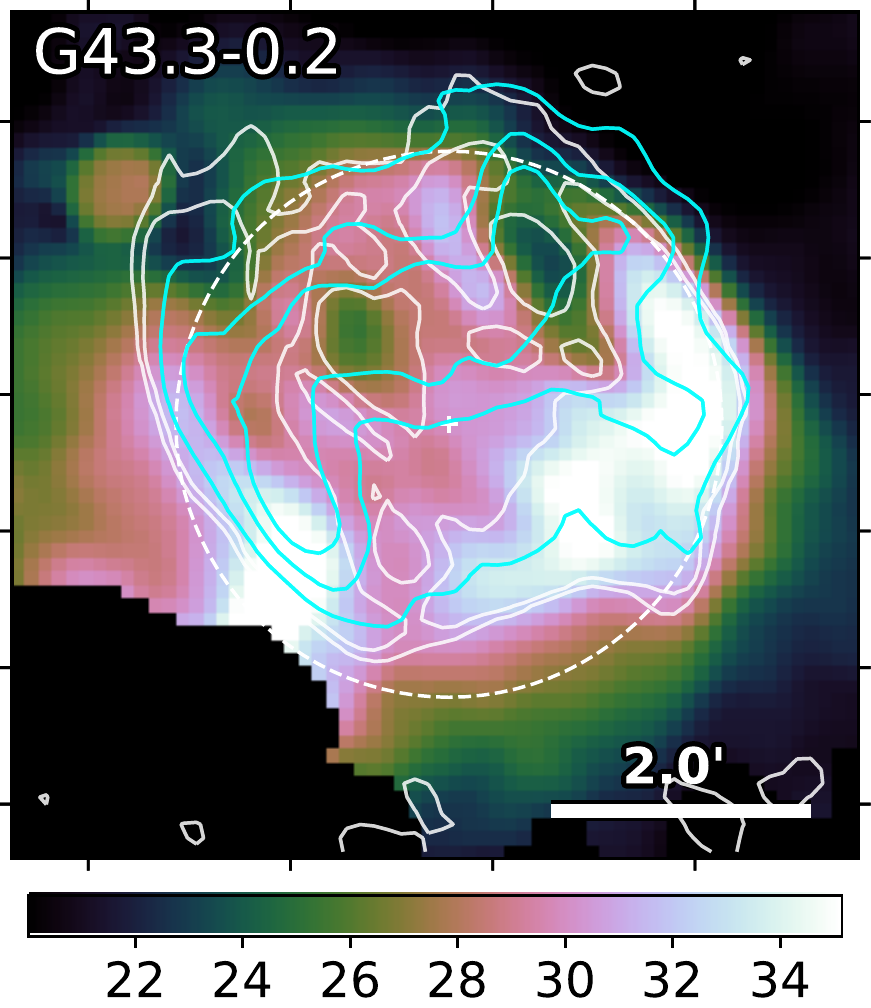}
	\includegraphics[width=0.24\linewidth]{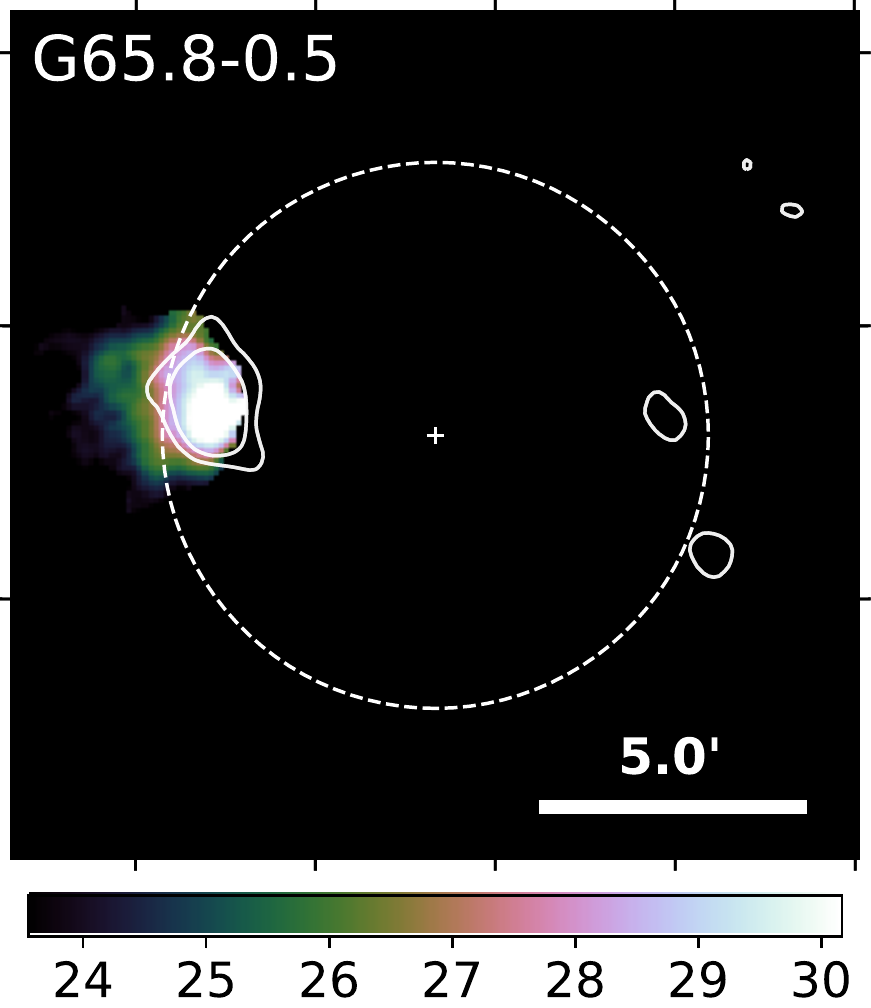}
	\includegraphics[width=0.24\linewidth]{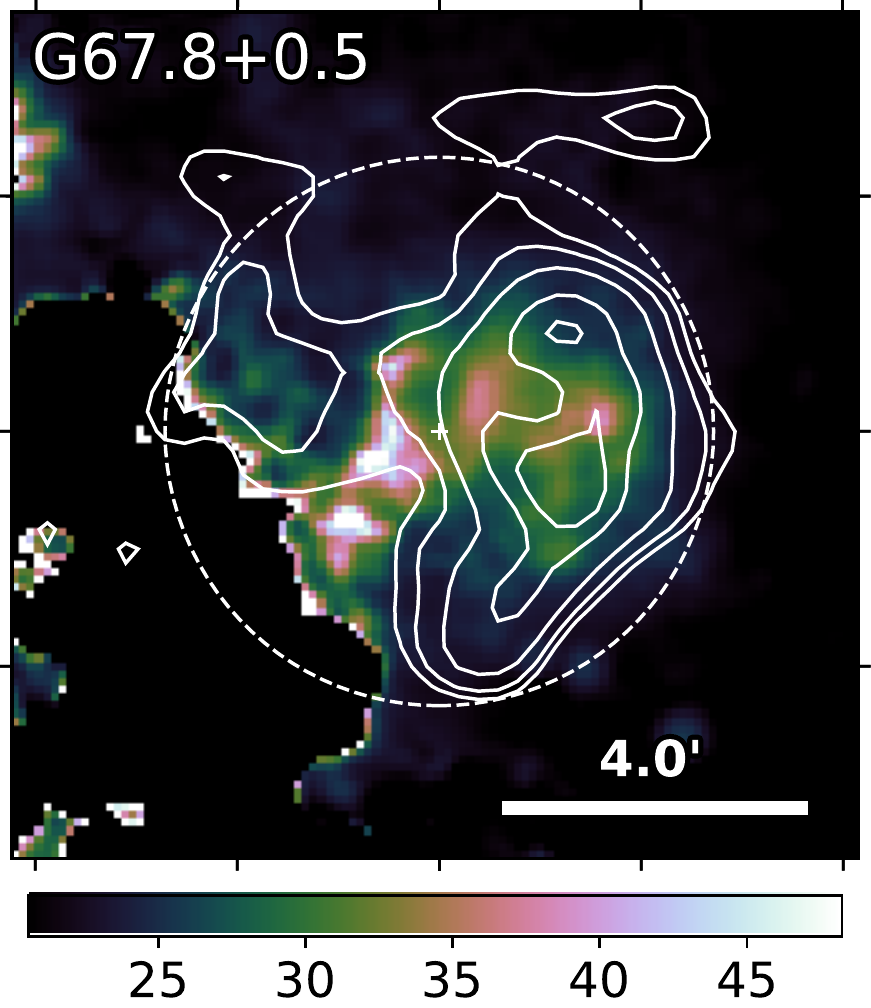}
	\includegraphics[width=0.24\linewidth]{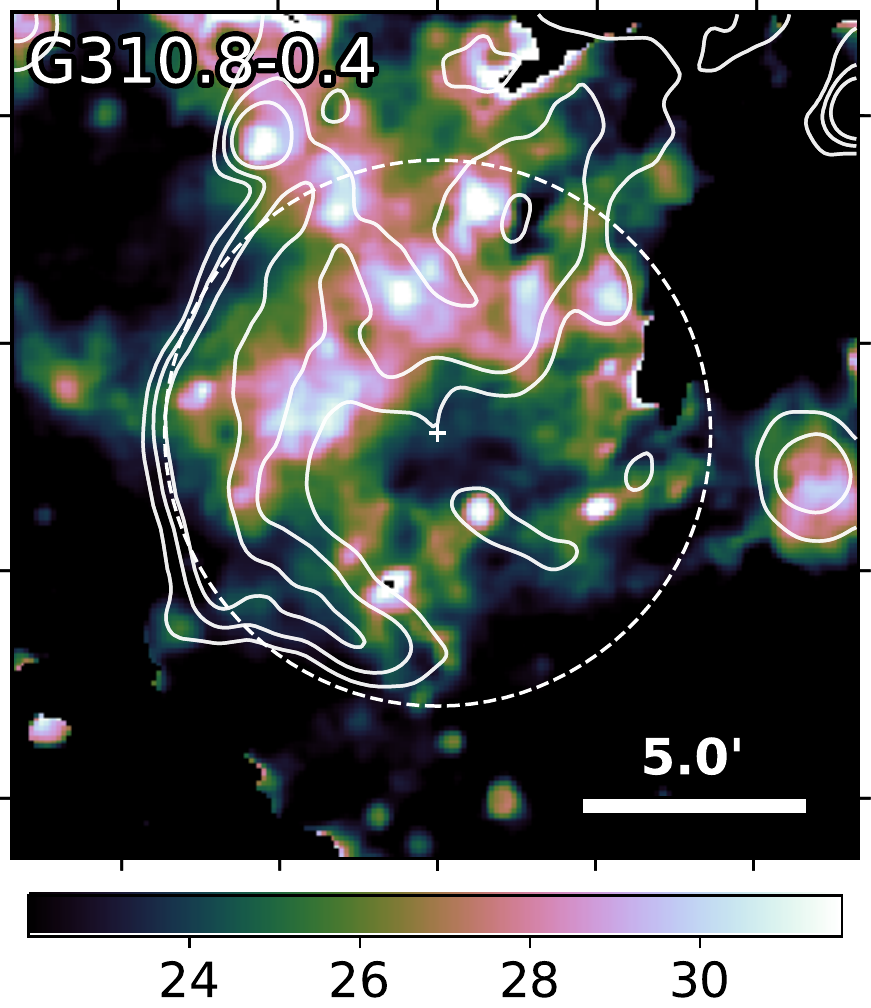}
	\includegraphics[width=0.24\linewidth]{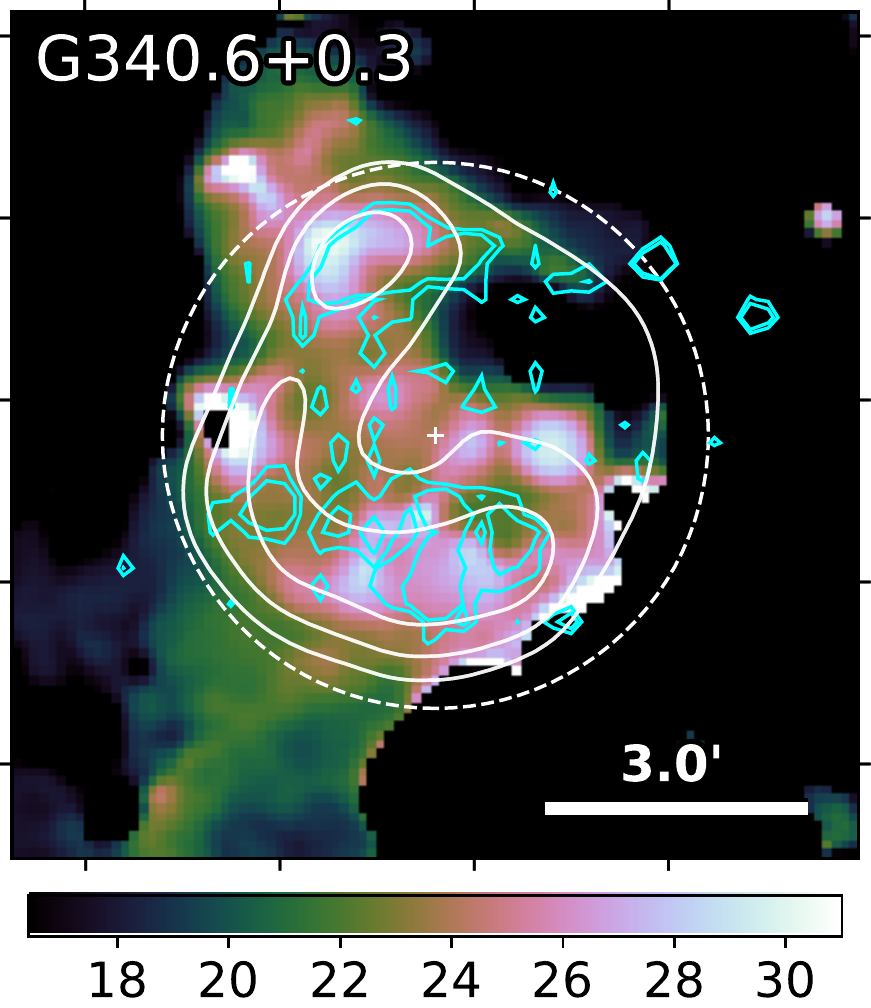}
	\includegraphics[width=0.24\linewidth]{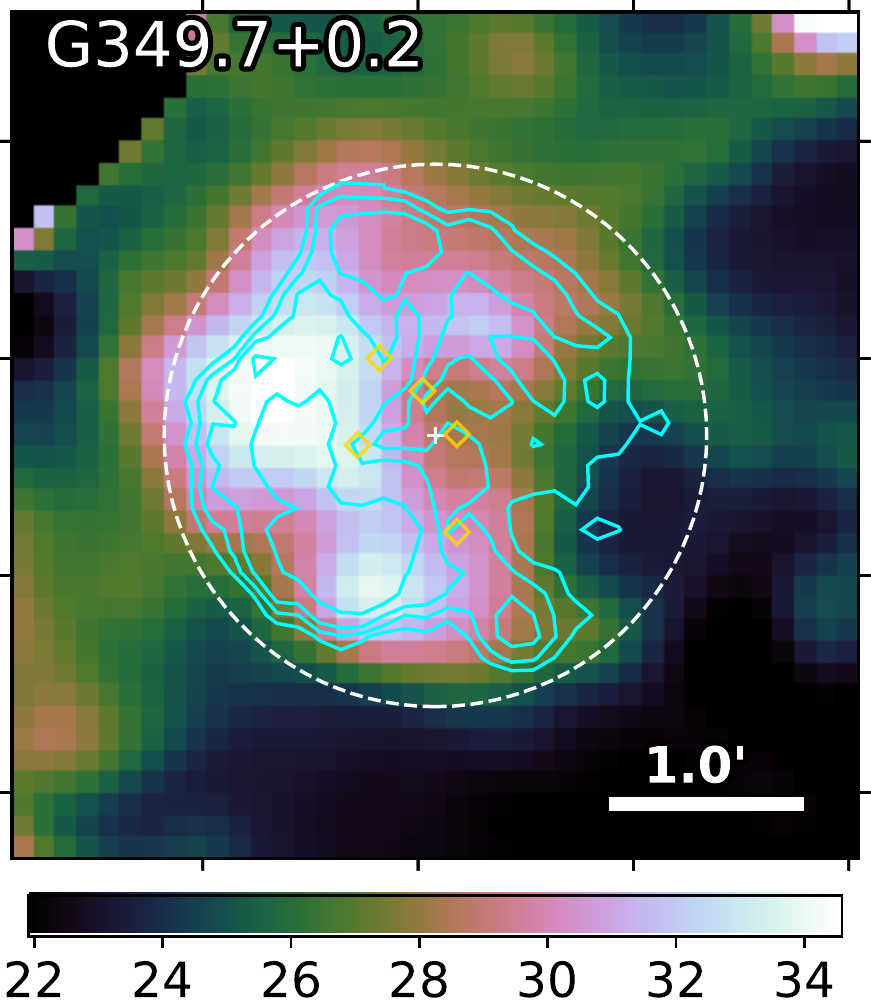}
	\caption{Dust temperature maps of our subset of detected level 1 sources derived using \autoref{eqn:dusttemp}.  The images have been convolved to 6\,arcsec to improve the signal.
	VLA GPS 20\,cm contours (white) are overlaid onto images of G8.3$-$0.0, G11.1$+$0.1, G16.4$-$0.5, G20.4$+$0.1, G21.5$-$0.1, and G43.3$-$0.2.
	NVSS 1.4\,GHz contours (white) are overlaid onto images of G65.8$-$0.5 and G67.8$+$0.5.
	MOST 843\,GHz contours (white) are overlaid onto images of G310.8$-$0.4 and G340.6$+$0.3.
	Chandra contours (cyan) are overlaid onto the images of G43.3$-$0.2, G340.6$+$0.3, and G349.7$+$0.2.
	In some cases over-subtraction of the 160\,\micron\ background introduces hot artefacts at the edges of masked areas.
	}
	\label{fig:dusttempmaps}
\end{figure*}

We present the dust temperature maps in \autoref{fig:dusttempmaps} with individual notes discussing the map for each SNR provided in Appendix~B2.
An example of the temperature maps derived for the SNR G8.3$-$0.0 before and after background subtraction is shown in \autoref{fig:example_t_sub} to illustrate the significant underestimation of the dust temperature (thus leading to an overestimation of the dust mass) if the background level is not properly accounted for.
Before subtraction only a narrow range of dust temperatures are seen in the SNR ($T_{d}\,<\,26$\,K), whereas after subtraction dust is seen at $26\,<\,T_{d}\,<\,30$\,K.
See also the recent reductions of the proposed SN dust mass in the Crab Nebula due to more sophisticated measures of the background level \citep{Nehme2019, DeLooze2019}.  We attempt to interpret the dust temperatures in Section~\ref{subsec:CollisionalHeating}.

We note that the Nebuliser routine used to subtract the background finds medium to large ISM variations, but may not subtract smaller scale complicated structure. Therefore, there may be some contamination remaining, which will act to artificially reduce the temperature. For two sources (G43.3$-$0.2 and G349.7$-$0.2) we find lower dust temperatures than \citet{Koo2016} derived using the 24\,--\,70\,\micron\ flux ratio, this may be due to differences in the methods of analysis and/or line contamination in the 24\,\micron\ waveband.
Conversely, in some regions the background may be over-subtracted where there are small scale variations, this seems to be more of an issue for the 160\,\micron\ images and this leads to hot artefacts in the maps.

\begin{figure*}
	\centering
	\includegraphics[scale=0.45]{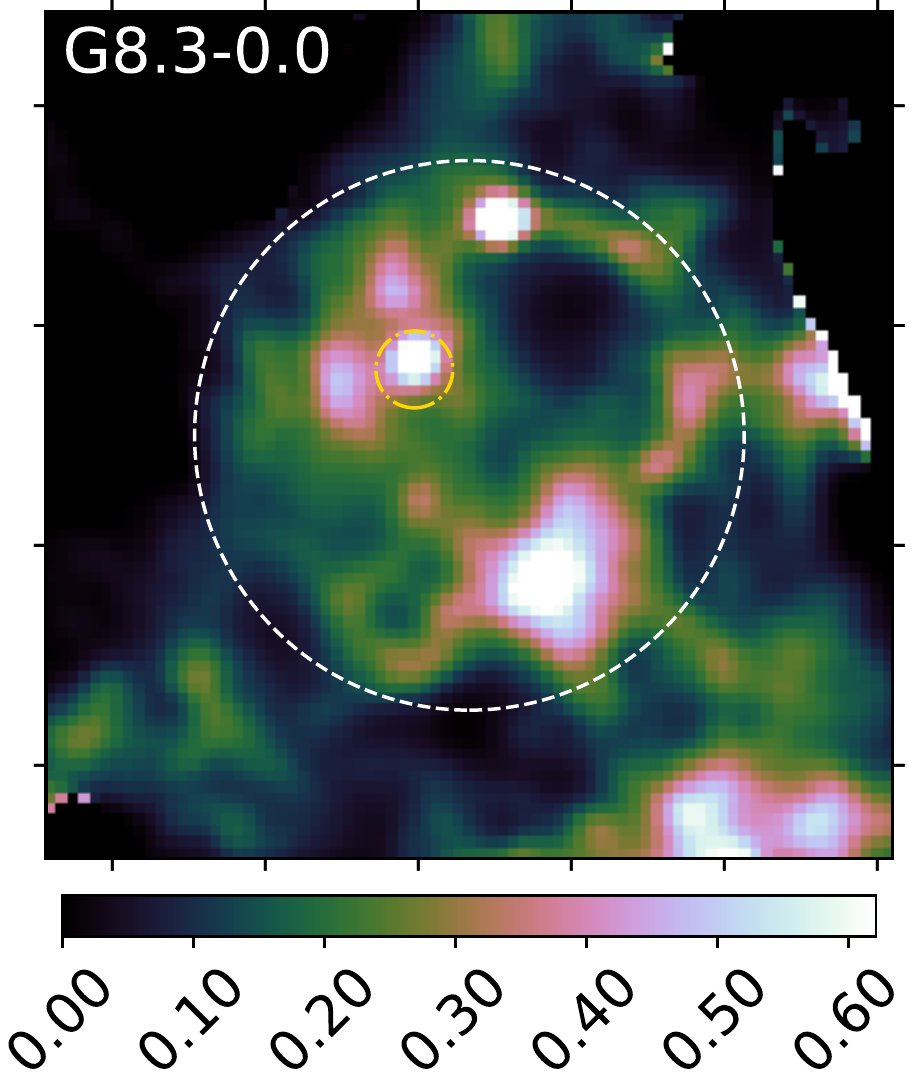}
	\includegraphics[scale=0.45]{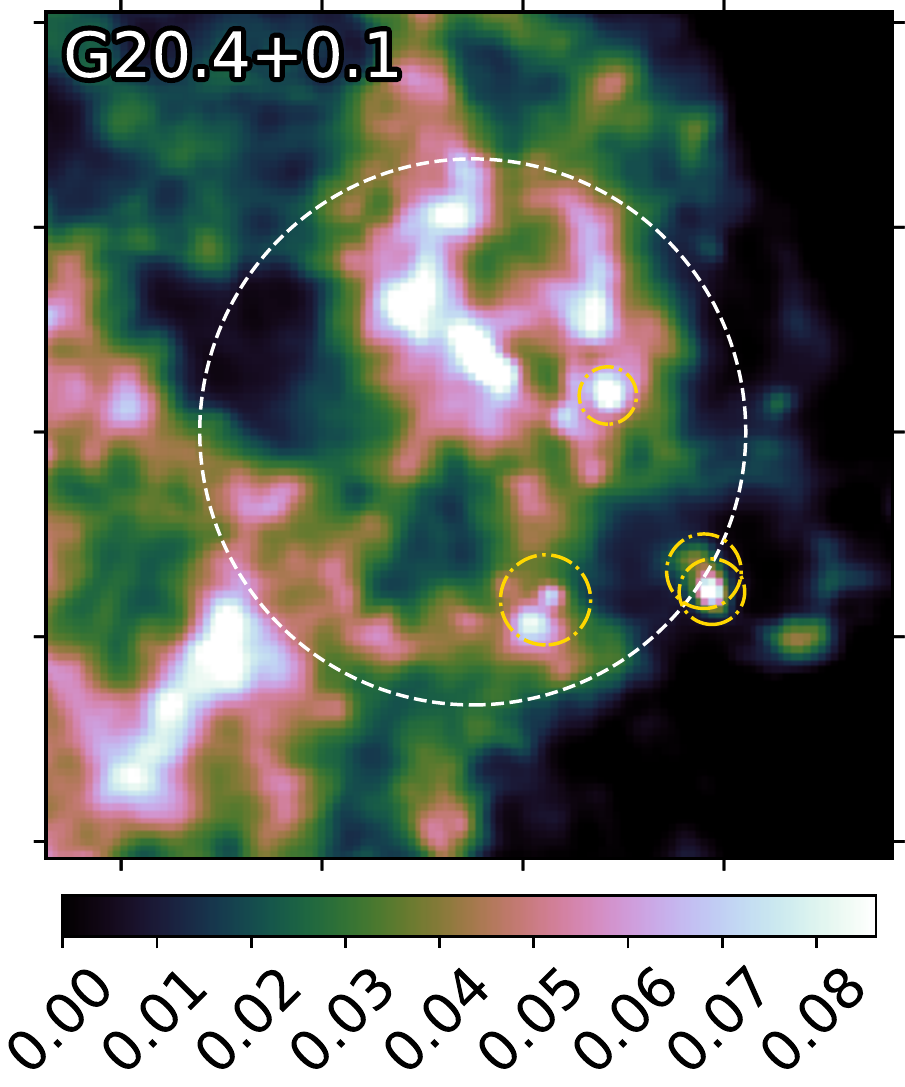}
	\includegraphics[scale=0.45]{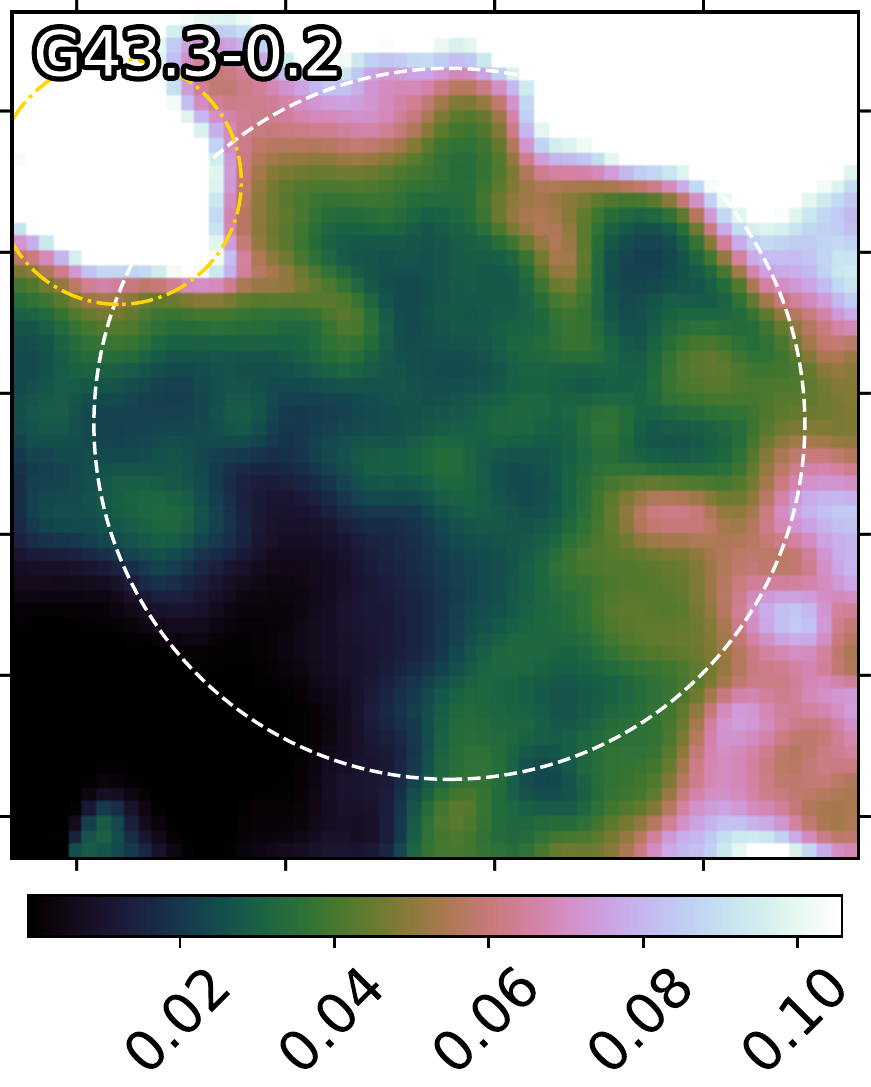}
	\includegraphics[scale=0.45]{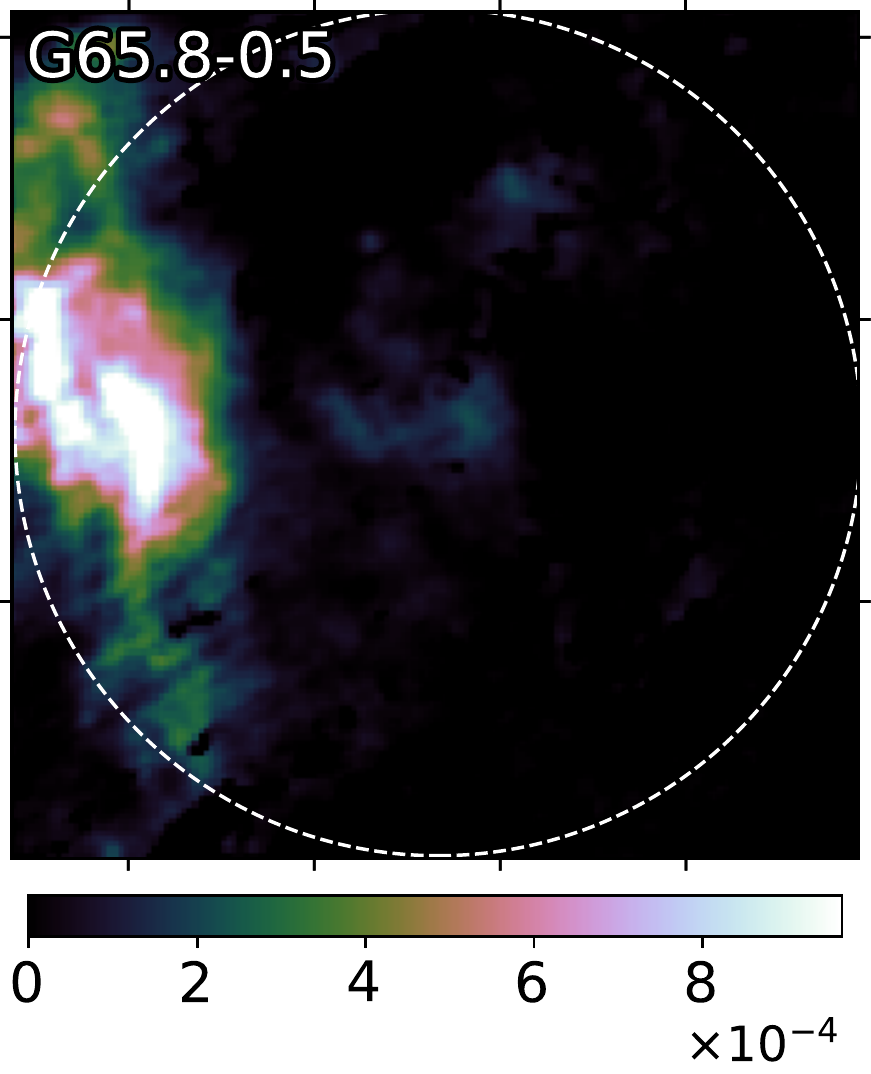}
	\includegraphics[scale=0.45]{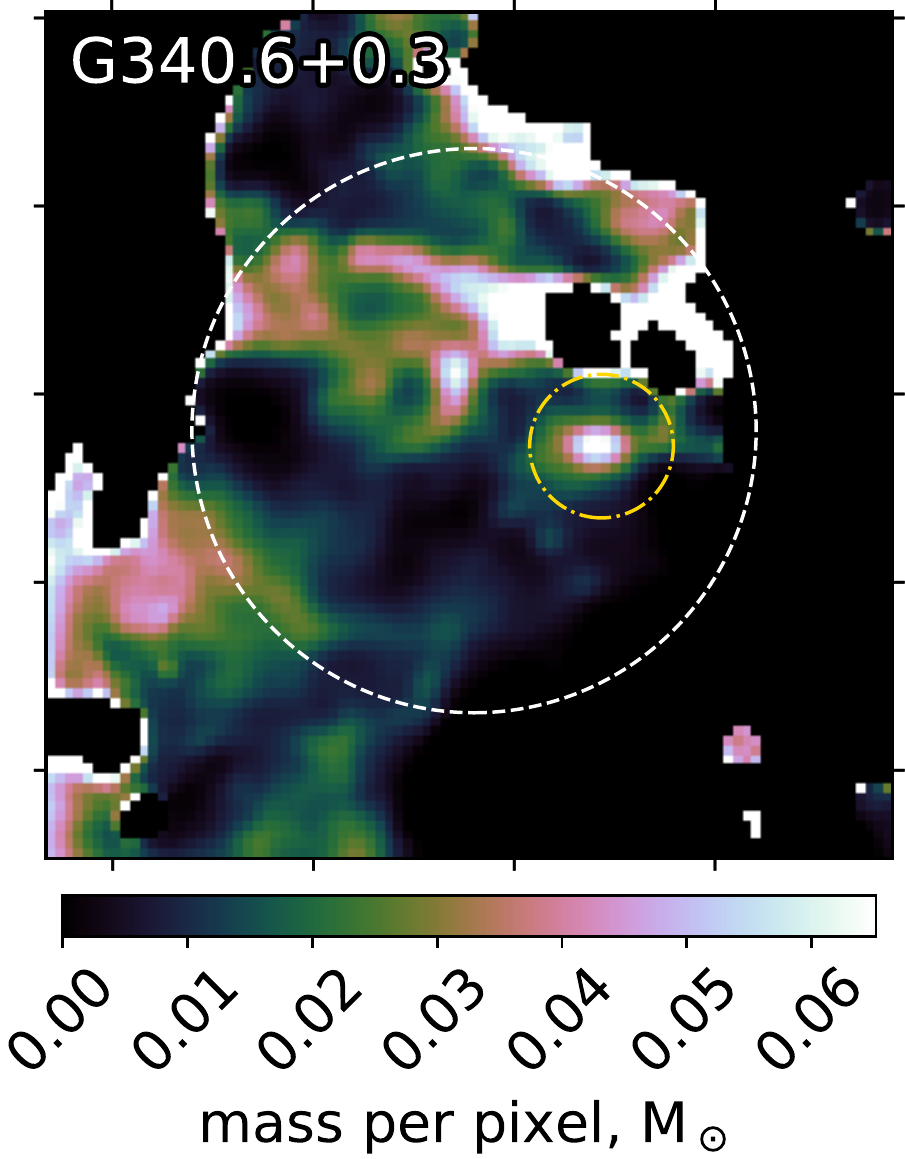}
	\includegraphics[scale=0.45]{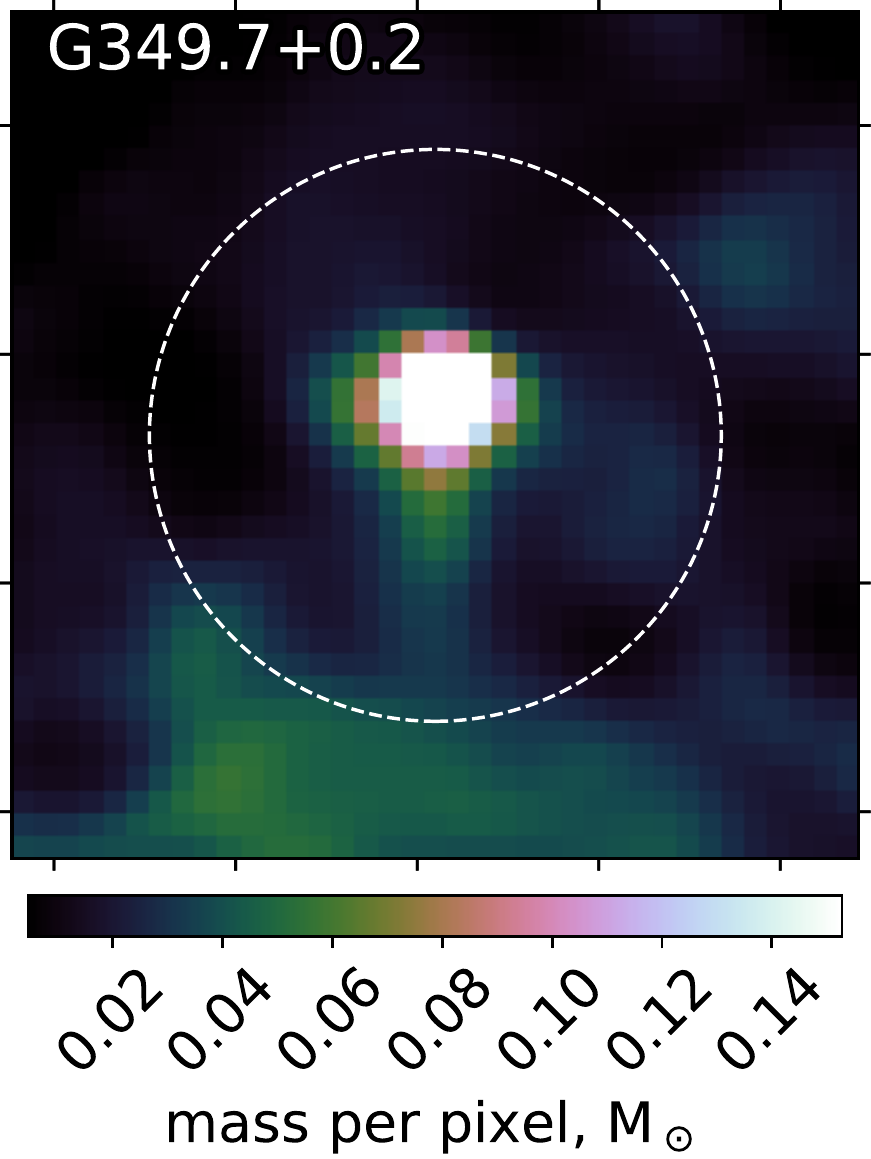}
	\caption{Dust mass maps for our subset of detected level 1 sources, derived using the background subtracted 70\,\micron\ flux and the temperature maps in \autoref{fig:dusttempmaps}. The estimated dust mass for each SNR, as given in \autoref{tab:SNRFluxes}, is estimated from within the white circle for all pixels which have a 70\,\micron\ flux above a threshold level, after masking unrelated sources (shown by the gold circles).
	}
	\label{fig:dustmassmaps}
\end{figure*}
\begin{table} 
	\csvreader[tabular= l c c c,
				late after last line=\\\hline,
				table head=\hline
				SNR & Distance (kpc) & Dust Mass (M$_\odot$) & Refs \\
				\hline\hline] 
	{SNR_Fluxes_Nebuliser.csv}{SNR=\snr, Flux_70=\fluxone, Flux_160=\fluxtwo, Flux_250=\fluxthree, Flux_350=\fluxfour, Flux_500=\fluxfive, Distance=\dist, Sigma_Clipped_Mass=\mass, Temperature=\temp, Ref_Num=\refnum} 
	{\snr & \dist & \mass & \refnum} 
	\\[1.5pt]
	\caption{
	$^*$Sources which may be confused with a H{\sc ii} region \citep{Anderson2017, Gao2019}.
	{\it Distance references: 1\,--\,\citet{Kilpatrick2016}, 2\,--\,\citet{Ranasinghe2018a}, 3\,--\,\citet{Ranasinghe2018b}, 4\,--\,\citet{Shan2018}, 5\,--\,\citet{Kothes2007}, 6\,--\,\citet{Tian2014}.}
	}
	\label{tab:SNRMasses}
\end{table}

Next we determine dust masses for the SNRs from this sample for which there are estimated distances in the literature.
\autoref{tab:SNRFluxes} shows that, even for our clearest detections, the SPIRE fluxes for our SNRs are very uncertain and we cannot confidently use these fluxes to estimate a dust mass via modified blackbody fits to the IR-submillimetre SED for these sources. Nevertheless, all of our level\,1 sources are detected at 70\,\micron, we therefore use these maps combined with the temperature information in \autoref{fig:dusttempmaps} to estimate the dust mass within each source, giving the dust maps in \autoref{fig:dustmassmaps}.
When estimating the mass from these sources there are 2 issues to resolve. First, there are large areas with relatively small dust mass which, when summed, contribute a large mass to our final estimate. Second, in the case of G43.3$-$0.2 there is a region to the north of the SNR which has a low temperature in \autoref{fig:dusttempmaps} because of low 70\,\micron\ emission in this region (relative to the 160\,\micron\ flux), resulting in a large dust mass. To overcome both issues we set a 1.5\,$\sigma$ threshold on the 70\,\micron\ background subtracted flux, below which pixels are excluded from our dust mass estimate (lower flux thresholds are required for G43.3$-$0.2 and G340.6$+$0.3).
Using the white apertures shown in \autoref{fig:dustmassmaps} we find the dust masses given in \autoref{tab:SNRMasses}.

We estimate unexpectedly large dust masses for 3 of our SNRs (G8.3$-$0.0, G20.4$+$0.1, and G340.6$+$0.3). In two of these cases the estimated distance is extremely large; if this is overestimated it will increase our dust mass, i.e. increasing the estimated distance by a factor of 2 would increase the estimated dust mass by a factor of 4. However, we also note that closer remnants will have interacted with a smaller volume of the ISM, and consequently would have conversely swept up less dust mass as their shock front moves out. It is therefore likely that there is another factor contributing to this issue.
As discussed previously, our images are highly confused and, despite careful background subtraction, there may still be contaminating ISM which acts to reduce the estimated temperatures in \autoref{fig:dusttempmaps}. In the case of G8.3$-$0.0, increasing the temperature of each pixel by 2\,K or 10\,K reduces our dust mass from 340\,M$_\odot$ to $\sim$\,190 or 35\,M$_\odot$. Evidently, our dust mass estimates are highly sensitive to small changes in the temperature and we expect that our results are greatly dependent on contaminating ISM.

Previous IRAS surveys by \citet{Arendt1989} and \citet{Saken1992} estimated dust masses for W49B (G43.3$-$0.2) of 127.7 and 500\,M$_\odot$ and for G349.7$+$0.2 of $<$\,25.1 and 4900\,M$_\odot$, which are considerably larger than those that we estimate (although we are consistent with the G349.7$+$0.2 mass estimated by \citet{Arendt1989}). It is clear that background subtraction is crucial in finding accurate dust masses and even small variations in the method of subtraction can give vastly different results. With the higher angular resolution of {\it Herschel} compared with IRAS we make drastic improvements in reducing the ISM contamination.

\subsection{The Nature of the Centre of G351.2$+$0.1}
\label{sec:g351}
In the previous Section, we did not produce a dust temperature and mass map for the SNR G351.2$+$0.1 due to the large gradient in interstellar dust seen in the immediate vicinity of the SNR shell (\autoref{fig:G351.2+0.1Image}). However the inner central/radio bright region where we discovered dust emission is observed at high S/N in all of the Herschel bands. This potentially indicates the discovery of FIR emission coincident with a possible Crab-like compact object. To determine the nature of this emission across the NIR\,--\,radio regime and because of its brightness in all bands, for this source, we create a spectral energy distribution (SED) of the centre. To determine the nature of this emission across the NIR\,--\,radio regime, we create a spectral energy distribution (SED) of the centre. We note that we can fit an SED for this source, as the emission in all of the {\it Herschel} bands observed at the location of the central radio core appear to be associated with the same source.

Photometry of the compact source from 3.6\,--\.350\,\micron\ (\autoref{fig:G351.2+0.1Image}) was derived using a circular aperture with diameter 1.09\,$^\prime$ centred at $\alpha = 17^\text{h}22^\text{m}24.8^\text{s}, \delta = -36^\circ11^\prime06''$. In order to minimize the effect of the second dust source just to the east of the radio core (the red `blob' in the bottom right panel of \autoref{fig:G351.2+0.1Image}), the background emission was derived using an annulus centred at the same co-ordinate with diameter 2.84\,$^\prime$ and width 1.00\,$^\prime$. Numerous point sources in the {\it Spitzer} bands were masked out and corrections were applied to the IRAC bands to account for the extended diffuse emission seen in the images following the instructions in the {\it Spitzer} IRAC handbook\footnote{\url{https://irsa.ipac.caltech.edu/data/SPITZER/docs/irac/iracinstrumenthandbook/29/}}. The {\it Spitzer} IRAC fluxes were dereddened following the reddening law of \citet{Indebetouw2005} where $\rm A_{K_S} = 2.5\,mag$\footnote{\url{https://irsa.ipac.caltech.edu/workspace/TMP_wYHAkz_6484/DUST/17_22_27_-36_11_00.v0001/extinction.html}}.
Errors in the fluxes were determined by combining calibration errors with the error determined in the background level.  The photometry is listed in \autoref{tab:g351_photometry} alongside the VLA photometry from \citet{Becker1988}, and the spectral energy distribution (SED) is shown in \autoref{fig:g351_sed}. The synchrotron power law for the compact radio core is shown with slope $+0.27$ \citep{Becker1988}.  Following C19, we fit the SED with two modified blackbodies to derive a cool dust temperature and a dust mass.  We have assumed that the dust emissivity index $\beta$ is kept constant at a value of 1.9.  We find a best-fit ($\chi^2$) model with temperature $45.8\,\rm K$ (see also \autoref{fig:dusttempmaps}) and dust mass $0.18\,{\rm M_{\odot}}(d/11\,\rm kpc)^{-2}$ (where $d$ is the distance from \citealp{Dubner1993}).  We also run a monte carlo analysis by perturbing the observed fluxes within their uncertainties 1000 times and refitting the SEDs. The median SED fit from this analysis produces a cold dust component with $T_{\rm d}\,=\,31.9\pm 1.5\,\rm\,K$ and $M_{\rm d}\,=\,1.1\,\pm\,0.3\,{\rm\,M_{\odot}}(d/11\,\rm kpc)^{-2}$, and a hot dust component with $T_{\rm d}\,=\,242.1\pm 6.8\,\rm\,K$ and $M_{\rm d}\,=\,(3.9\,\pm\,0.4)\,\times\,10^{-6}\,{\rm\,M_{\odot}}(d/11\,\rm kpc)^{-2}$.

One potential issue is whether the compact source is a H{\sc ii} region instead of a SNR. Although the dust temperature we derive is hotter than the average values observed in H{\sc ii} regions (typically 15\,--\,30\,K \citep{Anderson2012}), we note that some H{\sc ii} have observed dust temperatures up to 40\,K \citep{Povich2007,Anderson2012_bubble}.  We next compare NIR\,--\,radio colours with those observed in H{\sc ii} regions since different emission mechanisms should result in different predicted colours in the NIR\,--\,FIR and FIR-radio. \cite{Reach2006} suggests that the {\it Spitzer} IRAC colours (3.8\,--\,8\,$\mu$m) can be used as a diagnostic between SNRs, the ISM (H{\sc ii} regions or their associated photodissociation regions, PDRs), shocked molecular gas or ionised gas as well as synchrotron sources.  Recent \spitz\ and \hersc\ studies of H{\sc ii} regions suggest NIR\,--\,FIR colour plots can also be used as a diagnostic for their identification \citep{Goncalves2011,Anderson2012,Anderson2012_bubble,Paladini2012}. Similarly, the 8\,$\mu$m and 843\,MHz ratios can offer an alternative diagnostic \citep{Cohen2007}.  Next we test whether the fluxes derived for G351.2$+$0.1 can reveal whether we are observing dust from an unrelated H{\sc ii} region or a SNR.

\autoref{tab:g351_tests} shows various colours predicted or observed in H{\sc ii} regions in comparison to those measured for G351.2$+$0.1.  At face value, the IR and radio colours show this source is entirely consistent with interstellar material. However there are some caveats with this.
Although the IRAC colours for this source are consistent with the ISM regions in \citet{Reach2005} (see also \autoref{tab:g351_tests}), they are similar to that observed in the PWNe G21.5$-$0.9 \citep{Zajczyk2012}.  Comparing with the ${\rm log}\left(F_{\rm 8}/F_{\rm 24}\right)$ versus ${\rm log}\left(F_{\rm 70}/F_{\rm 24}\right)$ colour plot of \citet{Goncalves2011} (their Figure 3), G351.2$+$0.1 lies along the slope of colours observed in Galactic SNRs, consistent with dust emission with a colour temperature ($T_{24/70}$) of $\sim 45\,\rm K$ for $\beta=2$, and well above the different trend line observed for H{\sc ii} regions. Although the colours are consistent with the narrow range of observed IR-FIR values found in  H{\sc ii} regions \citep{Paladini2012}, these overlap considerably with observed values for SNRs which are observed to have much wider range of colours due to the variety of morphologies, ages, emission mechanisms in the latter. Comparing with colours for other known dusty PWNe e.g. G54.1$+$0.3 and the Crab Nebula (see \autoref{tab:g351_tests}), although G54.1 satisfies none of the H{\sc ii} diagnostic tests, the Crab does satisfy some.  It is not clear how the presence of ejecta dust in SNe would affect these tests, and whether, given the wide range of properties of SNRs, it is even possible to use colours to distinguish between source types.  We conclude therefore that we cannot rule out that the dust observed in the centre of G351.2 is unrelated interstellar material, but nor can we rule out a SN ejecta origin.

\begin{table}
	\begin{tabular}{cccc}
	\hline
	\multicolumn{2}{c}{NIR - FIR} & \multicolumn{2}{c}{Radio} \\
	Wavelength & Flux & Frequency & Flux \\
	\micron & Jy & GHz & Jy \\
	\hline
	3.6 &  0.143 $\pm$ 0.022 & 15 & 0.0124 $\pm$ 0.0002 \\
	4.5 &  0.095 $\pm$ 0.014 & 5 & 0.0090 $\pm$ 0.0003 \\
	5.8 &  0.579 $\pm$ 0.087 & 1.5 & 0.0064 $\pm$ 0.0010 \\
	8.0 &  1.590 $\pm$ 0.239 & & \\
	24  &  1.155 $\pm$ 0.041 & & \\
	70  & 26.960 $\pm$ 2.716 & & \\
	160 & 20.256 $\pm$ 2.496 & & \\
	250 &  4.469 $\pm$ 2.631 & & \\
	350 &  0.292 $\pm$ 1.966 & & \\ \hline
	\end{tabular}
	\caption{Background subtracted flux measurements for the compact source at the centre of G351.2$+$0.1. IRAC fluxes have been dereddened \citep{Indebetouw2005} and point sources in the {\it Spitzer} IRAC and MIPS 24\,$\mu$m images were masked out.}
	\label{tab:g351_photometry}
\end{table}

\begin{table*}
	\begin{tabular}{lccccccc}
	\hline
	\multicolumn{1}{c}{Colour} &	\multicolumn{1}{c}{H{\sc ii} region test} &	\multicolumn{2}{c}{G351.2$+$0.1}  & \multicolumn{2}{c}{G54.1$+$0.3} & \multicolumn{2}{c}{Crab} \\
		\multicolumn{1}{c}{} &	\multicolumn{1}{c}{} &	\multicolumn{1}{c}{Value}  & \multicolumn{1}{c}{Consistent with H{\sc ii}?} & \multicolumn{1}{c}{Value}  & \multicolumn{1}{c}{H{\sc ii}?} & \multicolumn{1}{c}{Value}  & \multicolumn{1}{c}{H{\sc ii}?}  \\ \hline
	$ \frac{3.6}{8} | \frac{4.5}{8} | \frac{5.8}{8} | I_8$$^{a}$ & $0.04 | 0.05 | 0.35 | 1$	& $0.09 | 0.06 | 0.36 | 1$ &\checkmark  & .. & ..& .. & ..\\
	& & & & & &  &\\
	${\rm log}\left(F_{\rm 160}/F_{\rm 24}\right)^{b}$ & $>0.8$ & $1.24$  & \checkmark & $0.10$ &  $\times$ & $0.45$ & $\times$\\
	${\rm log}\left(F_{\rm 160}/F_{\rm 70}\right)^{b}$ & $>-0.2$ & $-0.12$ & \checkmark & $-0.5$ & $\times$ & $-0.12$ & \checkmark  \\
	${\rm log}\left(F_{\rm 70}/F_{\rm 24}\right)^{b}$ & $>0.8$ & $1.37$ & \checkmark & $0.6$ & $\times$ & $0.58$ & $\times$\\
  ${\rm log}\left(F_{\rm 24}/F_{\rm 8}\right)^{b}$ & $<1.0$ &	$-0.14$ & \checkmark & $1.59$ & $\times$ & $0.52$ & \checkmark \\
	${\rm log}\left(F_{\rm 24}/F_{\rm 70}\right)^{c}$ & $ -1.5$ to $-1.0$ & $-1.37$ & \checkmark &  $-0.58$ & $\times$ & $-0.12$ & $\times$ \\
	& & & & & &  &\\
  $F_{\rm 8}/S_{\rm 843\,MHz}$$^{d}$ & $27 \pm 10$ & $282$ & ? & 1.4 & $\times$ & $0.02$ & $\times$ \\
	\\ \hline
	\end{tabular}
	\caption{Summary of the various colour tests to resolve whether the dust source associated with the radio source observed in G351.2$+$0.1 originates from a H{\sc ii} region or a SNR. Unless specified, all the subscripts refer to the wavelength in microns. References are $^{a}$ - \citet{Reach2006}, $^{b}$ - \citet{Anderson2012}, $^{c}$ -  \citet{Paladini2012}, $^{d}$ - \citet{Cohen2007}. Also shown are the results of the same tests for the PWNe G54.1$+$0.3 (fluxes are taken from \citealt{Temim2017} and \citealt{Rho2018}) and the Crab Nebula \citep[from][]{DeLooze2019}.
	}
	\label{tab:g351_tests}
\end{table*}

\begin{figure}
	\centering
  \includegraphics[width=1.0\linewidth, trim = 0.1cm 0.5cm 0.1cm 0.3cm, clip]{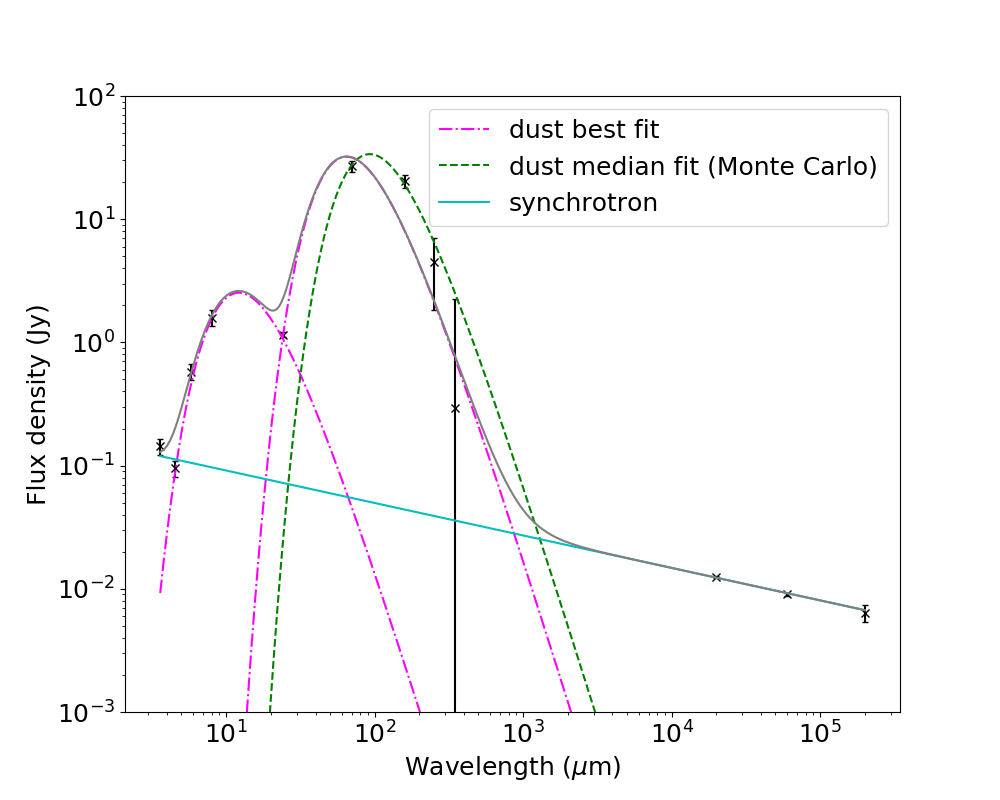}{}
	\caption{Spectral energy distibution of G351.2$+$0.1 from NIR-radio wavelengths.  Fluxes are listed in \autoref{tab:g351_photometry}. The synchrotron emission assuming a power law slope of $+0.27$ \citep{Becker1988} extrapolating from the VLA radio fluxes of the compact core is shown by the solid (blue) line. The best-fit ($\chi^2$) hot and cool modified blackbody fits to the photometry are shown by the dot-dashed curves (magenta) and the combination of all emission sources is shown in the solid (grey) curve.  The dotted (green) curve is the resultant median modified blackbody fit from 1000 SEDs derived by normally distributing the original observed fluxes within their errors.
	}
	\label{fig:g351_sed}
\end{figure}

\section{Discussion}
\label{sec:furtherwork}

\subsection{Types of sources detected} \label{detectiontypesdiscussion}
Here we discuss the detection rates found in our {\it Herschel} study.  This work finds signatures of dust in \numOne\ (\percOne\,per\,cent) SNRs in the Galactic Plane (detection level of 1) compared to the \CnumOne\ detected in C19 (\CpercOne\,per\,cent detection rate).  In total, we estimate a FIR detection rate of 21\,per\,cent of Galactic SNRs with $\mid\,b\,\mid\,\leq\,1^{\circ}$ and classed as a level 1 detection. \autoref{tab:DetectionSummary} and \autoref{fig:DetectionSummary} give a summary of the types of SNRs detected in this study. Of our new FIR detections, we observe dust emission from the shell/outer shock region of \numShell\ SNRs, and from the inner ejecta region (interior to the reverse shock) of \numCentral\ sources. We detect \numTypeOnea\ Type Ia and \numCoreCollapse\ core collapse SNe. Combining with C19, this gives {\it Herschel}-detected SN dust structures in 13 core-collapse SNe, including 4 {\it confirmed} PWNe, and 2 Type Ia's in the HiGAL survey.
Including the other {\it Herschel} discovered dusty remnants in the Milky Way that were not covered by the HiGAL survey, that is Cas A, the Crab Nebula, Tycho, Kepler and G292.0$+$1.8 (hereafter, we name this group the historical dusty SNRs), the final statistics for dust features related to SN structures in the Galaxy are therefore 16 core collapse SNe, including 6 PWNe, and 4 Type Ias (\autoref{tab:DetectionSummary}).  Based on these 20 sources where the SN explosion type is known, our dust-detected sample is made up of 80\,per\,cent core-collapse SNRs, and 20\,per\,cent Ia SNRs, closely mirroring the observed rates of core-collapse and Type Ia SN explosions (\citealp{vandenbergh1993,The2006}) with the former occuring at rates of $\sim 2-3$ SNe per century \citep{vandenbergh1991,Dragicevich1999,Diehl2006,Li2011}.

The detection of SN-related dust features in the {\it Herschel} images of the Type Ias G306.3$-$0.9 (this work) and G344.7$-$0.1 (C19) could provide further insight into dust formation in Ia SNRs, for which there is not yet any evidence of ejecta dust forming in their remnants. Although at first glance, the IR emission seen in the {\it Herschel} images appears to originate from reserve shock-heated material, we saw in Section~\ref{subsec:IndividualResults} that for G306.3$-$0.9, the dust appears to be associated with material swept up by the blast wave, or unrelated interstellar clouds, {\em and not from supernova dust}.  A similar conclusion was reached by \citet{Gomez2012a} for the Type Ia SNRs Kepler and Tycho.

Compared with previous FIR surveys of Galactic SNRs by \citet{Arendt1989} and \citet{Saken1992} our higher resolution helps us to better distinguish SNRs from the surrounding ISM, giving us higher detections rates within the Galactic Plane.
In comparison with \citet{Saken1992}, we detect 7 sources in common (G31.9$+$0.0, G33.2$-$0.6, G43.3$-$0.2, G54.1$+$0.3, G304.6$+$0.1, G340.6$+$0.3, and G349.7$+$0.2) and upgrade 16 of their SNRs to level 1 {\it Herschel} detections from the equivalent of their IRAS levels 2, 3, and 4.
We also find 9 detections in common with \citet[G11.2$-$0.3, G31.9$+$0.9, G34.7$-$0.4, G43.3$-$0.2, G54.1$+$0.3, G304.6$+$0.1, G340.6$+$0.3, G348.7$+$0.3, and G349.7$+$0.2]{Arendt1989} and upgrade 15 sources to level 1 {\it Herschel} detections from their IRAS levels 2, 3, and 4.
Although there are a small number of detections in each study which, with {\it Herschel}, we downgrade from a level\,1 detection (8 from \citet{Saken1992} and 2 from \citet{Arendt1989}), we increase the detection rate of Galactic SNRs within each sample from 17 to 24\,per\,cent for the \citet{Saken1992} sample, and from 13 to 26\,per\,cent for the \citet{Arendt1989} sample (excluding sources which were not included in the \citet{Green2014} catalogue).

\begin{table*}
	\begin{tabular}{lcccc}
	\hline
	\multicolumn{2}{c}{Detection Type} & \multicolumn{3}{c}{Number Detected}  \\
	\multicolumn{2}{c}{} & \multicolumn{1}{c}{C19} & \multicolumn{1}{c}{This Work} & \multicolumn{1}{c}{Historical}
	\\\hline\hline
	SNR Region $\natural$ & Shell / outer shock region & \CnumShell & \numShell & 3 \\
	& Inner ejecta region *  & \CnumCentral & \numCentral & 3 \\
	& Confirmed PWN $\dag$ & \CnumPWNd  & \numPWNd & 1 \\
	& Interacting molecular cloud & 0 & \numInteracting & \\
	&&&& \\\hline
	Age (kyr) & $\leq$ 1 & \CnumLessOne & \numLessOne & 3 \\
	& 1 \textless Age $\leq$ 10 & \CnumOnetTen & \numOnetTen & 2 \\
	& 10 \textless Age $\leq$ 20 & \CnumTentTwenty & \numTentTwenty & \\
	& \textgreater 20 & \CnumTwentyPlus & \numTwentyPlus & \\
	& Unknown & \CnumAgeUnknown & \numAgeUnknown & \\
	&&&& \\\hline
	SN Type & Type Ia & \CnumTypeOnea & \numTypeOnea & 2 \\
	& Core collapse & \CnumCoreCollapse & \numCoreCollapse & 3 \\
	& Unknown & \CnumTypeUnknown & \numTypeUnknown & \\ \hline

	\end{tabular}
	\caption{Summary of the level 1 detected sample in this work and C19.  We also include the previously detected {\it Herschel} Galactic SNRs as `historical remnants': Tycho, Kepler, Cas A, the Crab Nebula and G292.0$+$1.8.
	* Sources from which we detect FIR emission from the inner region are:
	G11.2$-$0.3, G16.4$-$0.5, G21.5$-$0.9, G21.5$-$0.1, G29.7$-$0.3, G34.7$-$0.4, G54.1$+$0.3, and G344.7$-$0.1 (C19), and
	G306.3$-$0.9, G350.1$-$0.3, G351.2$+$0.1, and G357.7$-$0.1 (this work).
	\dag SNRs for which there is evidence that the detected central region is associated with the confirmed PWN. We note that G351.2$+$0.1 could be associated with a PWN but the compact object has radio, NIR\,--\,FIR detection only and other sources cannot be ruled out (see Section~\ref{subsec:IndividualResults}).
	}
	\label{tab:DetectionSummary}
\end{table*}

\begin{figure}
	\includegraphics[width=1.0\linewidth, trim = 0cm 0.2cm 0cm 0cm, clip]{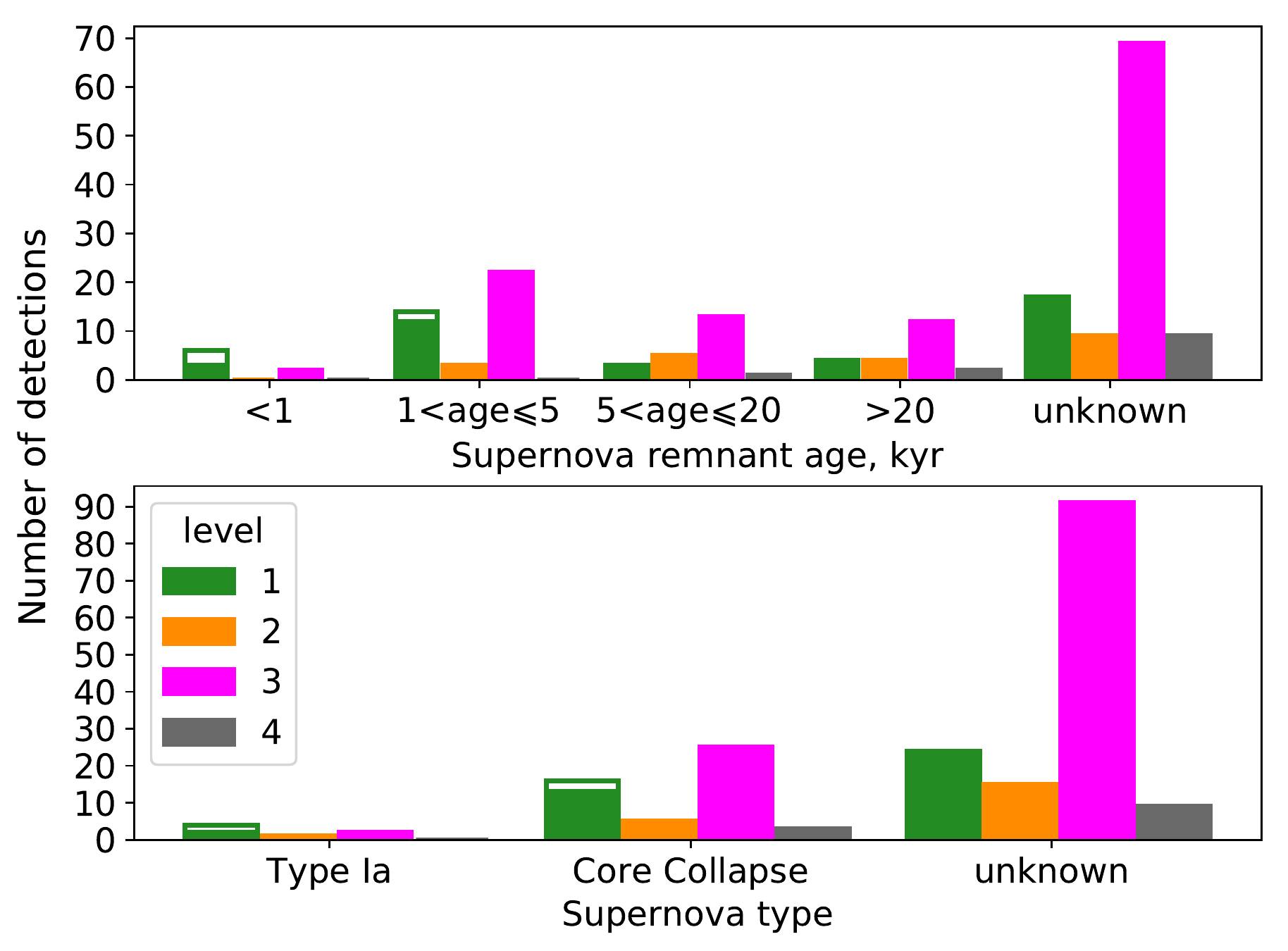}
	\includegraphics[width=1.0\linewidth, trim = 0cm 0cm 0cm 0cm, clip]{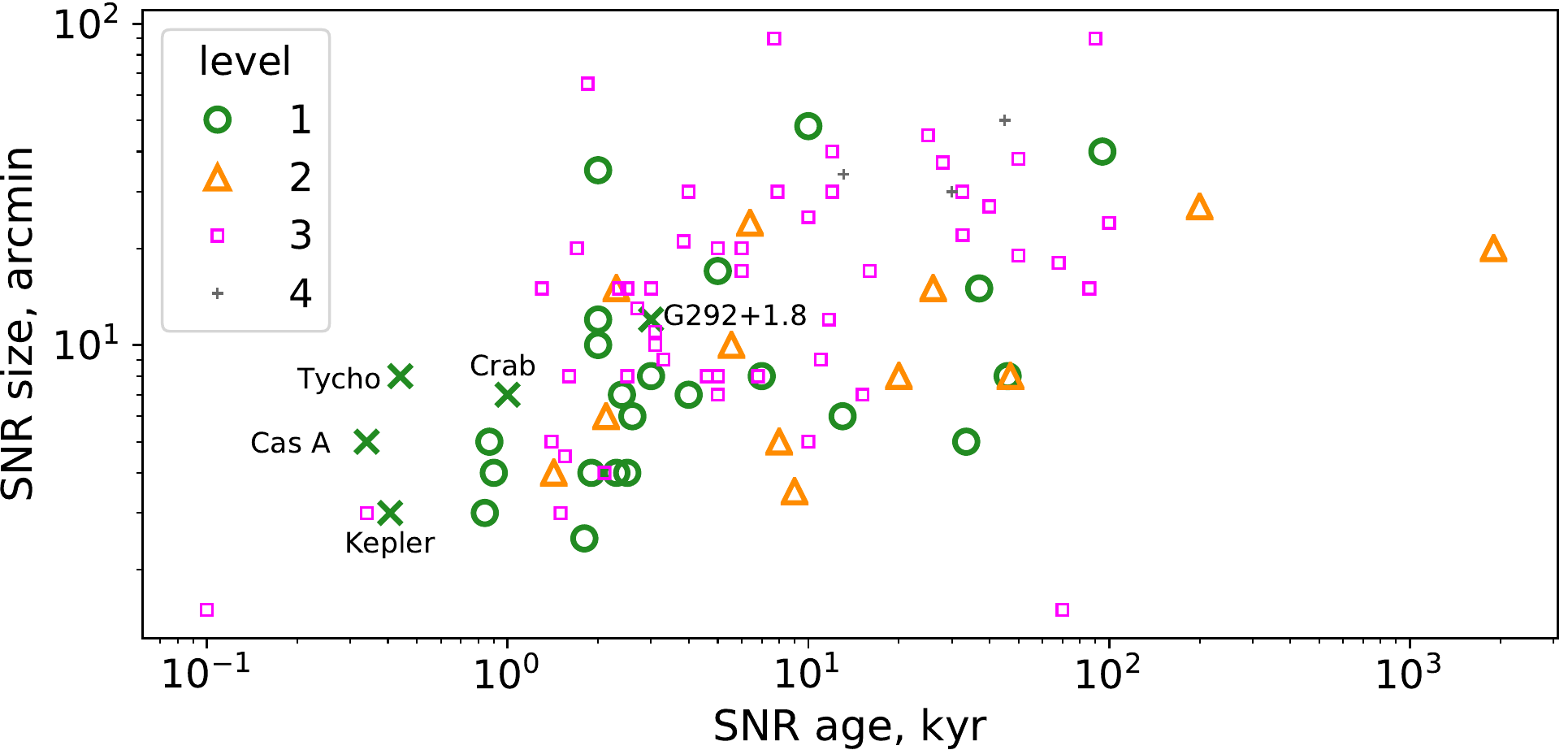}
	\caption{As in C19, we show a summary of the source types detected in the sample, including those in C19.
	{\it Top}: Filled histograms compare the number of sources with different ages for a given detection classification in HiGAL. The unfilled bar includes the previous {\it Herschel} detections of Galactic SNRs not covered by the HiGAL survey: Cas A, the Crab Nebula, Tycho, Kepler and G292.0$+$1.8.
	{\it Middle}: As top panel, but showing the number of sources with different SN types for a given detection classification.
	{\it Bottom}: The detection level compared with the size and age of SNR; the crosses indicate the properties of the previous historical SNRs.  The age is unknown for 104 SNRs from our sample.
	}
	\label{fig:DetectionSummary}
\end{figure}

\begin{figure*}
	\includegraphics[width=0.8\linewidth, trim=2.5cm 0.8cm 3.5cm 0.9cm]{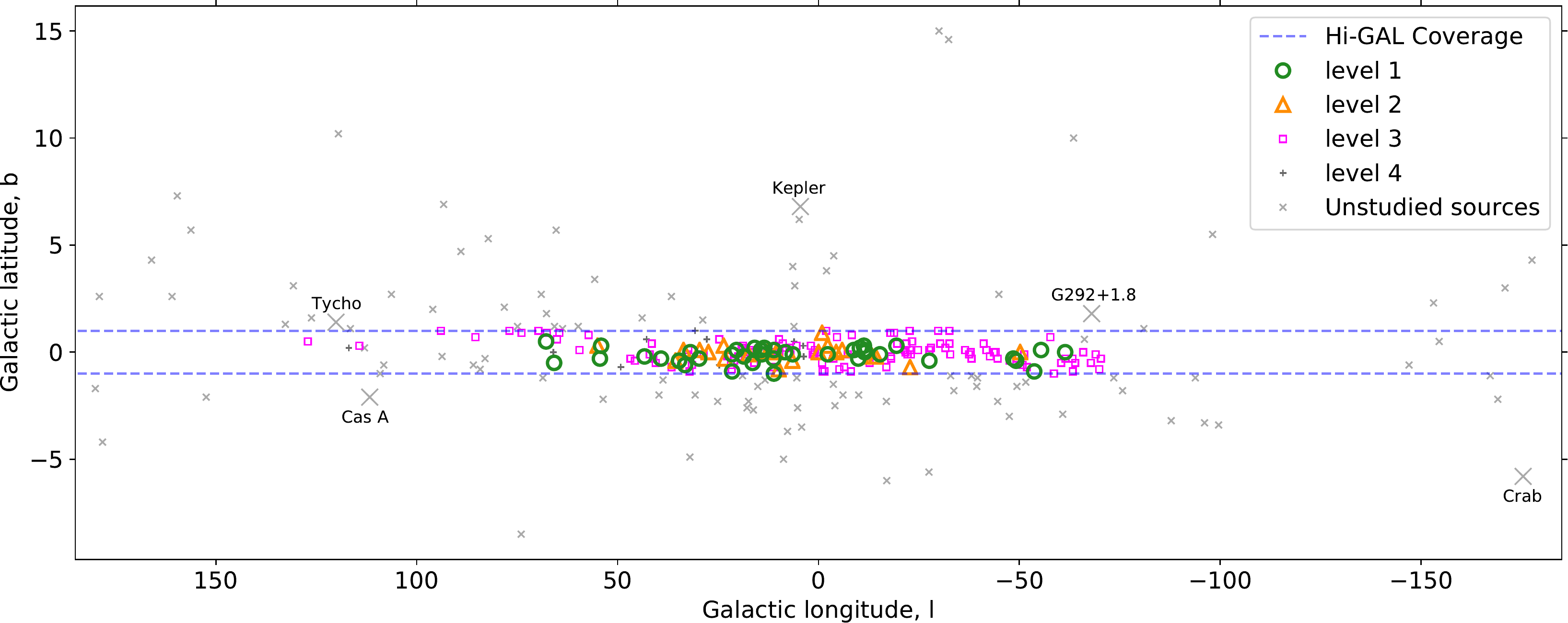}{}
	\includegraphics[width=0.8\linewidth, trim=3.2cm 0.5cm 3.6cm 0.5cm]{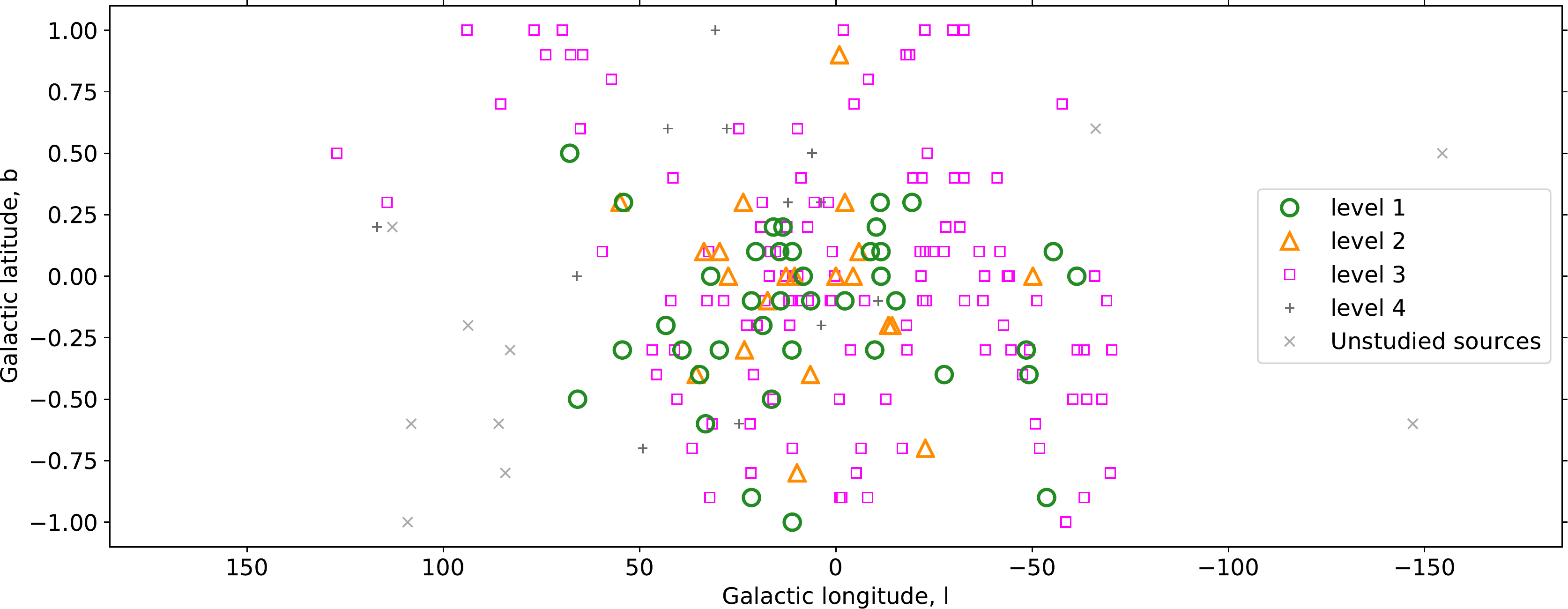}{}
	\caption{Location of Galactic SNRs from Green's catalogue as a function of their assigned detection number, 1--4 (green circles, orange triangles, magenta squares, and grey plus signs respectively) and those not covered in the Hi-GAL survey (i.e. not studied here) shown by the grey crosses. The dashed blue lines indicate the extent of the area surveyed by Hi-GAL. {\it Top}: All SNRs from Green's catalogue. {\it Bottom}: Sources within the Galactic Plane with $\mid b \mid \leq 1$  (including \citealp{Chawner2019}).
	Hi-GAL data is unavailable for 10 sources within the Galactic Plane.
	}
	\label{fig:sample_locations}
\end{figure*}

An interesting question is whether there exists a trend with how dusty a SNR is and its age \citep[e.g.][]{Gall2014,Otsuka2010}. (Ages are listed in Table~A1 where known, see references therein.) Not including those without estimated ages, \autoref{fig:DetectionSummary} shows that the highest proportion of our detected sources are young, $\leq\,5$\,kyrs, which may be because the SNR is still compact so that surface brightness is higher and therefore more easily `seen' in the comparison of the multicolour {\it Herschel} images. The historical dusty SNRs also lie within this age range (\autoref{fig:DetectionSummary}), indeed these are mostly clustered towards younger ages than the majority of the det=1 SNRs in our HiGAL survey (where the youngest det=1 SNR is the core-collapse SNR G29.7$-$0.3, Kes 75). This suggests that the previous historical dusty SNRs were biased to the youngest (and potentially dustiest) objects, whereas the sample in this work has detected dust signatures associated with SNRs at a wider range of ages up to $10^2$\,kyr.

However, for all SN ages and types we are more likely to find level 3 detections, where there is only unrelated FIR emission in the region. Although this may imply that there is no dust within these SNRs, we are also greatly affected by confusion in cases where sources may have a similar dust temperature to ISM dust or already starting to mix with the ISM (i.e. larger/older), wherein it becomes more difficult to separate ejecta and unrelated material. We therefore expect that the detection levels quoted here are a lower limit on the number of SNRs within the Galactic Plane which contain dust.

\autoref{fig:sample_locations} shows the location of all of the SNRs studied within the Galactic Plane with {\it Herschel} as part of the HiGAL survey. We find that the majority of the SNRs with clear dust signatures (detection level = 1) are located towards the central regions of the Galactic Plane in both longitude and latitude. This distribution follows the number density of SNRs in general, which are concentrated towards low latitudes and in the inner regions of the Galaxy (\citealt{Green2014}, \autoref{fig:sample_locations}, top panel).

The distribution of the `dust-detected' sources in this work can be explained due to our sample being dominated by the remnants of core-collapse SNe, which make up the majority of SNRs in the Galaxy. Core-collapse remnants are observed to be found at lower latitudes, closely linked to star forming regions concentrated in the thin disc of the Galaxy \citep{Hakobyan2016}.  Instead, Type Ia's, which constitute only a small fraction of our det=1 sample with known SN types, are observed to lie at a wider range of latitudes, with Galactic scale heights of $\sim 2$ times that of core-collapse remnants \citep{Hakobyan2017} due to their association with the older stellar population. We may therefore be biased against finding dust in Type Ia SNRs in this work.

In order to interpret the distribution of dusty SNRs in the Galaxy, we need to first note our selection effects. Although we start from a blind {\it Herschel} survey of the sky with HiGAL, and as such the dust information is `complete' within $b \pm 1^{\circ}$, due to the high levels of confusion in the FIR we have to rely on the known locations of SNRs, where we have used the \cite{Green2004} catalogue.  Since this is the fundamental selection effect for this work, here we briefly mention its completeness.  \citet{Green2015} comments that the Galactic SNR catalogue is likely missing intrinsically faint radio SNRs, SNRs where the physical size is small (diameters $<1.5{\arcmin}$), and more distant SNRs with smaller angular sizes, as well as many sources having uncertain distances due to the use of the $\Sigma-D$ relation. Therefore our sample will also suffer from the same selection effects. We note that our det=1 sampled follows the same $l,b$ distribution as his radio-bright SNR subsample (with a surface brightness cut $(1\,\rm GHz) > 10^{-20}\,\rm W\,m^{-2}\,Hz$) which they argue suffers from fewer selection effects than the whole Galactic radio SNR catalogue. However, the overlap between our detections and the bright SNR sample can be attributed to the fact that we rely on the radio images (where available) to identify SN-related dust features.

We include the historical dusty SNRs for reference in \autoref{fig:sample_locations}. These lie outside the inner Galactic Plane, and yet still have associated dust features (and indeed a significant amount of ejecta dust).  Again this implies that the detection levels quoted here are likely lower limits, since there are $\sim$100 more SNRs in the catalogue of \citet{Green2014} not included in the region surveyed by HiGAL (\autoref{fig:sample_locations}, top panel).

Our previous study (C19) covered $10\,\leq\,\mid\,l\,\mid\,\leq\,60^{\circ}$. Although this study covers the entire Galactic longitude range, the majority of additional sources are within $\mid\,l\,\mid\,\leq10^{\circ}$. This region is heavily contaminated with ISM dust, resulting in a lower detection rate in this study compared with C19. 
An additional factor causing a lower detection rate in this work compared to C19 could be the lack of {\it Spitzer} data in some regions; in particular {\it Spitzer} MIPS 24\,$\mu$m has better spatial resolution than {\it Herschel}, which helps to identify warm SN dust structures more confidently. In our study, the majority of the SNRs are classified as `heavily confused' in these regions. We remind the reader that the requirement of ancillary X-ray, radio or optical data in order to indentify SN features against the surrounding ISM/CSM, and the non-uniformity of the availability and quality of this data may have further resulted in a bias in our HiGAL Galactic Plane SNR sample.

\begin{figure*}
	\includegraphics[width=\linewidth]{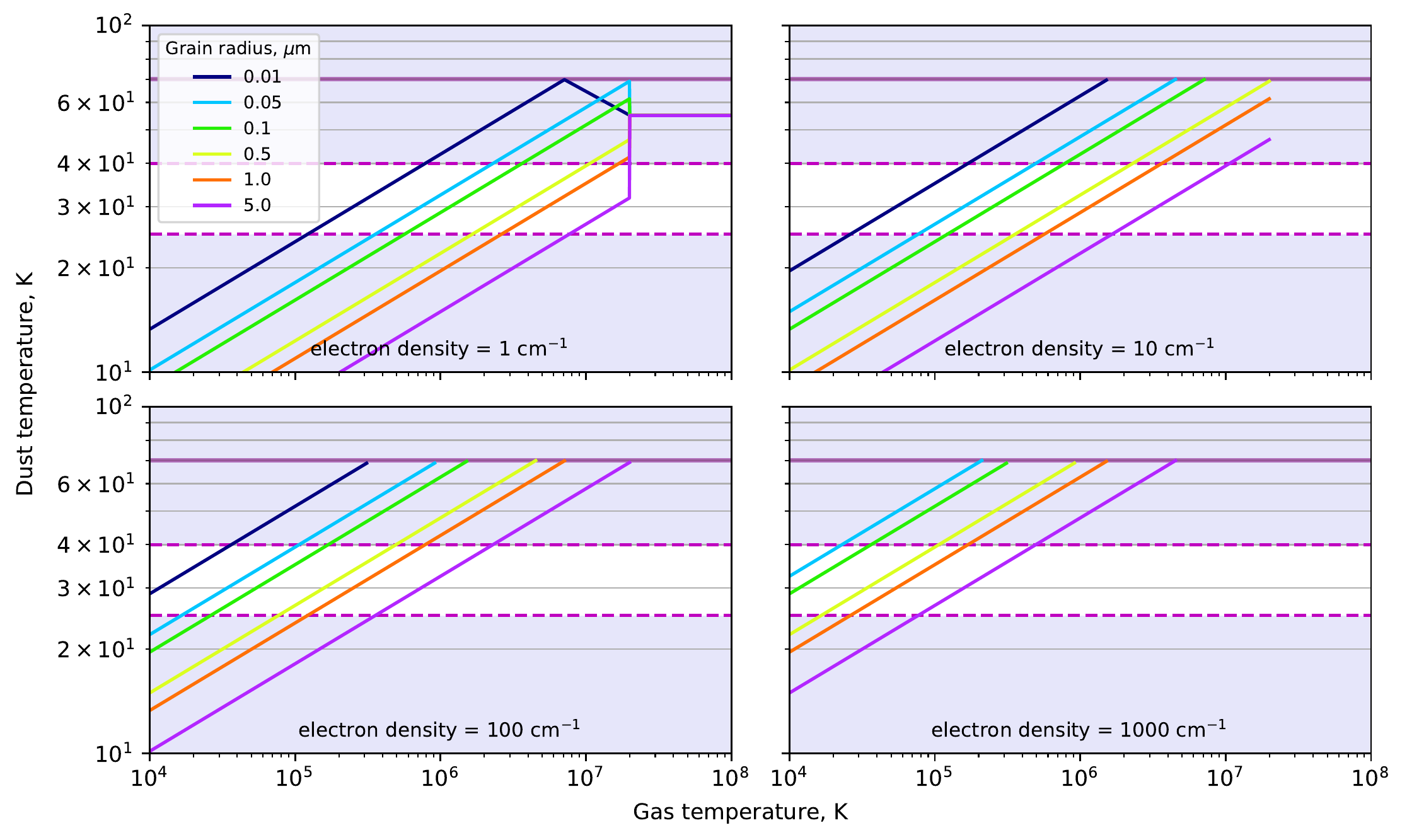}
	\caption{
	Temperature of collisionally heated dust given by \autoref{eq:collisionalheating} for electron densities of forward shock regions and molecular clouds assuming graphite and silicate dust grains. The white region between the dashed magenta lines gives the range of temperatures observed in our SNRs in \autoref{fig:dusttempmaps}. The solid purple line is at 70\,K, above which this relation for collisional heating no longer applies (hence the lines end at $T_d\,=\,70\,K$).
	}
	\label{fig:CollisionalHeatingGraph}
\end{figure*}

In summary, we are more likely to detect dust in remnants where we have good quality ancillary radio, X-ray, optical or {\it Spitzer} 24\,$\mu$m images, that are located at  low galactic latitudes due to both the higher concentration of core-collapse SNRs and bright radio SNRs in these regions, with PWNe due to the additional heating source that raises the SN dust temperatures above the ambient ISM level (similarly for shock heated dusty shells).

\subsection{Interpreting the SNR Dust Temperatures: is warm dust in SNRs collisionally heated?} \label{subsec:CollisionalHeating}
In Section~\ref{sec:tempmaps} we identified warm dust in shocked regions of X-ray and radio emission, suggesting that these interactions are heating this material. We compare models of shock heating to determine the gas properties required to heat the dust to the temperatures seen in \autoref{fig:dusttempmaps}.
In the case of collisional dust heating, the dust temperature in X-ray emitting plasmas can be given by the following expression for $T_{d} \lesssim 70$\,K \citep{Dwek1987}:

\begin{equation} \label{eq:collisionalheating}
	T_{d}^{4} = \frac{(\frac{32}{\pi m_{e}})^{0.5} n_{e} (k T_{gas})^{1.5} h(a, T_{gas})}{4 \sigma <Q(a, T_{d})>}
\end{equation}

where $n_e$ is the electron density (cm$^{-3}$), $a$ is the grain radius ($\mu$m), $T_{gas}$ is the gas temperature (K), $h(a, T_{gas})$ is the grain heating efficiency ($\sim$~1 for most grain sizes), $<Q>$ is the Planck averaged dust absorption coefficient and we assume $<Q(a, T_{d})> = 0.16 a T_{d}^{1.94}$ for graphite and silicates \citep{Dwek1992}. As indicated by \citet{Dwek1987} the dust temperature provides useful constraints on the allowable combinations of gas density and temperature behind the shock.

In \autoref{fig:CollisionalHeatingGraph} we consider 3 cases of electron density similar to that observed in shocked regions of CCSNRs \citep[e.g. for Cas A $n_e\,=\,16$\,cm$^{-3}$ and $n_e\,=\,61$\,cm$^{-3}$ in the hot and cold components respectively,][]{Willingale2003}, and one case of a typical molecular cloud density ($n_e\,\sim\,1000$\,cm$^{-3}$).
If the SNR dust in \autoref{fig:dusttempmaps} is collisionally heated we can rule out small grains ($a\,\lesssim\,0.1$~\micron) for all densities other than 1\,cm$^{-3}$. This would require unexpectedly cool electron gas to give the range of dust temperatures ($\sim25\,-\,40\,$K) where X-ray measurement suggest that typical shocked gas (or electron) temperatures are $\gtrsim\,10^{6}$\,K \citep[e.g.][]{Koo2016}.
For our lowest density ($n_e\,=\,1\,$cm$^{-3}$) the full range of SNR temperatures is available to us with grains of radius 0.5, 1.0, or 5.0\,\micron\ at reasonable gas temperatures of $T_g\,>\,10^6$. Smaller grains could provide the observed results, although this is less likely as the gas temperature would be lower than expected for shocked gas.
For the densest region we find that the surrounding gas must be cool ($T_g\,\lesssim\,10^{5}$\,K) and we must have relatively large dust grains ($a\,>\,1\,$\micron): it is unlikely that this describes our SNRs.
It is possible that these emitting dust grains are located in post-shock regions.

\section{Conclusions} \label{sec:Summary}
We searched for far-infrared (FIR) counterparts of known supernova remnants (SNRs) in the entire Galactic Plane, as surveyed by {\it Herschel} at 70\,--\,500\,$\mu$m, to supplement the first-look catalogue of 71 SNRs studied by \citet{Chawner2019}. Of \numStudied\ sources studied here, we find that \numOne\ (\percOne\,per\,cent) have a clear FIR detection of dust associated with the SNR. When combined with C19, this gives a total of 39 FIR detected sources out of 190 known remnants with $\mid\,b\,\mid\,\leq\,1^{\circ}$ across the entire Galactic Plane (a detection rate of 21\,per\,cent): with dust signatures detected in the remnants of 13 core-collapse supernovae (SNe), including 4 Pulsar Wind Nebulae (PWNe), and 2 Type Ia SNe. A further 24 are detected in sources with unknown types.

Additionally we have shown that:

\begin{itemize}
	\item We tend to detect dust in younger SNRs.
	\item We confirm the detection of ejecta dust within the core-collapse SNR G350.1$-$0.3, as seen in {\it Spitzer} 24\,$\mu$m and now {\it Herschel} 70$\mu$m images. This adds to a sample of only $\sim$\,10 SNRs from which ejecta dust has been observed and indicates that SNe can form dust grains from elements synthesised by the SN explosion.
	We see no evidence of cooler dust in the ejecta nor dust located at the location of the compact central object.
	\item We suggest that the dust features previously proposed to be related to the G357.7$+$0.3 SNR (seen previously with {\it Spitzer}) are dust bubbles associated with young star forming regions, and not with the SNR itself.
	\item We reveal dust associated with G351.2$+$0.1. We propose that the FIR emission from the latter source originates from a PWN (though we cannot conclusively rule out a H{\sc ii} region). We estimate a dust temperature and dust mass for G351.2$+$0.1 of 45.8\,K and $M_d\,=\,0.18\,\rm\,M_{\odot}$.
	\item We identify warm dust in several SNRs with temperatures between $\sim$\, 25 and 40\,K which, if collisionally heated, indicates that there must be relatively cool shocked plasma, or the dust must be made up of large grains ($a\,>\,1\,$\micron). However we caution that there could still be contamination in our apertures with unrelated dust along the line of sight, which we have shown could result in dust temperatures that are biased low. We see that the largest source of uncertainty in a FIR catalogue of SNRs is the result of significant confusion with unrelated dust in the Milky Way.
	\item We estimate that 6 of our sample contain considerable masses of dust, although ISM contamination may bias our dust masses high.
\end{itemize}

\section*{Acknowledgements}
We are grateful to Aya Bamba, and Hiroya Yamuguchi for providing X-ray data for G298.6$-$0.0 and G344.7$-$0.1 respectively, also to Ryan Lau and Matthew Hankins for providing SOFIA data for Sgr A East. We thank Felix Priestley for informative discussions on this topic.

HC, HLG, and PC acknowledge support from the European Research Council (ERC) in the form of Consolidator Grant {\sc CosmicDust} (ERC-2014-CoG-647939).
MJB acknowledges support from the ERC in the form of Advanced Grant SNDUST (ERC-2015-AdG-694520).
MM acknowledges support from an STFC Ernest Rutherford fellowship (ST/L003597/1). {\it Herschel} is an ESA space observatory with science instruments provided by European-led Principal Investigator consortia and with important participation from NASA.
IDL gratefully acknowledges the support of the Research Foundation Flanders (FWO).

This research has made use of data from the HiGAL survey (2012hers.prop.2454M, 2011hers.prop.1899M, 2010hers.prop.1172M, 2010hers.prop.358M) and Astropy\footnote{\url{http://www.astropy.org}}, a community-developed core Python package for Astronomy \citep{astropy2013,astropy2018}.  This research has made use of the NASA/ IPAC Infrared Science Archive, which is operated by the Jet Propulsion Laboratory, California Institute of Technology, under contract with the National Aeronautics and Space Administration.  The scientific results reported in this article are based in part on data obtained from the Chandra Data Archive.

\bibliographystyle{mnras}
\bibliography{library}

\bsp	
\label{lastpage}
\end{document}